\documentclass[a4paper,11pt]{article}
\pdfoutput=1 % if your are submitting a pdflatex (i.e. if you have
\usepackage{jcappub} % for details on the use of the package, please
\usepackage{simpler-wick}
\usepackage{simplewick}
\usepackage[normalem]{ulem}
\usepackage[T1]{fontenc} % if needed
\usepackage{bm}

\newcommand{\vx}{{\bm{x}}}

\usepackage{scalerel,tikz}
\usetikzlibrary{svg.path}
\definecolor{orcidlogocol}{HTML}{A6CE39}
\tikzset{orcidlogo/.pic={
 \fill[orcidlogocol] svg{M256,128c0,70.7-57.3,128-128,128C57.3,256,0,198.7,0,128C0,57.3,57.3,0,128,0C198.7,0,256,57.3,256,128z};
 \fill[white] svg{M86.3,186.2H70.9V79.1h15.4v48.4V186.2z}
 svg{M108.9,79.1h41.6c39.6,0,57,28.3,57,53.6c0,27.5-21.5,53.6-56.8,53.6h-41.8V79.1z M124.3,172.4h24.5c34.9,0,42.9-26.5,42.9-39.7c0-21.5-13.7-39.7-43.7-39.7h-23.7V172.4z}
 svg{M88.7,56.8c0,5.5-4.5,10.1-10.1,10.1c-5.6,0-10.1-4.6-10.1-10.1c0-5.6,4.5-10.1,10.1-10.1C84.2,46.7,88.7,51.3,88.7,56.8z};
}}
\newcommand\orcidicon[1]{\href{https://orcid.org/#1}{\mbox{\scalerel*{
\begin{tikzpicture}[yscale=-1,transform shape]
\pic{orcidlogo};
\end{tikzpicture}
}{|}}}}

\usepackage{booktabs}

\def\araa{\ref@jnl{ARA\&A}}
 
\newcommand\ba{\begin{eqnarray}}
\newcommand\ea{\end{eqnarray}}
\newcommand\be{\begin{equation}}
\newcommand\ee{\end{equation}}

\newcommand{\bl}{\bm{l}}
\newcommand{\bL}{\bm{L}}
\newcommand{\veck}{{\bf k}}
\newcommand{\lin}{{\rm lin}}
\newcommand{\Ncen}{N_{\mathrm{cen}}}
\newcommand{\Ngal}{N_{\mathrm{gal}}}
\newcommand{\Nsat}{N_{\mathrm{sat}}}

\newcommand\gsim{ \lower .75ex \hbox{$\sim$} \llap{\raise .27ex \hbox{$>$}} }
\newcommand\lsim{ \lower .75ex \hbox{$\sim$} \llap{\raise .27ex \hbox{$<$}} }

\newcommand{\vl}{{\bm{l}}}
\newcommand{\vL}{\bm{L}}

\title{\boldmath A halo model of extragalactic contamination to CMB lensing, delensing, and cross-correlations}
\author[a,b]{A. Baleato Lizancos~\orcidicon{0000-0002-0232-6480},}
\author[c,d]{W. Coulton,}
\author[c,d,e]{A. Challinor~\orcidicon{0000-0003-3479-7823},}
\author[c,d]{B. D. Sherwin,}
\author[f]{and Y. Mehta~\orcidicon{0009-0006-5846-6016}}
\affiliation[a]{Berkeley Center for Cosmological Physics, Department of Physics, University of California, Berkeley, CA 94720, USA}
\affiliation[b]{Lawrence Berkeley National Laboratory, One Cyclotron Road, Berkeley, CA 94720, USA}
\affiliation[c]{Kavli Institute for Cosmology Cambridge, Madingley Road, Cambridge, CB3 0HA, UK}
\affiliation[d]{DAMTP, University of Cambridge, Cambridge, CB3 0WA, UK}
\affiliation[e]{Institute of Astronomy, Madingley Road, Cambridge, CB3 0HA, UK}
\affiliation[f]{School of Earth and Space Exploration, Arizona State University, 781 Terrace Mall, Tempe, AZ 85287, U.S.A.}

\emailAdd{a.baleatolizancos@berkeley.edu}

\abstract{CMB lensing reconstructions are a sensitive probe of the growth of structure across cosmic time and a key tool to sharpen investigations of the very early Universe via `delensing'. At present, a large fraction of this information is drawn from the temperature anisotropies, which are ultimately also the most informative when reconstructing lenses on arcminute scales and smaller. But extragalactic foreground emission from galaxies and clusters can contaminate these reconstructions, limiting our ability to use information from small-scale temperature anisotropies. We develop analytic predictions of the biases from the thermal Sunyaev-Zel'dovich and cosmic infrared background to CMB lensing auto- and cross-correlations with low-redshift matter tracers, as well as $B$-mode delensing, based on a halo model for the dominant one- and two-halo contributions to the relevant foreground bi- and tri-spectra. The method is flexible enough to allow variations in cosmology, astrophysical modeling, experimental configurations and analysis choices, thus enabling an improved understanding of the uncertainties involved in current mitigation strategies. We find that the shape of the bias relative to the CMB lensing auto-spectrum signal is remarkably insensitive to changes in cosmological and astrophysical parameter values. On the other hand, the shape appears to depend on $\Omega_m$ for cross-correlations with low-redshift galaxies. We also clarify the ranges of redshifts and masses that simulations need to resolve in order to capture these effects accurately. Our code, \texttt{CosmoBLENDER}, is made publicly available.}

\begin{document}
\maketitle
\flushbottom

\section{Introduction}\label{sec:intro_extragalactic_lensing_biases}
The cosmic microwave background is made up of photons which, for the most part, last scattered with electrons during the era of cosmic recombination; they have been travelling, virtually undisturbed, ever since. Along their journey, however, the photons' paths are deflected by the gravitational influence of the matter distribution of the Universe --- an effect known as `gravitational lensing' (see Ref.~\cite{ref:lewis_challinor_review} for a review). This affects, in crucial ways, the statistical properties of the ensemble: while the unlensed CMB and the lensing potential are both very well approximated as statistically-isotropic, Gaussian random fields, lensing introduces statistical anisotropy in the form of correlations between angular scales.

Measurements of these correlations from the data can in turn be used to reconstruct the lensing potential on the sky --- an integral along the line of sight of the matter distribution responsible for the deflections, going all the way to the last-scattering surface at redshift $z\sim 1100$. In this way, CMB observations can be used to investigate phenomena whose imprint only becomes significant after recombination ---  dark energy, the sum of the neutrino masses, modifications of gravity, and more. With only information from last-scattering, these models would be highly degenerate with their alternatives~\cite{ref:efstathiou_bond_99, ref:smith_cmbpol}. And, unlike other leading cosmological probes, CMB lensing opens a window that is largely insensitive to the poorly-understood physics of galaxy formation, baryonic processes, or non-linear gravitational collapse.

Internal reconstructions of CMB lensing can also be used to delens the CMB. The main application is delensing $B$-mode polarization, an essential step towards unveiling primordial gravitational waves (see, e.g.,~\cite{Seljak:2003pn, ref:polarbear_delensing_19,  baleato_20_internal} and references therein), but delensing temperature and $E$-mode anisotropies can also lead to improved cosmological constraints~\cite{ref:green_17, ref:hotinli_et_al_22}.

For these reasons, CMB lensing reconstructions have become  central to the goals of the CMB community over the past two decades.
The first measurements of the angular power spectrum were obtained by the ACT Collaboration using temperature data~\cite{Das:2011ak}, followed shortly thereafter by SPT~\cite{vanEngelen:2012va} and Planck~\cite{Ade:2013tyw}. A combination of temperature anisotropy observations from the latter two experiments was also used by Ref.~\cite{Omori:2017tae} to reconstruct lensing maps, which were then cross-correlated with DES measurements of galaxy weak lensing, obtaining a high-significance detection of lensing effects~\cite{ref:omori_19}. As the resolution and sensitivity of observations improved, information from CMB polarization was steadily incorporated into analyses and used to improve the reconstruction signal-to-noise, as in Refs.~\cite{Story:2014hni, ref:sherwin_et_al_16, Planck2018:lensing, ref:wu_et_al_19}; in parallel, the POLARBEAR and BICEP2/Keck Array Collaborations measured the lensing power spectrum using only polarization data~\cite{ref:polarbear_14_lensing_ps, B2Keck:2016afx}. More recently, Planck produced almost full-sky maps of the lensing potential, reporting a $40\sigma$ detection of lensing~\cite{Planck2018:lensing}. A similar overall significance was achieved by ACT by producing much deeper lensing maps over approximately
$10\,000\,\text{deg}^2$ of the sky~\cite{ref:qu_et_al_23, ref:madhavacheril_et_al_23}. In parallel, SPT-3G has also produced competitive results from observations of a $1500\,\text{deg}^2$ patch from the South Pole~\cite{ref:pan_et_al_23, ref:millea_et_al_21_spt, geCosmologyCMBLensing2024a}. Going forward, ground-based telescopes probing the microwave sky with arcminute-scale resolution promise to improve on this. The Simons Observatory~\cite{ref:SO_science_paper} will soon produce reconstructions that are signal-dominated for multipoles $L\lesssim250$ over 40\% of the sky on a region similar to that covered by ACT. Longer term, CMB-S4~\cite{ref:s4_science_book} will do the same for $L\lesssim1000$~\cite{ref:s4_science_book}. Before then, the South Pole Observatory (SPO) will achieve similar lensing depth to CMB-S4, but only over a few percent of the sky.

In addition to the information obtained from their auto-spectra, CMB lensing reconstructions are also powerful probes in cross-correlation with lower-redshift tracers of the large-scale structure (see, e.g.~\cite{ref:schmittfull_and_seljak}), allowing for a tomographic decomposition of the lensing map. The most obvious application of CMB lensing cross-correlations is, perhaps, in constraining cosmological parameters; for example, by cross-correlating with galaxy lensing~\cite{ref:hand_15, ref:liu_and_hill_15, ref:kirk_16, ref:harnois_deraps_16, ref:harnois_deraps_17_cmb_cross_galaxy_lensing, ref:miyatake_17, ref:omori_19a, ref:namikawa_et_al_19, ref:marques_20, ref:robertson_20, shaikhCosmologyCrossCorrelationACTDR42024} or with biased tracers of the mass distribution. The latter include cross-correlations of CMB lensing with the number density of galaxies~\cite{ref:smith_07, ref:hirata_08, ref:bleem_12, ref:allison_15, ref:bianchini_15, ref:giannantonio_16, ref:peacock_18, ref:omori_19,ref:wilson_and_white, ref:krowlewski_20, ref:darwish_21, ref:krowlewski_et_al_21, ref:kitanidis_and_white_21, ref:white_et_al_22, modiModelingCMBLensing2017, quAtacamaCosmologyTelescope2024, farrenAtacamaCosmologyTelescope2024, kimAtacamaCosmologyTelescope2024, sailerCosmologicalConstraintsCrosscorrelation2024, chenCosmologicalAnalysisThreeDimensional2022}, quasars~\cite{ref:sherwin_12, ref:geach_13}, the cosmic infrared background (CIB)~\cite{ref:holder_13, ref:planck_13_cib, ref:van_engelen_15, ref:lenz_19}, galaxy groups~\cite{ref:madhavacheril_15} and galaxy clusters~\cite{ref:baxter_18, ref:madhavacheril_20}, which can be used to study the biasing properties of those tracers, but also to break the degeneracy between bias and the growth rate of structure. Cross-correlations with the Sunyaev-Zel'dovich effects~\cite{ref:van_waerbecke_14, ref:hill_spergel_14, ref:hojjati_17} contain information about the distribution and properties of gas in halos. Cross-correlations with biased tracers on large scales can be used to probe primordial non-Gaussianity via the scale-dependence of halo bias~\cite{ref:dalal_08}. Furthermore, cross-correlations can partially get around the impairing reliance --- see Ref.~\cite{ref:allison_et_al_15} --- on knowledge of the optical depth to reionization when constraining the sum of the neutrino masses~\cite{ref:yu_et_al_18}. Finally, CMB lensing reconstructions can also be used to calibrate the mass of galaxy clusters and from them place constrains on cosmology via the mass function (e.g., Ref.~\cite{zubeldia_cosmological_2019})\footnote{CMB lensing reconstructions have, of course, many more uses. Since they are sensitive to a set of systematic effects that is for the most part different from optical lensing surveys, another possibility is to use them to calibrate the multiplicative shear bias of galaxy weak-lensing surveys~\cite{ref:vallinotto_12, ref:das_shear_cal, ref:liu_et_al_16, ref:schaan_shear_cal, ref:singh_shear_cal, ref:harnois_deraps_17,ref:robertson_20}. Though CMB lensing reconstructions can be subject to additive biases (the likes of $N^{(0)}, N^{(1)}$, etc.), the standard weights employed guarantee that it will be correctly normalised --- i.e., that it will have no multiplicative bias. Calibrating multiplicative bias this way promises to be particularly effective at high redshift, where it is also most useful, since it is at high redshift that simulations of the galaxy population are most uncertain, making image-based calibration difficult. At the very least, CMB lensing will be a powerful complementary tool to cross-check other calibration methods; see Ref.~\cite{ref:mandelbaum_review} for a review.}.

\subsection{CMB lensing reconstruction biases and their mitigation}
The quadratic estimator of Ref.~\cite{ref:hu_okamoto_02} is the lowest-order (in the small deflection angles), optimally-weighted lensing estimator. In essence, this tool extracts information from off-diagonal elements of the covariance between pairs of CMB fields. The power spectrum of such a lensing reconstruction is therefore a linear functional of the trispectrum of the lensed CMB anisotropies, which is a particularly sensitive probe of the lensing effect~\cite{ref:bernardeau_97}. If sensitive enough observations of the CMB polarization are available, the statistical power of a temperature-based reconstruction can be expanded by combining it with other quadratic combinations of different pairs of fields, in a way that minimises the variance of the reconstructed map~\cite{ref:maniyar_et_al_21, ref:hirata_seljak_pol}.

Much work has been devoted to studying how auto- and cross-correlations of lensing reconstructions obtained using quadratic estimators are complicated by systematic effects. Some particularly critical areas are the $N^{(0)}, N^{(1)}$ and higher-order biases of lensing~\cite{ref:kesden_et_al_03,ref:hanson_et_al_11, shenCMBLensingPower2024}; the coupling to instrumental effects~\cite{ref:hanson_et_al_09, ref:smith_cmbpol, ref:mirmelstein_et_al_20, ref:nagata_and_namikawa_21, ref:fabbian_peloton_21}; the challenges posed by limited sky coverage~\cite{vanEngelen:2012va, Namikawa:2013, ref:benoit_levy_13}; the impact of non-Gaussianity from Galactic foregrounds~\cite{ref:fantaye_et_al, ref:core_lensing, ref:beck_et_al_20} or from the very early Universe~\cite{ref:lesgourgues_et_al_05, ref:merkel_et_al_13}; and biases due to the non-linear evolution of structure and post-Born lensing~\cite{ref:boehm_et_al_16, ref:beck_et_al_18, ref:boehm_et_al_18, ref:pratten_and_lewis, ref:fabbian_et_al_19}. In addition to characterizations of possible biases, these recent efforts have provided a useful set of tools to mitigate them; with those methods in hand, lensing reconstructions are rather clean, particularly when only polarization data is used, in which case there is effectively no contamination from extragalactic sources (e.g.,~\cite{quImpactMitigationPolarized2024a}).

However, for as long as a significant fraction of the lensing reconstruction signal-to-noise comes from the temperature anisotropies, contamination from extragalactic foregrounds will pose a major challenge. This will be the case for current and soon-upcoming experiments such as AdvACT~\cite{ref:advACT} or the Simons Observatory~\cite{ref:SO_science_paper}, but also for future, ultra-low-noise experiments such as the proposed CMB-HD~\cite{ref:sehgal_et_al_19} striving to measure small-scale lensing modes with $L\sim O(10,000)$~\cite{ref:nguyen_et_al_19}, where temperature is always the dominant source of information. The challenge lies in the fact that at microwave frequencies, there is substantial emission from radio-bright galaxies, from the thermal and kinetic Sunyaev-Zel'dovich effects (tSZ and kSZ, respectively), and from the CIB\footnote{Since these foregrounds are polarised by a smaller fraction than the CMB is, polarization-based lensing reconstructions are more robust to their impact than temperature-based ones are~\cite{ref:smith_cmbpol}.}.

The contributions of extragalactic foregrounds to the observed anisotropies have acutely non-Gaussian statistics --- Refs.~\cite{ref:wilson_et_al_12, ref:crawford_et_al_14,ref:planck_13_compton_polyspectra} and \cite{ref:crawford_et_al_14, ref:planck_13_cib} have measured the tSZ and CIB bispectra, respectively ---  which can potentially confuse the lensing estimators. Early work by Ref.~\cite{ref:amblard_et_al_04} using simulations suggested that the kSZ alone could bias the power spectrum of lensing reconstructions by $10\text{--}200\%$. Similar warnings were issued in Refs.~\cite{ref:smith_07, ref:bleem_12, vanEngelen:2012va} about biases arising from the CIB. These estimates were subsequently refined and extended using simulations by Refs.~\cite{ref:van_engelen_15, ref:osborne_et_al, ref:sailer_et_al, 2025JCAP...09..048D}, who harnessed the improved understanding of the microwave sky enabled by the observations of SPT, ACT, and Planck. They showed that a wide range of extragalactic foregrounds --- tSZ, CIB, radio sources --- can bias CMB lensing auto- and cross-spectra at the level of several percent, even after detectable point sources have been masked. Reference~\cite{ref:ferraro_hill_18} then completed the picture by providing a detailed study of biases associated with the kSZ. More recently, Ref.~\cite{ref:baleato_and_ferraro_22} showed that if unaccounted for, similar biases can severely hinder efforts to delens $B$-mode polarization, and provided a prescription for modelling the bulk of the effects. These extragalactic contaminants --- in particular, the tSZ and the CIB --- are probably the greatest impediment to extracting lensing information from the small-scale temperature anisotropies of the CMB.

Several techniques have been suggested in the literature to mitigate lensing biases from extragalactic foregrounds. One approach, used in the analyses of Refs.~\cite{ref:sherwin_et_al_16, ref:wu_et_al_19}, is to subtract from measured spectra the bias templates calculated by Ref.~\cite{ref:van_engelen_15} using the simulations of Ref.~\cite{ref:sehgal_sims_10}. If the templates were accurate, this approach would have the benefit of not incurring any loss in signal-to-noise. In reality, the templates are known only approximately, so the biases are, at best, only partially corrected for.

Another option, known as bias hardening~\cite{ref:osborne_et_al, ref:namikawa_et_al_13, ref:sailer_et_al}, entails using knowledge of the statistics of the foregrounds (in particular, their power spectrum and bispectrum) to engineer quadratic estimators that optimally extract the foreground contribution at leading order. It is then possible to form a linear combination of this foreground reconstruction and the lensing estimators that extensively reduces the bias. This method is very effective at mitigating the contaminants while only incurring a small noise penalty; as such, it featured in the analysis of Planck~\cite{ref:planck_15_lensing, Planck2018:lensing} and ACT data~\cite{quAtacamaCosmologyTelescope2024}. Strictly-speaking, however, the current implementations of bias-hardening only null the `primary bispectrum' bias (we will explain this nomenclature shortly) in the limit where the foreground bispectrum is indestinguishable from one where a foreground leg has been replaced with CMB lensing, and the `trispectrum' bias if the sources are Poisson-distributed with identical profiles; lastly, they are in general not effective at mitigating the `secondary bispectrum' bias~\cite{ref:sailer_et_al}.

Multi-frequency cleaning techniques (see, e.g.,~\cite{ref:tegmark_efstathiou_96, ref:tegmark_et_al_03, ref:remazeilles_et_al_11, ref:remazeilles_contrained_ilc,ref:sailer_et_al_21, ref:darwish_et_al_21}) will be useful in mitigating these biases, but they will be handicapped by the fact that for the foreseeable future, the high-resolution observations required for lensing reconstruction will only be available at a handful of frequency channels accessible from the ground. It has recently been suggested that applying frequency cleaning to only the gradient leg\footnote{This terminology makes reference to the real-space formulation of the quadratic estimators, equation~\eqref{eqn:real_space_tt_qe}.} of the quadratic estimator might incur a smaller loss in lensing signal-to-noise than doing it to both legs (at least on the scale of clusters)~\cite{ref:madhavacheril_and_hill_18, ref:darwish_21}. The reason for this is that contributions from the primary CMB to the local gradient of CMB anisotropies saturate by $l\approx 2000$ due to Silk damping, so that there is little cost associated with increasing the noise on scales smaller than that.

Finally, it is worth mentioning the foreground-immune estimators of Refs.~\cite{ref:schaan_ferraro_18, ref:qu_et_al_2023_shear}. By performing a harmonic expansion of the weights of the quadratic estimator, one can show that most of the signal-to-noise is in the first two moments --- the monopole and the quadrupole. In the limit of large-scale lenses, these correspond to extracting the convergence and shear, respectively. Importantly, to the extent that the source emission profiles are azimuthally-symmetric, only the monopole is affected by foregrounds. Consequently, the shear-only estimator can be used to obtain lensing reconstructions that are highly robust to foregrounds; although it has only around $ 60\%$ of the signal-to-noise of the full quadratic estimator, this method can incorporate smaller-scale data while limiting biases.

\subsection{What this work provides}

In this paper, we present a tool, complementary to the ones just described, to model the biases to CMB lensing reconstructions arising from the tSZ and the CIB, which are expected to be the main sources of bias. We do so analytically by developing a halo model for the dominant one- and two-halo contributions to the relevant bi- and trispectra, and propagating them through the lensing reconstructions\footnote{The reason we do not study the kSZ for now is that it receives important contributions from material outside halos; extensions of our method are therefore required to properly deal with the kSZ. Radio galaxies, on the other hand, largely behave as Poisson sources, so they can be understood with even simpler treatments.}. 

The result is a flexible framework where we can quickly calculate these biases as a function of experimental parameters such as sensitivity, resolution, frequency channels, filtering and point-source masking schemes. This can help optimise experiment design and analysis choices. Moreover, we can evaluate predictions in different cosmological and astrophysical models, enabling a new understanding of the uncertainties implicit to our characterization of these effects, which at present is derived from a very limited set of simulations of the microwave sky~\cite{Sehgal:2016eag, ref:websky, ref:wagoner_et_al_67}, each with a fixed cosmology and foreground model. This is particularly important at a time when our standard understanding of feedback processes that redistribute baryonic matter within halos is being challenged by observations~\cite{schaanAtacamaCosmologyTelescope2021, amonNonlinearSolutionS82022, hadzhiyskaEvidenceLargeBaryonic2024, efstathiouPowerSpectrumThermal2025}. In addition, our calculations can help inform the next generation of simulations by placing requirements on simulation hyperparameters, such as redshift ranges and mass resolutions. Combining these with a newfound understanding of the intricacies of the dominant clustering configurations that engender the biases can inform mitigation techniques, such as bias hardening, and shed light on their expected regime of validity.

Our calculation framework can also be used to predict the expected biases when cross-correlating CMB lensing reconstructions with low-redshit probes of the large-scale structure of the Universe. In this paper, we will provide one such example where CMB lensing is cross-correlated with a galaxy number density tracer implemented via a halo-occupation-distribution (HOD) prescription. Finally, the framework can also be used to predict the impact of extragalactic foregrounds on delensing of $B$-mode polarization when using a $TT$ quadratic estimator reconstruction as the matter tracer. This is important not only to minimise potential bias, but also to determine the analysis choices that will maximize the sensitivity to a primordial component after delensing~\cite{ref:baleato_and_ferraro_22}.

This paper is organised as follows. We begin with an overview of the tSZ effect and the CIB in section~\ref{sec:extrag_fgs}. This initial section is meant to be pedagogical and can be skipped by a reader familiar with these phenomena.  Our new contributions begin in section~\ref{sec:intro_to_lensing_biases}. First, we introduce the extragalactic foregrounds biases to CMB lensing reconstruction auto- and cross-correlations, as well as $B$-mode delensing. Then, in section~\ref{sec:analytic_framework}, we describe the analytic framework for computing those biases using the halo model. (An extensive introduction to the halo model and its application to the CIB is provided in appendix~\ref{sec:halo_model}.) In section~\ref{sec:results}, we present key results of our calculations, beginning with CMB lensing auto-spectra in section~\ref{sec:results_auto}, followed by cross-spectra in section~\ref{sec:results_cross} (including a subsection on the origin of the zero-crossing of the primary bispectrum bias) and delensing in section~\ref{sec:results_delensing}. In section~\ref{sec:hyperparam_requirements} we investigate mass resolution and redshift requirements for the computations to converge. This is followed by an investigation of the impact of varying cosmological and astrophysical parameters in section~\ref{sec:results_varying_params}. We compare our analytic calculations to measurements from simulations in section~\ref{sec:comparison_with_websky}.
Finally, in section~\ref{sec:outlook}, we conclude and summarise our findings. Our code, named \texttt{CosmoBLENDER} (Cosmological Biases to LENsing and Delensing due to Extragalactic Radiation), is made publicly available on \texttt{Github}\footnote{\url{https://github.com/abaleato/CosmoBLENDER}}.

Unless stated otherwise, we assume a flat $\Lambda$CDM cosmology similar to \texttt{Planck 2018}~\cite{ref:planck_params_18} with $\Omega_{\mathrm{cdm}}=0.31$, $\Omega_{\mathrm{b}}=0.049$, $n_{\mathrm{s}}=0.96$, $A_{\mathrm{s}}=2.08\times 10^{-9}$,  $\tau=0.05$, $H_0 = 68\,\mathrm{km\,s}^{-1}\,\mathrm{Mpc}^{-1}$ and massless neutrinos.

\section{Extragalactic foregrounds}\label{sec:extrag_fgs}
\subsection{The thermal Sunyaev-Zel'dovich effect}\label{sec:tsz}
A fraction of the CMB photons we observe will have followed trajectories that took them through galaxy clusters --- massive, gravitationally-bound regions with many tens of galaxies in a volume only a few megaparsecs in diameter. These objects have masses ranging from the $10^{12}\,\mathrm{M}_{\odot}$  of ``galaxy groups'' to the $10^{15}\,\mathrm{M}_{\odot}$ of the most massive clusters. In addition to galaxies, these clusters contain dark matter, and, crucially for our purposes, gas.

This gas is mostly hydrogen and helium, with traces of metals, and it is smoothly distributed across the intra-cluster medium (ICM). X-ray measurements have determined it to be very hot, with temperatures of order $10^8$\,K, consistent with the temperature expected if the gas particles are in virial equilibrium in the gravitational potential of the cluster~\cite{ref:kravtsov_and_borgani_12}.

At such high temperatures, the gas is largely ionised, and the hot, free electrons inverse-Compton-scatter with CMB photons, which are comparatively much colder (around $0.3\,$meV at $z=0$, compared to electron energies of a few keV in the ICM of clusters). On average, photons receive an energy boost, which results in a $y$-type distortion of the CMB frequency spectrum; the \emph{thermal Sunyaev-Zel'dovich effect}, or tSZ~\cite{ref:sunyaev_zeldovich_72} (for reviews, see, e.g.,~\cite{ref:rephaeli_review_95, ref:carlstrom_et_al_02, ref:mroczkowski_et_al_19}). Photons are thus systematically shifted from the Rayleigh--Jeans side to the Wien side of the spectrum, which translates to a temperature decrement at frequencies below 217\,GHz, and an increment at frequencies higher than this threshold. It is thanks to this effect that clusters can be easily identified in observations of the CMB temperature anisotropies on arcminute scales, appearing as shadows at low frequencies, or as bright spots at high frequencies. Though in principle there can also be a signature in polarization, the effect is expected to be small, of $O(1\%)$ of the amplitude in intensity~\cite{ref:sazonov_sunyaev_99, ref:carlstrom_et_al_02}.

The resulting CMB temperature deviation due to a cluster of mass $M$ and redshift $z$ in the direction of $\bm{\theta}$ is\footnote{Equation~\eqref{eqn:T_change_due_to_tsz} ignores relativistic corrections, which could only be relevant for the most massive clusters in the Universe, with masses in excess of a few times $10^{14}M_\odot$ and able to support electron energies of $k_{\mathrm{B}}T_{\mathrm{e}}\gtrsim10\,$keV~\cite{ref:itoh_et_al_98, ref:challinor_lasenby_98, ref:nozawa_et_al_06}. We do not include relativistic corrections in our analytic treatment for two reasons: first, because these objects are likely to be masked in actual analyses, and second, because the `Battaglia' profile, the cornerstone of our calculation, was fit to simulations that ignored such corrections. The \texttt{WebSky} and \texttt{Agora} simulations, against which we validate our calculations in section~\ref{sec:comparison_with_websky}, do not feature them either.}
\begin{equation}\label{eqn:T_change_due_to_tsz}
    \frac{\Delta T (\bm{\theta}, M, z)}{T_{\text{CMB}}}  = g_{\nu}y(\bm{\theta}, M, z) \,.
\end{equation}
Here, we have split the effect into a characteristic frequency signature, $g_{\nu}= x/\tanh(x/2)-4$, with $x=h\nu/k_{\mathrm{B}}T_{\text{CMB}}$ ($k_{\mathrm{B}}$ is Boltzmann's constant), multiplied by the Compton-$y$ parameter,
\begin{equation}
    y(\bm{\theta}, M, z)  = \frac{\sigma_{\mathrm{T}}}{m_{\mathrm{e}}c^2} \int_{\mathrm{l.o.s.}} P_{\mathrm{e}}\left(\sqrt{l^2 + d_{\mathrm{A}}^2 |\bm{\theta}|^2}, M, z\right)\mathrm{d}l\,,
\end{equation}
which is an integral of the electron pressure, $P_{\mathrm{e}}$, along the line of sight ($d_{\mathrm{A}}$ is the angular diameter distance). In a fully-ionised medium with primordial hydrogen fraction $X_{\mathrm{H}}=0.76$, the electron pressure is related to the gas pressure, $P_{\mathrm{th}}$, by $P_{\mathrm{th}} = P_{\mathrm{e}} (5X_{\mathrm{H}}+3)/[2(X_{\mathrm{H}}+1)]=1.932 P_{\mathrm{e}}$. Typical values for the $y$-parameter associated with clusters are $y\sim10^{-4}$, so the tSZ effect can produce anisotropy of order 1\,mK in the Rayleigh-Jeans part of the CMB spectrum; this is comparable in amplitude to the primary CMB anisotropies~\cite{ref:mroczkowski_et_al_19}.

It is worth noting that free electrons residing outside clusters, in filaments of the cosmic web, can also produce tSZ emission (their contribution may in fact have recently been detected at moderate significance~\cite{ref:de_graaff_et_al_19, 2019MNRAS.483..223T, 2020A&A...637A..41T, 2025arXiv250708561L}). However, since electron pressure is proportional to the product of electron number density and temperature, the tSZ contribution from the `field' is tiny compared to that originating from clusters, the densest and hottest large-scale structures in the Universe. The predominance of the most massive clusters is captured by the scaling relation % $\Delta T\propto M^{5/3}$
$\Delta T\propto M$, where $M$ is the mass of the cluster (see, e.g.,~\cite{ref:kravtsov_and_borgani_12}).

Measurements of the tSZ effect can be used to extract valuable cosmological information, either from its angular clustering statistics~\cite{ref:komatsu_seljak_02}, or by relating tSZ-detected clusters to the underlying halo mass function\footnote{As we review in appendix~\ref{sec:hmf}, the mean number of clusters can be predicted as a function of redshift and mass in different cosmologies; this can then be contrasted with the abundance of clusters detected via their tSZ signature. The comparison requires relating the tSZ amplitude to cluster mass --- for instance by calibrating against X-ray observations (e.g.,~\cite{ref:planck_16_sz_cosmo}) or lensing of galaxies (e.g.,~\cite{ref:bocquet_et_al_2024_clusterlensing}) or the CMB lensing (e.g.,~\cite{ref:zubeldia_challinor_19}) --- and obtaining cluster redshifts by means of follow-up observations. Since the temperature decrement they induce is independent of their distance to us, clusters can be identified out to high redshifts, granted the angular resolution is sufficient. In principle, this could enable the detection of clusters out to the highest redshifts where they exist; in practice, the numbers are indeed growing rapidly, with SPT, Planck and ACT each having catalogued upwards of 600~\cite{Bleem:2014iim}, 1200~\cite{ref:planck_16_sz_cluster_catalog} and 4000~\cite{ref:hilton_et_al_21} SZ-detected clusters, respectively. These have enabled tight constraints on the growth of structure as function of time; see Ref.~\cite{ref:zubeldia_challinor_19} and references therein.}. In this work, however, we will be concerned with the tSZ effect not as the signal of interest, but as a source of bias to CMB lensing reconstructions. We will strive to model those biases analytically, using the halo model described in appendix~\ref{sec:halo_model}. It is well known --- and we will corroborate this --- that the most massive clusters have an outsize impact on these lensing biases~\cite{vanEngelen:2013rla}. For this reason, clusters are typically masked before the CMB maps are fed through the lensing reconstruction pipeline. In our calculations, we will often mimic this approach by restricting the maximum mass of the halos allowed to feature in the calculations.

Our ability to carry out this modeling will hinge on being able to parametrise accurately the pressure profile of clusters as a function of cluster mass and redshift. The profile is in principle sensitive to the rich gastrophysics of the ICM: AGN and supernova feedback, radiative cooling and star formation are but a few of the phenomena that affect it. In practice, the sensitivity to feedback processes and phenomena happening at the core of clusters depends on the mass and redshift of the halos that matter most (e.g.,~\cite{ref:komatsu_seljak_02, efstathiouPowerSpectrumThermal2025}).

Analytic efforts to characterize the pressure profile have been popular in the literature. These approaches necessitate several assumptions: spherical symmetry, some type of hydrostatic equilibrium possibly allowing non-thermal pressure terms\footnote{Thermal pressure is the dominant type of pressure in the majority of clusters, especially those that have not undergone any recent merges~\cite{ref:pratt_et_al_19} (and the only type, in fact, if the ICM gas is completely thermalised); but non-thermal contributions are also possible. There is growing evidence, coming from hydrodynamical simulations, that the main source of non-thermal pressure in clusters is turbulence, accounting for 10--30\% of the total pressure at $R_{500}$~\cite{ref:nelson_et_al_14, ref:shi_et_al_16}, the radius within which the average density is 500 times the mean matter density of the universe.}, and a relationship between pressure and density to populate halos with baryons given an underlying dark matter profile~\cite{ref:komatsu_seljak_02, ref:shaw_et_al_10}. Semi-analytic models share with analytic ones the underlying assumptions of hydrostatic equilibrium and some pressure--density relation, but they replace mass functions with simulations of the dark matter distribution onto which baryons are then pasted~\cite{ref:sehgal_sims_10, ref:trac_et_al_11}.

However, there is growing evidence that these fundamental assumptions do not accurately describe real clusters: hydrostatic equilibrium is violated in clusters that have undergone mergers and in the outskirts of all clusters; and fixing a pressure--density relation is also inaccurate, as this relation is in fact a function of radius, and it is affected by non-thermalised bulk flows which provide pressure support in the outer parts of the cluster but do not contribute to the tSZ~\cite{ref:battaglia_et_al_12}.

These factors are naturally taken into account in both empirical and simulation-based characterizations of the pressure profile. Observational constraints have been important in recent times (see, e.g.,~\cite{ref:arnaud_et_al_10, ref:pointecouteau_et_al_21}), revealing an apparent universality of pressure profiles in mass and redshift. However, it is difficult to obtain data extending to large cluster radii. For this reason, simulations enjoy a particularly prominent position in the literature. Here, we harness the work of Ref.~\cite{ref:battaglia_et_al_12}, who fit the average, normalised thermal pressure profile, $\bar{P}_{\rm th} = \langle P_{\mathrm{th}}/P_{200}\rangle$,\footnote{Here, $P_{200} = 200 G M_{200} \rho_{\mathrm{crit}}(z) f_{\mathrm{b}} / (2R_{200})$, with $f_{\mathrm{b}} = \Omega_{\mathrm{b}}/ \Omega_{\mathrm{m}}$, is the self-similar amplitude for pressure~\cite{ref:kaiser_86, ref:voit_05, ref:bertschinger_85, ref:fillmore_goldreich_84}.} in hydrodynamical simulations --- featuring AGN and supernova feedback, radiative cooling and star formation processes --- with a generalised NFW profile of the form
\begin{equation}\label{eqn:battaglia_profile}
    \bar{P}_{\mathrm{fit}} = P_0 \left(x/x_c\right)^{\gamma} \left[ 1 + \left(x/x_c\right)^{\alpha}\right]^{-\beta}\quad,\quad x\equiv r/R_{200}\,.
\end{equation}
In addition to the parameters of the standard NFW profile, $\alpha$ and $\gamma$, this functional form introduces a core scale parameter, $x_c$, an amplitude, $P_0$, and a power-law index, $\beta$, which parametrises the fall-off of the profile. Reference~\cite{ref:battaglia_et_al_12} found that the parametric form above sufficed to fit the simulations with better than 5\% accuracy, out to a radius $R\lesssim2R_{200}$, and over a wide range of cluster masses and redshifts; performing better than the naïve profile prediction associated with the self-similar solution to the equations of gravitational collapse (which ignore the radiative gas physics).

However, Ref.~\cite{ref:battaglia_et_al_12} also noticed a significant dependence of the best-fit parameters on mass and redshift. This dependence can be incorporated into the model by fixing $\alpha=1.0$ and $\gamma=-0.3$, and casting the remaining parameters as separable functions of mass and redshift,
\begin{equation}\label{eqn:generalising_battaglia_profile}
    A = A_0 \left(\frac{M_{200}}{10^{14}\mathrm{M}_{\odot}}\right)^{\alpha_{\mathrm{m}}} \left(1+z\right)^{\alpha_z},
\end{equation}
where $A$ can be any of $P_0$, $\beta$ or $x_c$. The best-fit parameters determined by Ref.~\cite{ref:battaglia_et_al_12} are shown in table~\ref{tab:battaglia_fit_params}. We will refer to this combination of fitting form and parameters as the `Bataglia profile'.

\begin{table}
    \centering
    \begin{tabular}{l l l l }
        \toprule
        Parameter & $A_0$ & $\alpha_m$ & $\alpha_z$ \\ \midrule
        $P_0$ & 18.1 & 0.154 & -0.758\\ 
        $x_c$ & 0.497 & -0.00865 & 0.731 \\
        $\beta$ & 4.35 & 0.0393 & 0.415\\
        \bottomrule
    \end{tabular}
    \caption[Mass and redshift fit parameters for the pressure profile of Ref.~\cite{ref:battaglia_et_al_12}]{Mass and redshift fit parameters for the pressure profile parametrisation of equations~\eqref{eqn:battaglia_profile} and~\eqref{eqn:generalising_battaglia_profile}, as obtained by Ref.~\cite{ref:battaglia_et_al_12}.}\label{tab:battaglia_fit_params}
\end{table}

With these parameter values, the predicted pressure profiles were found to deviate from the simulated ones by less than $10\%$ over a wide range of redshifts and masses. There was also good agreement at the level of the tSZ power spectrum between analytic calculations based on the Battaglia profile and simulations (see figure~5 of~\cite{ref:battaglia_et_al_12}).  Recently, the Battaglia profile has also passed several observational tests: X-ray observations of low-$z$ clusters~\cite{ref:arnaud_et_al_10}, observations of massive clusters with Bolocam~\cite{ref:sayers_et_al_13} and Planck~\cite{ref:planck_sz_profiles}, and explorations of less massive galaxy groups using the stacked SZ signal from locally brightest galaxies (LBGs)~\cite{ref:greco_et_al_15}. Given its ongoing success, we will use the Battaglia profile to model the pressure profile of clusters in section~\ref{sec:intro_to_lensing_biases}, where we use an extension of the halo model to calculate biases to CMB lensing reconstrucions due to the tSZ effect.

\subsection{The cosmic infrared background}\label{sec:intro_to_cib}
Stars in actively-star-forming galaxies emit UV light. This radiation can be absorbed by dust grains in the intragalactic medium of those very galaxies, which in turn re-emit the energy in the infrared. The ensemble of infrared radiation produced in this way by galaxies across the redshift range where star formation takes place is called the cosmic infrared background (CIB). Since the dust is at a temperature of a few tens of Kelvin, CIB emission peaks in the sub-millimetre part of the electromagnetic spectrum and can be measured at CMB frequencies. Although theoretical predictions of its existence can be found as early as the 1950s (see~\cite{ref:hauser_dwek_01} for a historical review), the challenges posed by atmospheric absorption of infrared light, together with the faintness of the CIB relative to foreground emission from our Galaxy and the Solar System, meant that a direct detection~\cite{ref:puget_96} was not possible until the era of COBE~\cite{ref:smoot_92} and IRTS~\cite{ref:murakami_et_al_94}.

The CIB is highly isotropic on large scales, owing to its extragalactic origin, but some degree of anisotropy was always expected. Since the CIB originates from sources that are intrinsically discrete (though most of them are unresolved), any fluctuations in the number of galaxies along different lines of sight will result in angular fluctuations in the measured brightness on small scales. Moreover, galaxies trace the underlying matter distribution, so the CIB will display large-scale correlations, to the extent that they exist in the matter field~\cite{ref:haiman_knox_00, ref:knox_et_al_01}. These anisotropies have indeed been detected at high Galactic latitudes~\cite{ref:lagache_puget_00, ref:kashlinsky_odenwald_00, ref:matsumoto_00}, and mapped with precision by a number of experiments, most notably by Planck~\cite{ref:planck_13_cib, ref:gnilc, ref:mak_17} using its 353, 545 and 857\,GHz channels.

The CIB anisotropies depend upon the redshift–dependent spectral energy distributions (SEDs) of star-forming galaxies, the relationship between the luminosity of a galaxy and the mass of its host halo, and the clustering properties of galaxies relative to those of the dark matter (the galaxy 'bias' we address in appendix~\ref{sec:halo_bias}). By studying the CIB fluctuations, these interesting variables can be investigated much more extensively than would be possible using only individually-detected star-forming galaxies, which are difficult to observe at high redshift (see, e.g.,~\cite{ref:maniyar_18, ref:planck_13_cib, ref:hermes_viero_wang}). In order to derive constraints, a theoretical model for the CIB emission must be written down and contrasted with observations. In appendix~\ref{sec:halo_model_cib}, we describe a state-of-the-art CIB halo model which has featured in a number of recent investigations~\cite{ref:shang_et_al_12, ref:hermes_viero_moncelsi, ref:planck_13_cib}. Then, in section~\ref{sec:intro_to_lensing_biases}, we extend it to model bispectra and trispectra of the CIB, and calculate the biases that these induce on CMB lensing reconstructions.

We expect the CIB to bias CMB lensing reconstructions primarily through its correlation with lensing\footnote{As we will soon see, a non-Gaussian CIB with a non-vanishing trispectrum would bias CMB lensing reconstructions even if it were not correlated with lensing.}. Star formation happens in a fairly broad range of redshifts, peaking at approximately $z\sim 2$, but extending as far back as $z \sim 4$~\cite{ref:hermes_bethermin, ref:hermes_viero_wang, ref:hermes_viero_moncelsi}. This overlaps extensively with the distribution of structures responsible for the CMB lensing deflections, to the point that the CIB and CMB lensing are up to $80\,\%$ correlated \cite{ref:song_2003,ref:holder_13, ref:planck_lensing_cib_cross}.\footnote{For this reason, the CIB has played a key role in early implementations of delensing (see, e.g.,~\cite{ref:sherwin_15, ref:larsen_16, ref:bicep_delensing, ref:cib_delensing_biases}).}

\section[Impact of extragalactic foregrounds on CMB lensing science]{Impact of extragalactic foregrounds on CMB lensing science}\label{sec:intro_to_lensing_biases}
The temperature--temperature ($TT$) quadratic estimator of Ref.~\cite{ref:hu_okamoto_02} can be written in the flat-sky limit as
\begin{equation}\label{eqn:tt_qe}
    \hat{\phi} (\vL) = A^{TT}_{L} \int \frac{\mathrm{d}^2\bm{l}}{2\pi}T^{\mathrm{obs}}(\bm{l}) T^{\mathrm{obs}}(\vL-\bm{l}) g(\bm{l},\vL)\,,
\end{equation}
where $T^{\mathrm{obs}}(\bm{l}) $ is the Fourier transform of the observed CMB temperature, which comprises the lensed CMB signal, experimental noise, and foreground emission. We denote its angular power spectrum as $C_l^{TT,\mathrm{obs}}$. If we choose the weight function $g(\bm{l},\vL)$ to be
\begin{equation}\label{eqn:QE_weights}
    g(\bm{l},\vL) =  \frac{(\vL - \vl)\cdot \vL \tilde{C}^{TT}_{|\vl-\vL|} + \vl\cdot \vL \tilde{C}^{TT}_l}{2C_l^{TT,\mathrm{obs}} C_{|\bl - \bL|}^{TT,\mathrm{obs}}}\,,
\end{equation}
and the normalisation $ A^{TT}_{L}$ as
\begin{equation}\label{eqn:QE_norm_TT}
    A^{TT}_{L} = \left[2 \int \frac{\mathrm{d}^2\bl }{(2\pi)^2} C_l^{TT,\mathrm{obs}} C_{|\bl - \bL|}^{TT,\mathrm{obs}} \left[g(\bm{l},\vL)\right]^2 \right]^{-1}\,,
\end{equation}
then the estimator has unit response to lensing at leading order, and its lowest-order variance is $N^{(0),TT}_L=A^{TT}_{L}$, which is guaranteed to be the minimum value possible for an estimator of $\phi$ that is quadratic in $T$~\cite{ref:hu_okamoto_02}. Note the use of the lensed power spectrum in the numerator of equation~\eqref{eqn:QE_weights}: this is an approximation to the non-perturbative weights calculated by Ref.~\cite{ref:lewis_11}.

The auto-correlation of such a reconstruction takes the form
\begin{align}\label{eqn:lensing_auto}
    \langle \hat{\phi}(\vL)\hat{\phi}(\vL') \rangle = A^{TT}_{L} A^{TT}_{L'} \int &  \frac{\mathrm{d}^2\bm{l}'}{2\pi} \frac{\mathrm{d}^2\bm{l}''}{2\pi} g(\bm{l}',\vL) g(\bm{l}'',\vL') \nonumber \\
    & \times \langle T^{\mathrm{obs}}(\bm{l}') T^{\mathrm{obs}}(\vL-\bm{l}') T^{\mathrm{obs}}(\bm{l}'') T^{\mathrm{obs}}(\vL'-\bm{l}'') \rangle \,;
\end{align}
from this, it is possible to extract the angular power spectrum of the reconstruction,
\begin{equation}
 \langle \hat{\phi}(\vL)\hat{\phi}(\vL') \rangle \nonumber \\
= \delta^{(2)}(\vL' + \vL)\left(C^{{\phi}{\phi}}_{L}+\Delta C^{{\phi}{\phi}}_{L} \right)\,,
\end{equation}
which is an estimator of the lensing potential angular power spectrum, $C^{{\phi}{\phi}}_{L}$, once the biases denoted as $\Delta C^{{\phi}{\phi}}_{L}$ have been accounted for.

The 4-point function in equation~\eqref{eqn:lensing_auto} contains both connected and disconnected terms. The disconnected ones contribute to the $N^{(0), TT}$ bias (see, e.g.~\cite{ref:cooray_kesden_03, ref:hanson_et_al_11}); removing these should, ideally, include contributions from the foreground power.

As for the connected ones, there are several possibilities. One of them is a pure lensing component,
\begin{equation}\label{eqn:pure_lensing_trispec}
    \langle T^{\mathrm{obs}}(\bm{l}') T^{\mathrm{obs}}(\vL-\bm{l}') T^{\mathrm{obs}}(\bm{l}'') T^{\mathrm{obs}}(\vL'-\bm{l}'') \rangle \supset \langle \tilde{T}(\bm{l}') \tilde{T}(\vL-\bm{l}') \tilde{T}(\bm{l}'') \tilde{T}(\vL'-\bm{l}'') \rangle_{\mathrm{c}} \, ,
\end{equation}
where $\tilde{T}$ denotes the lensed CMB temperature anisotropies. Some of the couplings making up this trispectrum constitute the signal of interest, $C_L^{\phi\phi}$; others, those whose response is non-trivially related to the lensing potential, will contribute to the $ N^{(n), TT}$ biases with $n>0$ (see Ref.~\cite{ref:hanson_et_al_11} and references therein). We will not be concerned with any of these pure-lensing terms.

Instead, we shall be interested in couplings that involve emission from extragalactic foregrounds. These can also give rise to bias, and, to date, are not as well characterized. We describe these schematically now, and leave a careful analysis of each term to the following sections.

Let us denote the $TT$ quadratic estimator, applied to an observed map, $T^{\mathrm{obs}}$, as
\begin{equation}
    \hat{\phi}[T^{\mathrm{obs}}, T^{\mathrm{obs}}] = \hat{\phi}[\tilde{T} + n + s,\tilde{T} + n + s]\,.
\end{equation}
Here, $n$ is the instrument noise (which we assume to be Gaussian) and $s$ refers to one of possibly several foregrounds, which we take to be correlated with the underlying matter distribution. Throughout this work, we ignore the fact that the foregrounds are themselves lensed, since this has only a small effect on the tSZ~\cite{ref:wu_04} and the CIB~\cite{ref:schaan_et_al_18}.

The connected 4-point function of $T^{\mathrm{obs}}$ contributes to the power spectrum of a lensing reconstruction, schematically, as\footnote{Here and throughout, we will ignore the small correlation between the unlensed CMB and the lensing potential or extragalactic foregrounds induced by the integrated Sachs-Wolfe~\cite{ref:sachs_wolfe_67} (ISW) and Rees-Sciama effects~\cite{ref:rees_sciama_68} (these are small corrections to begin with, and the ISW gets additionally down-weighted when performing the lensing reconstructions due to it being a large-scale effect). As a result, we need an even number of $\tilde{T}$s in every correlator. 
}
\begin{align}\label{eqn:auto_biases}
    \langle \hat{\phi}[T^{\mathrm{obs}}, T^{\mathrm{obs}}]\, \hat{\phi}[T^{\mathrm{obs}}, T^{\mathrm{obs}}] \rangle_{\mathrm{c}} = & \langle \hat{\phi}[\tilde{T},\tilde{T}]\,\hat{\phi}[\tilde{T},\tilde{T}]\rangle_{\mathrm{c}} + 2\,\langle \hat{\phi}[\tilde{T},\tilde{T}]\, \hat{\phi}[s,s]\rangle_{\mathrm{c}} \nonumber \\
    & + 4\, \langle \hat{\phi}[\tilde{T},s]\,\hat{\phi}[\tilde{T},s]\rangle_{\mathrm{c}} + \langle \hat{\phi}[s,s]\,\hat{\phi}[s,s]\rangle_{\mathrm{c}}\,.
\end{align}
The first term on the right-hand side arises from the trispectrum in equation~\eqref{eqn:pure_lensing_trispec}; as we just discussed, it contains the signal of interest, as well as a number of lensing-related bias terms. In addition, there are three other terms on the right, above, that can bias the estimate. The second and third terms on the right are usually called the primary and secondary bispectrum biases, respectively, for reasons that will become apparent soon. They arise due to the fact that the foregrounds are correlated with lensing (this has been measured directly for the CIB~\cite{ref:holder_13, ref:planck_lensing_cib_cross, ref:hanson_13} and the tSZ~\cite{ref:hill_spergel_14}). The last term, on the other hand, comes from the foreground trispectrum, so it would be present even if there was no such correlation.

If the foregrounds are indeed correlated with the large-scale structure, they will also bias cross-correlations with other tracers of the matter distribution --- for example, galaxy surveys. If we denote such a tracer as $\mathcal{G}$, then
\begin{equation}\label{eqn:cross_biases}
    \langle \mathcal{G}\, \hat{\phi}[T^{\mathrm{obs}}, T^{\mathrm{obs}}] \rangle = \langle \mathcal{G}\,\hat{\phi}[\tilde{T},\tilde{T}]\rangle + \langle \mathcal{G}\,\hat{\phi}[s,s]\rangle\,.
\end{equation}
Once again, the second term on the right biases estimates of the true lensing cross-correlation.

Note that, for clarity of exposition, the expressions above omit the possibility of $s$ differing across legs of the quadratic estimators, or between different estimators. In the discussions that follow, we allow for the more general possibility of mixed couplings arising from the fact that different foregrounds are correlated by virtue of tracing the same underlying matter distribution. Indeed, the tSZ and CIB correlation was measured by Ref.~\cite{ref:planck_cib_tsz_correlation} to be at a level of approximately $15\%$ at $l=3000$ for all-sky Planck data, and indirect evidence for such a positive correlation was also found by Refs.~\cite{ref:reichardt_12, ref:george_15, ref:choi_20, ref:reichardt_21}. Physically, a one-halo contribution arises if some of the galaxies comprising the CIB reside in the massive, nearby clusters that produce the tSZ effect; and a two-halo term is also possible --- even if there was no star formation in the clusters that produced the tSZ --- as long as there is some overlap in the redshift distributions of the tSZ and CIB~\cite{ref:addison_tsz_cib_corr, ref:maniyar_tsz_cib_corr}.

\subsection{CMB lensing auto-correlations}
Let us now unpack further the bias couplings in equation~\eqref{eqn:auto_biases}.

\subsubsection{Primary bispectrum bias}\label{sec:prim_bispec}
Extragalactic foregrounds trace the matter distribution, so they correlate with the lensing potential. This is the defining property of the `bispectrum' biases affecting CMB lensing measurements. Because of this correlation, there is a term in the power spectrum of lensing reconstructions that depends on
\begin{equation}\label{eqn:prim_bispec_coupling}
    \langle T^{\mathrm{obs}}(\bm{l}') T^{\mathrm{obs}}(\vL-\bm{l}') T^{\mathrm{obs}}(\bm{l}'') T^{\mathrm{obs}}(\vL'-\bm{l}'') \rangle \supset  2\, \langle {  {\tilde{T}}(\bm{l}')  {\tilde{T}}(\vL-\bm{l}')  {T^{s_1}}(\bm{l}'') {T^{s_2}}(\vL'-\bm{l}'') } \rangle_{\mathrm{c}} \, .
\end{equation}
In this coupling arrangement --- the `primary' one --- the two lensed CMB legs are input into one quadratic estimator, and the two foreground legs ($T^{s_1}$ and $T^{s_2}$) are arranged into the other one. (Notice that, from now on, we allow the foreground components, $\mathrm{s}_i$, to all be different.) By construction, the quadratic estimator produces an estimate of the lensing potential that is unbiased to leading order when averaging over the unlensed CMB. Hence, we can replace the quadratic estimator acting on the $\tilde{T}$ fields with $\phi$. We then see that couplings of this sort induce a bias on the power spectrum of CMB lensing reconstructions:
\begin{align}\label{eqn:prim_bispec_bias}
    \Delta C^{\phi 
    \phi}_{L} \supset & 2\,  A^{TT}_{L}\int \frac{\mathrm{d}^2\bm{l}}{(2\pi)^2} g(\bm{l},\bm{L}) b_{\phi s_1s_2}(-\bm{L},\bm{l},\bm{L}-\bm{l})\, ,
\end{align}
where we have defined the angular bispectrum,
\begin{equation}
    \langle \phi(\bm{l}_1) T^{s_1}(\bm{l}_2) T^{s_2}(\bm{l}_3) \rangle = \frac{1}{2\pi} \delta^{(2)}(\bm{l}_1 + \bm{l}_2 + \bm{l}_3)  b_{\phi s_1s_2}(\bm{l}_1,\bm{l}_2,\bm{l}_3) \,.
\end{equation}
Equation~\eqref{eqn:prim_bispec_bias} goes by the name of `primary bispectrum bias'.

\subsubsection{Secondary bispectrum bias}
Another possible contribution to the trispectrum of observed CMB temperature anisotropies is
\begin{equation}\label{eqn:second_bispec_coupling}
    \langle T^{\mathrm{obs}}(\bm{l}') T^{\mathrm{obs}}(\vL-\bm{l}') T^{\mathrm{obs}}(\bm{l}'') T^{\mathrm{obs}}(\vL'-\bm{l}'') \rangle \supset  4\, \langle {  {\tilde{T}}(\bm{l}')  T^{s_1}(\vL-\bm{l}')  {\tilde{T}}(\bm{l}'') {T^{s_2}}(\vL'-\bm{l}'') } \rangle_{\mathrm{c}} \, .
\end{equation}
Now, each of the two quadratic estimators takes in both a lensed CMB leg and a foreground leg --- a `secondary' coupling that gives the resulting bias the name of `secondary bispectrum bias'.

To calculate this term, we will expand perturbatively to leading order in lensing, obtaining
\begin{align}\label{eqn:secondary_bispec_bias}
    \langle \hat{\phi}(\vL)\hat{\phi}(\vL') \rangle \supset - 4 \delta^{(2)}(\vL + \vL') \, A^{TT}_{L} A^{TT}_{L'} \int &  \frac{\mathrm{d}^2\bm{l}'}{(2\pi)^2} \frac{\mathrm{d}^2\bm{l}''}{(2\pi)^2} g(\bm{l}',\vL) g(\bm{l}'',\vL') h(\bm{l}', -\bm{l}'') C_{l''}^{TT}\nonumber \\
    & \times b_{\phi s_1s_2}(\vl' + \vl'',\vL-\bm{l}',\vL'-\bm{l}'') + s_1 \leftrightarrow s_2\,,
    %& \times \langle \phi(\vl' + \vl'') T^{s_1}(\vL-\bm{l}') T^{s_2}(\vL'-\bm{l}'') \rangle \,,
\end{align}
where $h(\bm{l}, \bm{l}')\equiv \bm{l}' \cdot \left(\bm{l} - \bm{l}'\right)$.

\subsubsection{Trispectrum bias}\label{sec:trispectrum_bias}
Finally, we consider the case where all four quadratic estimator legs in the lensing power spectrum estimator take in foregrounds. This is associated with the coupling
\begin{equation}\label{eqn:trispec_coupling}
    \langle T^{\mathrm{obs}}(\bm{l}') T^{\mathrm{obs}}(\vL-\bm{l}') T^{\mathrm{obs}}(\bm{l}'') T^{\mathrm{obs}}(\vL'-\bm{l}'') \rangle \supset   \langle   T^{s_1}(\bm{l}')  T^{s_2}(\vL-\bm{l}')  T^{s_3}(\bm{l}'') T^{s_4}(\vL'-\bm{l}'')  \rangle_{\mathrm{c}} \,,
\end{equation}
which translates to an additive bias on the lensing power spectrum of the form
\begin{align}\label{eqn:trispec_bias}
    \Delta C^{\phi 
    \phi }_L \supset   (A^{TT}_{L})^2
     \int \frac{\mathrm{d}^2\bm{l}}{(2\pi)^2}\frac{\mathrm{d}^2\bm{l}'}{(2\pi)^2} g(\bm{l},\bm{L}) g(\bm{l}',-\bm{L}) t_{s_1 s_2 s_3 s_4}(\bm{l}',-\bm{L}-\bm{l}',\bm{l},\bm{L}-\bm{l})\,,
\end{align}
where we have defined the angular trispectrum,
\begin{equation}
    \langle T^{s_1}(\bm{l}_1) T^{s_2}(\bm{l}_2) T^{s_3}(\bm{l}_3) T^{s_4}(\bm{l}_4) \rangle = \frac{1}{(2\pi)^2}\delta^{(2)}(\bm{l}_1 + \bm{l}_2 + \bm{l}_3 + \bm{l}_4)  t_{s_1 s_2 s_3 s_4}(\bm{l}_1,\bm{l}_2,\bm{l}_3,\bm{l}_4) \,.
\end{equation}
We emphasise that this `trispectrum bias' exists also for foregrounds which are not correlated with the lensing potential ---  it is relevant, for example, to biases from galactic foregrounds~\cite{ref:beck_et_al_20}.

Reference~\cite{ref:van_engelen_et_al_14} notes that, in the limit that the sources are Poisson-distributed and sufficiently numerous, their distribution will approach that of a Gaussian random field by the central limit theorem. It is then easy to see, using Wick's theorem, that their 4-point function will only contribute to the $N^{(0)}$ bias, which can be easily and accurately calculated and removed. Though this limit no longer applies to the contribution from tSZ and CIB to current and upcoming CMB lensing analyses, it may well be sufficient to capture the contribution from radio sources\footnote{The contribution from these galaxies to microwave-frequency measurements can still appear to follow a Poisson distribution if the emission is dominated by a few, bright sources. This is, indeed, the dominant impact that extragalactic radio sources appear to have had on Planck's lensing analyses~\cite{ref:planck_15_lensing}, which could only resolve  --- and mask away --- sources brighter than 145\,mJy at 143\,GHz. Since the clustering contributions were estimated to be small given the limited angular resolution ($\ell_{\rm max}\approx 1800$), the bias was expected to be sourced largely by approximately-Poisson-distributed, bright galaxies just below the flux cut. It was therefore possible to estimate their contribution analytically and account for it in the analyses; it came out to be at around 2\% of the lensing power spectrum amplitude. For higher-resolution experiments, this simple treatment might not do. The infrastructure developed in this paper could then be extended to model those clustering contributions. Note that calculations of the shot-noise terms can be found in other works; e.g., Refs~\cite{ref:van_engelen_15, ref:lacasa_halo_model}.}. 

\subsection{CMB lensing cross-correlations}
As illustrated by equation~\eqref{eqn:cross_biases}, cross-correlations between CMB lensing reconstructions and any observable that traces the matter distribution, call it $\mathcal{G}$, will be affected by a coupling very similar to the primary bispectrum bias of section~\ref{sec:prim_bispec}. The calculation is the same (up to a factor of 2) once we make the replacement $\phi \rightarrow \mathcal{G}$.

\subsection{$B$-mode delensing}\label{sec:delensing}
Gravitational lensing converts $E$-modes into $B$-modes, inducing spurious variance that hinders the search for a primordial signal produced by inflationary gravitational waves. In order to mitigate this effect, polarization measurements can be `delensed' by subtracting the specific realisation of lensing $B$-modes present on the sky. Doing this requires having access to high-resolution $E$-mode measurements and some proxy of the matter distribution that caused the deflections. With these in hand, one can build a template meant to approximate the lensing $B$-modes present on the sky,
\begin{align}\label{eqn:template}
        \hat{B}^{\mathrm{lens}}(\bm{l}) &= \int \frac{d^2\bm{l}'}{2\pi} W(\bm{l},\bm{l}') \mathcal{W}^{E}_{l'} \mathcal{W}^{\phi}_{|\bm{l}-\bm{l}'|}   E^{\mathrm{obs}}(\bm{l}')\hat{\phi}(\bm{l}-\bm{l}')  \nonumber \\
        & \equiv \mathcal{B} \left[ E^{\mathrm{obs}} ,\hat{\phi}\right](\bm{l})\,,
\end{align}
where $E^{\mathrm{obs}}$ are the observed $E$-modes\footnote{Note that, as suggested in Ref.~\cite{ref:limitations_paper}, this template can be made significantly more accurate by using lensed rather than delensed (or unlensed) $E$-modes.}, $\hat{\phi}$ is an estimate of the lensing potential, $\mathcal{W}^{E}$ and $\mathcal{W}^{\phi}$ are Wiener filters for $E^{\mathrm{obs}}$ and $\hat{\phi}$, respectively, and 
\begin{equation}
        W(\bm{l},\bm{l}') = \bm{l}' \cdot (\bm{l}-\bm{l}') \sin 2(\psi_{\bm{l}} - \psi_{\bm{l}'}) \, .
\end{equation}
Here, $\psi_{\bm{l}}$ is the angle that the wavevector $\bm{l}$ makes with the $x$-axis used to define (positive) Stokes $Q$. In the equation above, we have defined the functional $\mathcal{B}[E^{\rm obs}, \phi^{\rm proxy}]$ to make explicit the dependence of the template on the observed $E$-mode fluctuations and the large-scale-structure tracer used as proxy of the convergence field. The delensed field can then be obtained by subtracting this template from observations\footnote{It is common practice to delens by including the lensing template as an additional `channel' in a multi-frequency, cross-spectral pipeline (e.g.,~\cite{ref:bicep_delensing}); this approach can be shown to be equivalent to a map-level subtraction in most scenarios~\cite{hertigSimonsObservatoryCombining2024b}.},
\begin{align}\label{eqn:Bdel}
        \hat{B}^{\mathrm{del}}(\bm{l}) = \tilde{B}^{\mathrm{obs}}(\bm{l}) -  \hat{B}^{\mathrm{lens}}(\bm{l})\,.
\end{align}

A prime method to obtain the lensing tracer used in equation~\eqref{eqn:template} is by means of internal reconstructions. However, Ref.~\cite{ref:baleato_and_ferraro_22} showed that when the tracer of choice is a temperature-based internal reconstruction one must proceed with care, because the power spectrum of delensed $B$-modes, $\Delta C_l^{BB, \mathrm{del}}$, will be biased. We can split the additional contributions into two kinds:
\begin{equation}\label{eqn:bias_TT}
        \Delta C_l^{BB, \mathrm{del}} = -2\,\Delta C_l^{\tilde{B} \times \hat{B}^{\mathrm{lens}}} + \Delta C_l^{\hat{B}^{\mathrm{lens}}\times \hat{B}^{\mathrm{lens}}} \,,
\end{equation}
the bias to the cross-spectrum of template and lensing $B$-modes, and the bias to the template auto-spectrum, respectively. Ref.~\cite{ref:baleato_and_ferraro_22} showed that the dominant contributions to these biases are of the form
\begin{equation}\label{eqn:bias_couplings_cross}
    \Delta C_l^{\tilde{B} \times \hat{B}^{\mathrm{lens}}} \supset \langle \tilde{B}(\bm{l}) \mathcal{B} \left[ \tilde{E} ,\hat{\phi}^{TT} \left[s,s\right]\right](\bm{l}')\rangle'_{\mathrm{c}} \,,
\end{equation}
and
\begin{align}\label{eqn:bias_couplings_auto}
    \Delta  C_l^{\hat{B}^{\mathrm{lens}}\times  \hat{B}^{\mathrm{lens}}} \supset & \, \wick[offset=1.5em]{ 2\, \langle \mathcal{B} [ \c {E}^{\mathrm{obs}}, \hat{\phi}^{\mathrm{TT}} [\tilde{T},\tilde{T}]](\bm{l})\,  \mathcal{B} [\c {E}^{\mathrm{obs}}, \hat{\phi}^{\mathrm{TT}} [s,s]](\bm{l}')\rangle'_{\mathrm{c}}} \nonumber \\
    & \wick[offset=1.5em]{ + 4\,  \langle \mathcal{B}[ \c {E}^{\mathrm{obs}}, \hat{\phi}^{\mathrm{TT}} [\tilde{T},s]](\bm{l})\, \mathcal{B} [ \c {E}^{\mathrm{obs}}, \hat{\phi}^{\mathrm{TT}} [\tilde{T},s]](\bm{l}')\rangle'_{\mathrm{c}}} \nonumber \\
    & \wick[offset=1.5em]{ + \langle \mathcal{B} [  \c {E}^{\mathrm{obs}}, \hat{\phi}^{\mathrm{TT}} [s,s]](\bm{l})\,  \mathcal{B} [  \c {E}^{\mathrm{obs}}, \hat{\phi}^{\mathrm{TT}} [s,s]](\bm{l}')\rangle'_{\mathrm{c}}} \,.
\end{align}
The primes following the angle brackets in the equations above denote the Dirac delta function $\delta(\bm{l}+\bm{l}')$, which we have removed to simplify notation. In addition, we adopt a notation convention where a Gaussian contraction of the fields connected by over-bars is to be taken first, and this is then to be multiplied by the connected $n$-point function of the remaining fields inside the angle brackets. Note that all of these terms vanish if the foregrounds are Gaussian; disconnected contributions from such Gaussian foregrounds are assumed to be included in $C_{l}^{BB, \rm del}$.

The biases in equations~\eqref{eqn:bias_couplings_cross} and~\eqref{eqn:bias_couplings_auto} are manifestly analogous to the primary and secondary bispectrum and trispectrum biases studied in previous sections. In fact, they can be obtained from our previous calculations as
\begin{align}\label{eqn:full_theory_cross}
\Delta C_l^{\tilde{B}\times \hat{B}^{\mathrm{lens}}} = \int \frac{d^2\bm{l}'}{(2\pi)^2} W^2(\bm{l},\bm{l}') & \mathcal{W}^{E}_{l'} \mathcal{W}^{\phi}_{|\bm{l}-\bm{l}'|} C^{EE}_{l'} \Delta C^{\phi \hat{\phi}}_{|\bm{l}-\bm{l}'|}\,,
\end{align}
and
\begin{align}\label{eqn:full_theory_auto}
        \Delta C_l^{\hat{B}^{\mathrm{lens}}\times \hat{B}^{\mathrm{lens}}} = \int \frac{d^2\bm{l}'}{(2\pi)^2} W^2(&\bm{l},\bm{l}') \left(\mathcal{W}^{E}_{l'} \mathcal{W}^{\phi}_{|\bm{l}-\bm{l}'|} \right)^2  C^{EE, \mathrm{tot}}_{l'}
        \Delta C^{\hat{\phi} \hat{\phi}}_{|\bm{l}-\bm{l}'|}\,,
\end{align}
where $\Delta C^{\hat{\phi} \hat{\phi}}$ and $\Delta C^{\phi \hat{\phi}}$ are, respectively, the foreground-induced biases to the reconstruction's auto- and cross-spectra with the true CMB lensing potential.

In summary, once we have calculated bispectrum and trispectrum biases to $C_{L}^{\hat{\phi}\hat{\phi}}$, we will be able to use equations~\eqref{eqn:full_theory_cross} and~\eqref{eqn:full_theory_auto} to predict the delensing biases that arise when $\hat{\phi}$ is used to delens $B$-modes.

\section{Halo model calculation of the biases}\label{sec:analytic_framework}
In the previous section, we saw that computing the biases requires knowledge of $b_{\phi s_1s_2}$ and $t_{s_1 s_2 s_3 s_4}$; that is, the angular bispectrum that the foregrounds make with the matter/tracer distribution and their trispectrum. Since both the CIB and the tSZ effect are produced by material that is (for the most part) in virialized structures, we should, in principle, be able to calculate these rather accurately using the halo model. (Note that the same cannot be said about other effects such as the kSZ, which receives significant contributions from the `field' external to halos.) In appendix~\ref{appendix:projected_higherpnt_fns}, we explain the details of our calculation, but see also appendix~\ref{sec:halo_model} for an introduction to the formalism. Our modelling of the tSZ hinges on the `Battaglia' pressure profile~\cite{ref:battaglia_et_al_12}, described in section~\ref{sec:tsz}. For the CIB, on the other hand, we extend the halo model of the CIB introduced by Ref.~\cite{ref:shang_et_al_12}, which allows for a mass-dependent luminosity; this is detailed in section~\ref{sec:halo_model_cib}.

We incorporate into our framework the possibility of foreground cleaning via an internal linear combination (ILC~\cite{ref:tegmark_et_al_03}), a scheme for combining observations at different frequencies such that the variance of the resulting field is minimised while retaining unit response to the primary CMB. We will also consider a slightly different flavour of the method where certain components are explicitly nulled (or `deprojected') --- the constrained ILC~\cite{ref:remazeilles_contrained_ilc, ref:sailer_et_al_21, ref:Abylkairov_et_al_20}). The optimal ILC weights are calculated taking into account CMB, extragalactic foregrounds, (white) instrument noise, and Galactic dust emission, using the publicly-available code \texttt{BasicILC}\footnote{\url{https://github.com/EmmanuelSchaan/BasicILC}}, which builds on the sky models of~\cite{ref:dunkley_et_al_13}. Further details are provided in appendix~\ref{appendix:fg_cleaning}.

When calculating the primary bispectrum and trispectrum biases, we restrict ourselves to the subset of one- and two-halo contributions that we expect to dominate on physical grounds. Indeed, we find that these terms are enough to obtain good qualitative agreement with the simulation-based characterization of Refs.~\cite{ref:van_engelen_et_al_14, ref:sailer_et_al, ref:osborne_et_al}, and measured here again in section~\ref{sec:comparison_with_websky}. However, it would be rather straightforward to include additional terms (e.g., three-halo contributions) should it become necessary in future.

Calculating the biases in equations~\eqref{eqn:prim_bispec_bias},~\eqref{eqn:secondary_bispec_bias} and~\eqref{eqn:trispec_bias} involves evaluating the quadratic estimators of lensing. Several implementations of these can be found in the literature, allowing for their efficient application to CMB maps. Later in this work, we will use the flat-sky version of the \texttt{QuickLens}\footnote{\texttt{QuickLens} can be found at \url{https://github.com/dhanson/quicklens}, though we provide an amended and extended version in \url{https://github.com/abaleato/Quicklens-with-fixes}.} code to reconstruct lensing maps from simulations; this code has been extensively tested --- for example, in the analysis of Ref.~\cite{ref:planck_15_lensing}.

In spite of the proved performance of publicly available quadratic estimator codes, the lensing reconstructions involved in our analytic calculations can be evaluated much faster thanks to the following insight: to the extent that the emission profiles of the foregrounds are azimuthally-symmetric\footnote{Note that this assumption is also often made in the context of bias hardening~\cite{ref:namikawa_et_al_13, ref:osborne_et_al, ref:sailer_et_al} and shear-only reconstructions~\cite{ref:schaan_ferraro_18}.} and the filtering is isotropic, the two-dimensional Fourier transforms involved in the lensing reconstruction reduce to one-dimensional Hankel transforms. These can be evaluated very efficently (in their discrete limit) using Gaussian quadratures\footnote{We also explore using the FFTlog algorithm~\cite{ref:talman_78, ref:hamilton_00} in appendix~\ref{appendix:fftlog}.} formulated as matrix multiplication and accelerated on GPUs using \texttt{JAX}; see appendix~\ref{appendix:qe_w_fftlog} for details. When comparing the bias calculations obtained in this way to the output of 2D codes such as \texttt{QuickLens}, we indeed find indistinguishable results. This is perhaps not suprising given that the assumption of spherical symmetry of the halos is implicit in our halo model calculations by virtue of having used the NFW and Battaglia profiles. However, we find that the 1D approach is up to four orders of magnitude faster, taking $O(1\,\mathrm{ms})$ per reconstruction on a modern laptop.

In order to harness this vast improvement in computational speed, we evaluate equations~\eqref{eqn:prim_bispec_bias} and~\eqref{eqn:trispec_bias} by performing a lensing reconstruction for every mass and redshift step in the Limber-approximated halo model calculations of the angular bispectrum and trispectrum --- that is, the projection, using equation~\eqref{eqn:limber_polyspectra}, of the 3D bi- and trispectra in appendices~\ref{appendix:hm_bispectra} and~\ref{appendix:hm_trispectra}. We find sufficient convergence of the integrals when 50 steps in redshift and 50 in mass are used, for a total computation time of $O(10\,\mathrm{s})$ per foreground on a modern laptop. Note that this approach is in contrast to what equation~\eqref{eqn:prim_bispec_bias} might appear to suggest: evaluating the angular bispectrum first, and only carrying out the lensing reconstruction once, at the very end.

Unfortunately, these faster lensing reconstructions cannot be applied to the secondary bispectrum bias, equation~\eqref{eqn:secondary_bispec_bias}. The reason is that the two reconstructions in the expression are not separable, and when one tries to evaluate them as nested operations, azimuthal symmetry is broken. In principle, equation~\eqref{eqn:secondary_bispec_bias} can be evaluated by first computing the angular bispectrum in the integrand, and then carrying out the outer integrals associated with the lensing reconstructions. However, we find an advantage in terms of evaluation speed if we instead perform the nested lensing reconstructions for each mass and redshift step in the halo model integrals, while harnessing the convolution theorem. This approach is explained in appendix~\ref{appendix:evaluating_secondary_bispec_bias}. Evaluation of these secondary couplings can still be done in order one minute on a laptop.

We make our code, \texttt{CosmoBLENDER}, publicly available\footnote{The code \texttt{CosmoBLENDER} (Cosmological Biases to LENsing and Delensing due to Extragalactic Radiation) can be found at \url{https://github.com/abaleato/CosmoBLENDER}.}. In what follows, we use it to present some example evaluations of extragalactic foreground biases to CMB lensing auto- and cross-correlations using our new code.

\section{Results}\label{sec:results}
\subsection{CMB lensing auto-correlations}\label{sec:results_auto}
As a demonstration of the capabilities of our analytic framework and code, we now present a breakdown of bias contributions to the auto-spectrum of CMB lensing reconstructions for two example scenarios. We will begin with a comprehensive case study of ACT-DR6-like reconstructions before studying the impact of foreground cleaning using multi-frequency observations in the context of the large-aperture telescope of SO. Our approximate characterization of these experiments is given in table~\ref{tab:experiment_details}.

We assume that the lensing reconstructions are obtained exclusively from the temperature anisotropies, ignoring the fact that roughly 20\% of the information in the fiducial ACT analysis comes from polarization. The fiducial ACT analysis also uses a point-source-hardened estimator~\cite{quAtacamaCosmologyTelescope2024, maccrannAtacamaCosmologyTelescope2023a, ref:sailer_et_al}, but here we focus on the standard QE. Let us assume for now that that maps from the two available frequencies are coadded in such a way that the total map noise variance is minimized. 

\begin{figure}
    \centering \includegraphics[width=\textwidth]{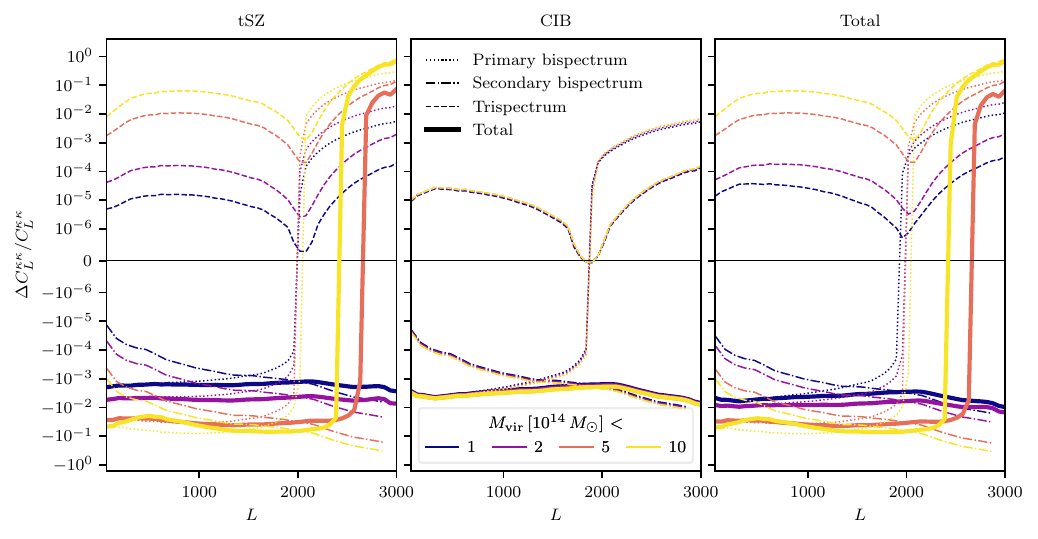}
    \caption{Fractional bias to the CMB lensing auto-spectrum associated with tSZ (left panel), CIB (middle) and their combination (right, including cross-terms) for an experiment similar to ACT DR6 with $\ell_{\rm max}=3000$. Different line styles correspond to different couplings of the four-point function, with the total shown as solid lines. Different colors correspond to different maximum halo mass cuts. While the trispectrum bias is always positive, the primary bispectrum bias is negative for $L<2000$, and the secondary bispectrum term is always negative. When stringent mass cuts are imposed (or in the case of the CIB, which is insensitive to the high masses varied here), the bispectrum terms dominate and the total bias is negative on the scales of interest. On the other hand, as higher mass halos are included, the bias becomes positive starting at high $L$ and with the zero-crossing moving to larger scales as the mass cut is raised.}
    \label{fig:tsz_cib_mixed_biases_to_auto_vs_Mcut}
\end{figure}

In figure~\ref{fig:tsz_cib_mixed_biases_to_auto_vs_Mcut}, we show our calculations of the trispectrum and bispectrum biases (including both primary and secondary couplings), split into contributions from tSZ or CIB as well as the total including mixed tSZ-CIB terms arising from the fact that these two foregrounds are correlated. Whenever we plot biases as a fraction of the signal, the latter will be computed using \texttt{CAMB}, which in turn uses a fitting function for the non-linear matter power spectrum from Ref.~\cite{ref:mead_et_al_15}. We assume that only temperature modes up to $\ell_{\rm max}=3000$ are used for the reconstructions.
\begin{table}[h!]
\centering
\begin{tabular}{lccc}
\toprule
Experiment & Frequency [GHz] & $\Delta_T$ [$\mu$K-arcmin] & $\theta_{\rm FWHM}$ \\
\midrule
ACT-like & 90.0  & 15.0 & 2.1 \\
    & 150.0 & 15.0 & 1.4 \\
\midrule
SO-like  & 27.3  & 52.0 & 7.4 \\
    & 41.7  & 27.0 & 5.1 \\
    & 93.0  & 5.8 & 2.2 \\
    & 143.0 & 6.3  & 1.4 \\
    & 225.0 & 15.0 & 1.0 \\
    & 278.0 & 37.0 & 0.9 \\
\bottomrule
\end{tabular}
\caption{Frequency channels, white noise levels, and Gaussian beam full width at half-maximum for the different CMB experiments we approximate.}
\label{tab:experiment_details}
\end{table}

In different colors, we show the outcome of varying the maximum mass of halos included in the calculation. This is mostly intended to reproduce the impact of masking tSZ-detected galaxy clusters, which is known to mitigate lensing biases\footnote{In the case of the CIB, halo mass can be connected to the flux limit of CMB experiments through a luminosity function, but we find that removing detectable sources has a negligible effect on the bias arising from the clustering terms described in this work.}. ACT has produced a 90\%-complete census of clusters above a virial mass of $M_{\rm vir}>(2\text{--}4)\times 10^{14}\,\mathrm{M}_{\odot}$ within its footprint~\cite{hiltonAtacamaCosmologyTelescope2021a}. For SO, the threshold is likely to be closer to the $M_{\rm vir}\approx10^{14}\,\mathrm{M}_{\odot}$ mark~\cite{ref:SO_science_paper, zubeldiaCosmocncFastFlexible2024a}. In the figure, we test the values $M_{\rm vir}<\{1\times10^{14}, 2\times10^{14}, 5\times10^{14}, 1\times10^{15}\}\,\mathrm{M}_{\odot}$. For the minimum mass, we set $M_{\rm vir}>10^{10}\,\mathrm{M}_{\odot}$, which is a conservative choice guaranteed to capture the bulk of the effects, as we show explicitly in section~\ref{sec:hyperparam_requirements}. In that section, we also determine requirements on the redshift coverage of our calculations; we adopt those insights in the calculations of this and subsequent sections.

In general, the trispectrum bias is positive and the secondary bispectrum contribution is negative. The primary bispectrum bias, on the other hand, is negative on large scales but becomes positive on smaller ones\footnote{The heuristic explanation for the negative sign of the primary bispectrum bias is that the additional small-scale power introduced by foregrounds is degenerate with the action of a large-scale underdensity that induces negative convergence and thus transfers CMB anisotropy power to smaller scales.}. For this experimental configuration, the transition happens ar $\ell\sim 2000$. We will investigate this sign change in greater detail in section~\ref{sec:results_cross}, finding that the zero-crossing depends almost exclusively on $\ell_{\rm max}$. There are therefore subtle cancellations between these different terms, each of which depends differently on the maximum-mass threshold.

If unmitigated, the tSZ biases can give an order unity positive bias to the CMB lensing auto-spectrum across the relevant scales, driven by the trispectrum term. However, this trispectrum scales as the fourth power of the Compton-$y$ parameter (which, in turn, grows with the mass of the halo as $y\propto M^{5/3}$ according to the virial theorem), so it is very sensitive to the upper mass cut, much more so than the primary bispectrum bias. Consequently, for the realistic mass cuts explored in figure~\ref{fig:tsz_cib_mixed_biases_to_auto_vs_Mcut}, the tSZ biases become much smaller: order a few percent for the ACT-like cuts, in agreement with previous characterizations on simulations~\cite{ref:sailer_et_al_21, maccrannAtacamaCosmologyTelescope2023a}. Such extensive suppression of the trispectrum means the negative bispectrum terms can dominate on larger scales, giving a negative correction overall (there is still a transition to positive bias at high multipoles, the exact location of which depends on the mass cut).

\begin{figure}
    \centering    \includegraphics[width=\textwidth]{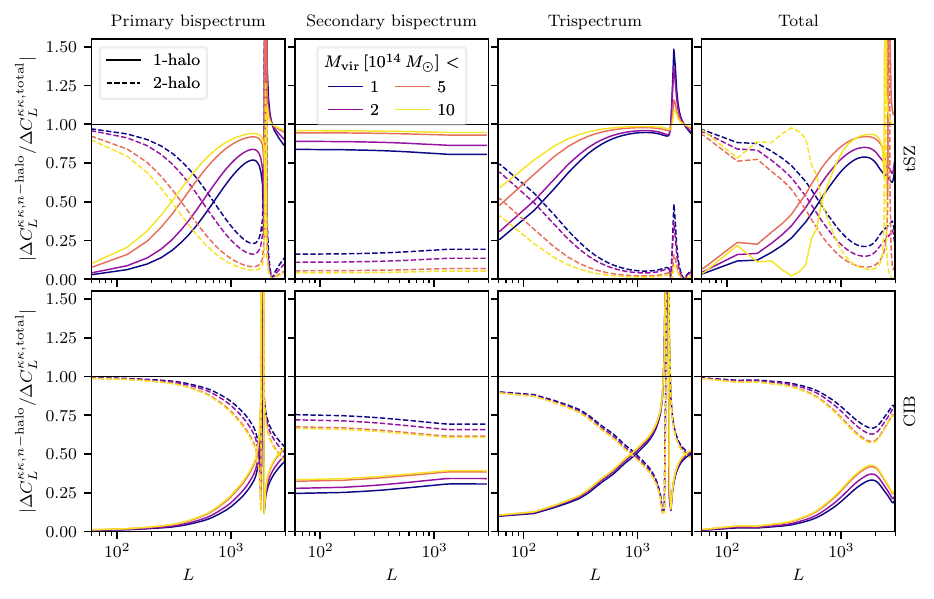}
    \caption{Fractional contribution of one-halo (solid) and two-halo terms (dashed) to each of the CMB lensing auto-spectrum bias couplings (different columns, with the total bias being shown in the rightmost column). Top and bottom rows correspond to the tSZ and CIB biases, respectively. Different colors show different cuts in the maximum halo mass allowed to feature in the calculation. At any given scale, higher-mass cuts shift the emphasis towards one-halo terms. Notably, for the ACT-like setup and masking scheme described here, two-halo terms dominate the total tSZ bias below $L<400$, with one-halo terms becoming more relevant at higher multipoles. The CIB bias is always dominated by two-halo terms.}
    \label{fig:tsz_cib_whichterm_biases_to_auto}
\end{figure}

If the most massive clusters are allowed to feature in the CMB maps from which the reconstructions are obtained, one-halo terms will constitute the dominant contribution (particularly via the trispectrum bias). However, for more typical maximum-mass cuts, the two-halo contributions become quite relevant. This is shown in figure~\ref{fig:tsz_cib_whichterm_biases_to_auto}, where we compare the relative importance of one- and two-halo terms for each of the couplings contributing to the CMB lensing auto-spectrum bias. At any given scale, masking progressively more and more of the most massive halos reduces the relative importance of one-halo contributions. By the time the $M_{\rm vir}$ cut is at a few times $10^{14}\,\mathrm{M}_{\odot}$, two-halo contributions are actually dominant below $L\lesssim 400$, with one-halo ones dominating at higher multipoles. This is largely due to the fact that the primary bispectrum bias is becoming more important, and it is relatively more sensitive to two-halo effects, particularly on large scales\footnote{Figure~\ref{fig:tsz_cib_whichterm_biases_to_auto} appears to show an exception to this trend between $100<L<500$ in the $M_{\rm vir}<10\times 10^{14}\,\mathrm{M}_{\odot}$ curve for the total tSZ bias. This is just a coincidence related to the fact that the total tSZ bias spectrum dips slightly at these multipoles (as seen also in figure~\ref{fig:tsz_cib_mixed_biases_to_auto_vs_Mcut}) by virtue of a larger trispectrum cancelling more extensively with the bispectrum terms.}. Note, finally, that the secondary bispectrum bias of tSZ is typically dominated by one-halo contributions.

Let us move on from tSZ to CIB biases, which are shown in the centre column of figure~\ref{fig:tsz_cib_mixed_biases_to_auto_vs_Mcut}. Note that these effects are intrinsically smaller than for the tSZ, in agreement with results from simulations (see section~\ref{sec:comparison_with_websky}). Any of the mass cuts considered here is enough to bring the bias to sub-percent level at this $\ell_{\rm max}=3000$. The same sign change around $L= 2000$ observed for the primary bispectrum bias produced by the tSZ is observed for the CIB, suggesting that this may be a product of the geometrical weighting inside the quadratic estimator rather than the emission processes themselves (we will argue this more rigorously in section~\ref{sec:sign_change}). Notice that mass cuts we are studying have much less of an impact here: this is because the vast majority of galaxies contributing CIB emission live in halos with mass below $10^{14}\,\mathrm{M}_{\odot}$; this agrees with the behaviour seen by Ref.~\cite{ref:van_engelen_et_al_14} in simulations. This increased reliance on lower-mass halos also emphasizes the importance of two-halo terms: the bottom row in figure~\ref{fig:tsz_cib_whichterm_biases_to_auto} shows that these terms dominate all the bias couplings across the relevant scales.

For completeness, we show, in the right-most column of figure~\ref{fig:tsz_cib_mixed_biases_to_auto_vs_Mcut}, the sum of tSZ and CIB biases, including terms arising due to the fact that they are correlated. For the particular experimental configuration shown here, the total is dominated by the tSZ contributions. But note that this may not be the case if observations are taken at different frequencies, if a different foreground mitigation scheme is applied, or if a more extensive halo-masking scheme is adopted.

\begin{figure}
    \centering
    \includegraphics[width=\textwidth]{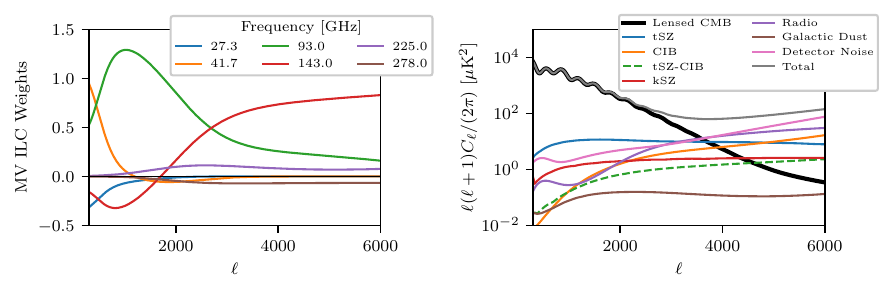}
    \caption{Minimum-variance ILC weights for our SO configuration (left) and power spectrum residuals from different sky components after using these weights to produce multi-freuency cleaned maps. The contribution from the correlation between tSZ and CIB is negative, so we plot it as dashed line.}
    \label{fig:ILC_summary}
\end{figure}

Let us therefore explore the impact of multi-frequency cleaning on these biases. To do so, we now consider an experiment with the approximate characteristics of SO, with the six frequency channels described in table~\ref{tab:experiment_details}. We adopt a maximum-halo mass cut of $M_{\rm vir}<2\times 10^{14}\,\mathrm{M}_{\odot}$, similar to the ACT DR6 scenario. As explained in section~\ref{sec:analytic_framework}, the ILC weights we compute using the \texttt{BasicILC}\footnote{\url{https://github.com/EmmanuelSchaan/BasicILC}} code are calculated in multipole space taking into account CMB, extragalactic foregrounds (kSZ, tSZ, radio and CIB), white instrument noise and Galactic dust emission, all taken from the sky models of~\cite{ref:dunkley_et_al_13}.\footnote{Note that there is a small mismatch
in the SED model used in the ILC and that used in the halo-based calculations throughout this paper. This is illustrative of current uncertainties in foreground
modeling.} These weights are shown in figure~\ref{fig:ILC_summary}, together with the power spectrum residuals from different sky components.

\begin{figure}
    \centering
    \includegraphics[width=\textwidth]{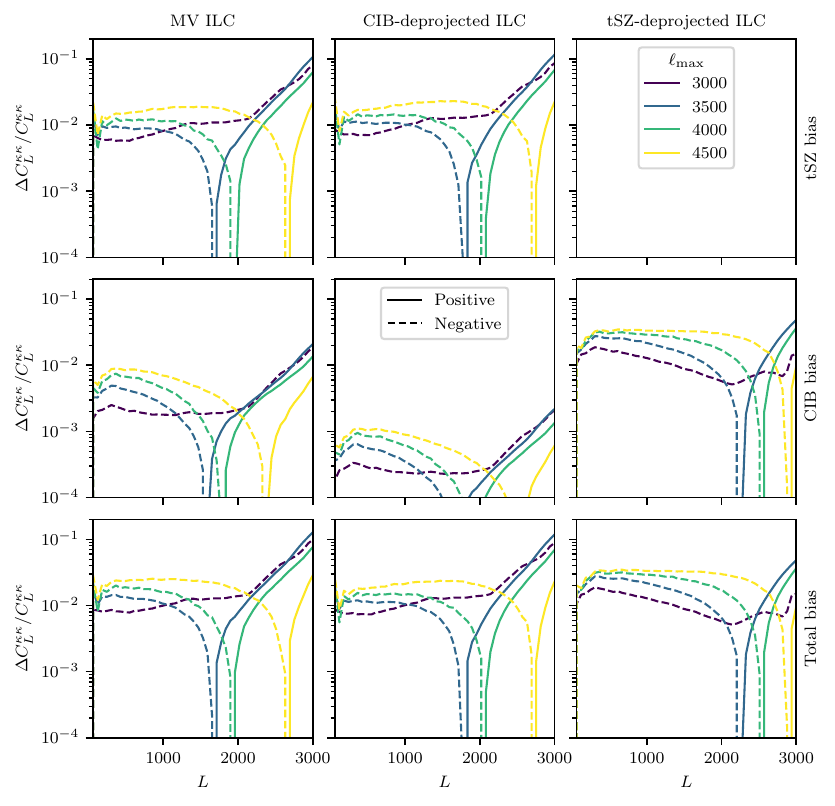}
    \caption{Fractional bias on the CMB lensing auto-spectrum for an SO-like experiment as a function of the maximum multipole used in the reconstructions ($\ell_{\rm max}$; different colors), multi-frequency cleaning prescription (different columns) and extragalactic foreground contribution (different rows). In all cases, halos with $M_{\rm vir}>2\times 10^{14}\,M_{\rm \odot}$ have been removed, which increases the importance of bispectrum terms over trispectrum ones (particularly for higher $\ell_{\rm max}$). Raising $\ell_{\rm max}$ increases the significance of the bias and induces a sign change that shifts to smaller scales with higher $\ell_{\rm max}$ (driven by the primary bispectrum bias; cf. figure~\ref{fig:zero_crossing_prim_bispec}). When cluster-masking is this extensive, tSZ biases are greatly reduced and there is no clear advantage in further deprojecting tSZ, as this would significantly boost CIB contributions.}
\label{fig:auto_biases_vs_fgcleaning}
\end{figure}

Figure~\ref{fig:auto_biases_vs_fgcleaning} shows the fractional biases on the lensing auto-spectrum as a function of multi-frequency cleaning prescription, $\ell_{\rm max}$ and extragalactic foreground type. The first thing worth noting is that all things being equal, an increase in $\ell_{\rm max}$ brings about a boost in the significance of the bias on large angular scales ($L\lesssim 1500$). We also notice that the bias develops a sign change in the multipole range plotted once $\ell_{\rm max}>3000$. This is due to the fact that a halo mass cut this stringent is very effective at neutralising the trispectrum bias (which, recall, is positive everywhere), emphasising the role of bispectrum biases. At low $L$, both bispectrum biases are negative. At higher multipoles, however, the primary bispectrum bias changes sign. Later, in section~\ref{sec:sign_change} we will show that the angular scale where this zero-crossing happens depends almost exclusively on $\ell_{\rm max}$, shifting to smaller scales as the latter is raised. We see a reflection of this in figure~\ref{fig:auto_biases_vs_fgcleaning}, where the overall bias chnages sign at an $L$ that increases with $\ell_{\rm max}$.

Let us now focus on a particular value of $\ell_{\rm max}$ and compare the impact of different multi-frequency foreground cleaning prescriptions. As expected, deprojecting either the tSZ or the CIB is a very effective way to mitigate contributions from the particular foreground in question\footnote{The reason `CIB-deprojection' does not exactly null this component in figure~\ref{fig:auto_biases_vs_fgcleaning} has to do with the fact that the CIB is a sum of grey bodies at different redshifts, which means that the CIB is not described by a single SED, or equivalently that there is decorrelation across observations of the CIB at different channels; see, e.g.~\cite{mccarthyComponentseparatedCIBcleanedThermal2024}.}. But naturally, doing so boosts the bias from the other foregrounds (and the reconstruction noise, though we do not show that explicitly here). For the extent of halo-masking we have assumed here, there is in fact no clear benefit from deprojecting either component. In particular, once this many tSZ-emitting clusters have been removed, there is no clear advantage in explicitly deprojecting tSZ, as this comes at the cost of much larger CIB biases.

A final takeaway from figure~\ref{fig:tsz_cib_mixed_biases_to_auto_vs_Mcut} is that with more extensive halo masking, contributions from the secondary bispectrum bias can eventually become important, particularly at high multipoles and for the CIB bias. This behaviour is also seen in figure~\ref{fig:auto_biases_vs_fgcleaning}, where, for $\ell_{\rm max}=3000$, the biases are always negative on the multipole range plotted. This is because they are dominated by the bispectrum biases; in fact, except for the largest angular scales, it is the secondary coupling that dominates and counters the increase in amplitude of the trispectrum\footnote{Contrary to the claim of Ref.~\cite{omoriAgoraMultiComponentSimulation2022}, which attributes the neutralization of the trispectrum term to the subtraction of $N^{(0)}$ and $N^{(1)}$.}. This regime would thus require particular care, as the secondary couplings of the bispectrum are notoriously difficult to harden against~\cite{ref:sailer_et_al}.

\subsection{CMB lensing cross-correlations}\label{sec:results_cross}
Our formalism also lets us calculate biases to cross-correlations between CMB lensing reconstructions and lower-redshift tracers of the large-scale structure. The relevant contributions all originate from (primary) bispectra and are described in appendix~\ref{appendix:hm_bispectra}. In this section, we focus on cross-correlations with a galaxy number density tracer as an example, though we note that the extensions to other probes such as cosmic shear are straightforward. 

Let us then consider the cross-correlation of luminous red galaxies (LRGs) similar to those observed by the Dark Energy Spectroscopic Instrument (DESI) with CMB lensing reconstructions from an experiment similar to the ACT DR6. The calculation entails placing central and satellite galaxies in halos following an HOD, with further details provided in appendix~\ref{appendix:hm_bispectra}. For definiteness, we use the redshift distribution of the DESI LRGs observed during the science validation run~\cite{collaborationValidationScientificProgram2023}, which extends between roughly $0.3<z<1.1$, peaking at $z\approx0.9$. We approximate the ACT DR6 observations following table~\ref{tab:experiment_details} and adopt an $\ell_{\rm max}=3000$ as in the fiducial DR6 lensing analysis~\cite{quAtacamaCosmologyTelescope2024}.

\begin{figure}
    \centering
    \includegraphics[width=\textwidth]{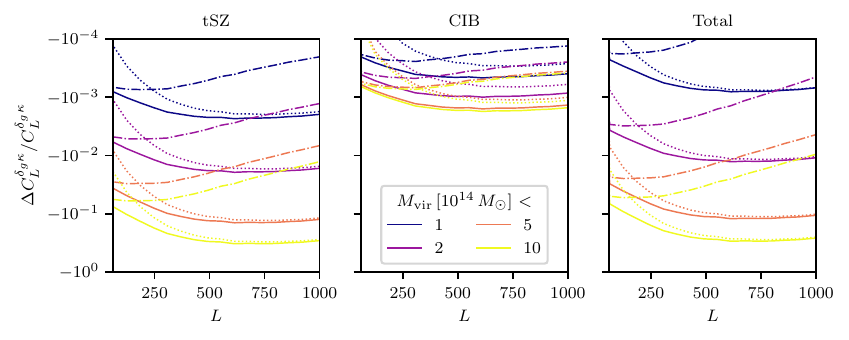}
    \caption{Fractional bias to the cross-correlation of DESI-like LRGs with ACT-DR6-like CMB lensing reconstructions, broken up into contributions from one-halo (dotted), two-halo (dot-dashed) terms as well as the their sum (solid). Different panels show the impact of different foreground contaminants, with the rightmost one showing the combination of tSZ and CIB, including contributions arising from the fact that they are correlated. This analysis assumes an $\ell_{\rm max}=3000$ and that no foreground cleaning is applied to the CMB maps besides a removal of structures more massive than the thresholds listed in the legend (identified, presumably, via their tSZ signature).  }
    \label{fig:cross_biases}
\end{figure}

Figure~\ref{fig:cross_biases} shows the amplitude of the foreground bias as a fraction of the signal and as a function of angular scale for this hypothetical cross-correlation measurement. Different colors correspond, once again, to different values of the maximum mass of halos that are allowed to feature in the input CMB maps. The lessons we learnt in the previous section around the primary bispectrum bias can in fact directly be ported to cross-correlations. We immediately notice that the tSZ bias is very sensitive to the removal of massive clusters. Note that ACT DR6 is expected to detect and remove halos above approximately $(2\text{--}4)\,\times 10^{14} \mathrm{M}_{\odot}$,\footnote{We estimate this as a moderate improvement on the mass detection limit reported in Ref.~\cite{hiltonAtacamaCosmologyTelescope2021a} for the earlier DR5.} at which point the bias is percent-level across all the relevant scales for this choice of $\ell_{\rm max}=3000$. The CIB contribution is less sensitive to the removal of massive clusters because the CIB is sourced primarily by dusty, star-forming galaxies inhabiting significantly lighter halos. Fortunately, however, this contribution is much smaller to begin with, as found also by Ref.~\cite{omoriAgoraMultiComponentSimulation2022}. This is likely because LRGs tend to populate rather massive halos that in turn are bright in tSZ. They are much less likely to be found alongside the young, star-forming kinds of galaxies that source the CIB; moreover, the overlap in redshift is also smaller.

Figure~\ref{fig:cross_biases} also breaks up the bias into one- and two-halo contributions. The relative importance of these two changes with the maximum mass threshold --- with the two-halo gaining relevance at any given scale the lower the maximum mass cut --- but in general (for this experimental configuration at least) it can be said that the two-halo term dominates the tSZ contribution below $L\lesssim 250$ and the CIB one below $L\lesssim 400$.

\begin{figure}
    \centering
    \includegraphics[width=\textwidth]{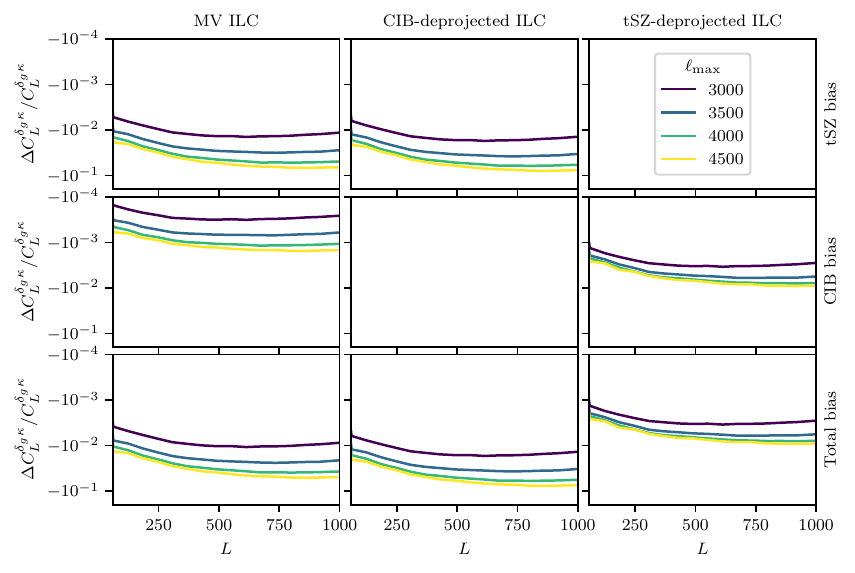}
    \caption{Fractional bias to the cross-correlation of DESI-like LRGs with SO-like CMB lensing reconstructions as a function of multi-frequency cleaning prescription (different columns), maximum multipole used in the reconstructions ($\ell_{\rm max}$; different colors) and extragalactic foreground (different rows). Halos with masses above $M_{\rm vir}>2\times 10^{14}\,\rm{M}_{\odot}$ are assumed to be masked away. Unlike for the lensing auto-spectrum, in this case deprojecting tSZ does lead to lower bias. The CIB bias with CIB deprojection is off scale though not exactly zero as is the case for the tSZ bias with tSZ deprojection.}
\label{fig:cross_biases_vs_fgcleaning}
\end{figure}

Figure~\ref{fig:cross_biases_vs_fgcleaning} then explores a wider range of $\ell_{\rm max}$ values and multi-frequency foreground cleaning prescriptions in the context of cross-correlating DESI LRGs with SO-like reconstructions (see table~\ref{tab:experiment_details} for details of how we parametrize this experiment). As expected, raising $\ell_{\rm max}$ increases the bias amplitude and alters its shape somewhat, becoming larger than percent-level per multipole once $\ell_{\rm max}\gtrsim 3000$. Folding in these additional small-scale modes would also reduce the size of the statistical error bars, leading to an even bigger increase in the statistical significance of the bias. In contrast to the case of the lensing auto-spectrum, we now see that deprojecting the tSZ is a rather effective way to reduce the overall bias. This is because before projection, the tSZ bias is much larger than CIB; deprojecting tSZ very effectively removes the tSZ bias but does not boost the CIB bias enough for there to be no overall benefit.

\subsubsection{Understanding the sign structure of the primary bispectrum bias}\label{sec:sign_change}
\begin{figure}
    \centering
    \includegraphics[width=\textwidth]{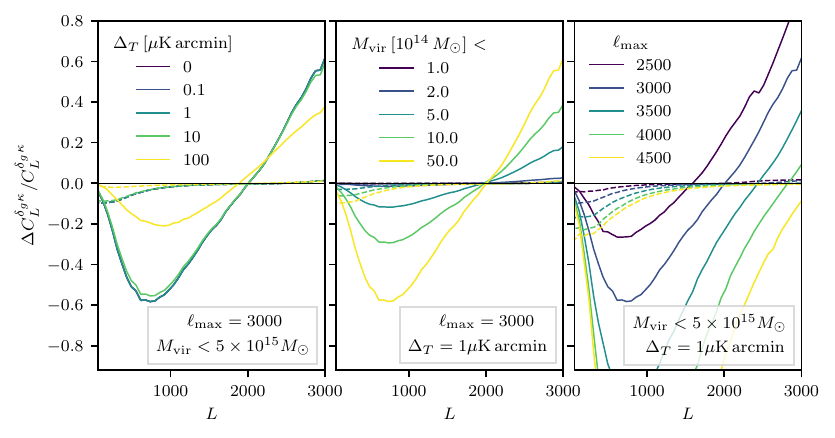}
    \caption{Sign structure of the primary bispectrum bias, relevant to both auto- and cross-correlations. We investigate this by varying the noise level (left panel), maximum halo mass (center) and $\ell_{\rm max}$ (right) in the cross-correlation of DESI-like LRG galaxies with reconstructions from an idealized CMB experiment with a point-like beam observing at 143\,GHz, and focusing on the dominant bias from tSZ. Solid and dashed lines show the one- and two-halo contributions, respectively. We find, without loss of generality, that the zero-crossing of the bispectrum bias at a fixed $\ell_{\rm max}$ does not change once noise levels drop below a few tens of $\mu$K\,arcmin, as the temperature anisotropies relevant for lensing reconstruction are all resolved by then. Similarly, contributions from different halo masses change the overall amplitude of the bias, but not the scale of zero-crossing. What does impact the location of the sign change is $\ell_{\rm max}$. This can all be attributed to the coupling structure of the quadratic-estimator weights.}
    \label{fig:zero_crossing_prim_bispec}
\end{figure}

Since cross-correlations are only sensitive to primary couplings of the bispectrum, they are an ideal testbed to study the sign change of the primary bispectrum bias, which also affects auto-correlations and delensing. Figure~\ref{fig:zero_crossing_prim_bispec} strives to understand this zero-crossing by studying how it is impacted by the level of white noise (left panel), maximum-halo-mass cuts (middle panel) and $\ell_{\rm max}$ (right panel). Without loss of generality, we focus on the bias sourced by tSZ. The first of the variables --- white noise --- seems to have no impact once it drops below around $50\,\mu\mathrm{K\,arcmin}$, that is, low enough that the intermediate-scale CMB temperature anisotropies needed to reconstruct these lenses are signal-dominated. The second variable, the maximum-mass cut, does have a signficant impact on the amplitude of the bias, but not on the location of the zero-crossing. On the other hand, $\ell_{\rm max}$ affects both the amplitude and the shape of the bias -- including the scale where it changes sign.

This behaviour can be understood from the QE weights of equation~\eqref{eqn:QE_weights}, which we quote here again for convenience:
\begin{equation}
    g(\bm{l},\vL) =  \frac{(\vL - \vl)\cdot \vL \tilde{C}^{TT}_{|\vl-\vL|} + \vl\cdot \vL \tilde{C}^{TT}_l}{2C_l^{TT,\mathrm{obs}} C_{|\bl - \bL|}^{TT,\mathrm{obs}}}\,.
\end{equation}
Let us also quote the expression for the dominant one-halo contribution to the primary bispectrum bias,
\begin{align}
    \Delta C^{\phi 
    \phi}_{L} \supset & 2\, A^{TT}_{L}\int \frac{\mathrm{d}^2\bm{l}}{(2\pi)^2} g(\bm{l},\bm{L}) b_{\phi s_1s_2}(-\bm{L},\bm{l},\bm{L}-\bm{l})\, ,
\end{align}
which is of course exactly analogous to the term that affects cross-correlations. The angular bispectrum in the integrand is a projection of the 3D bispectrum, which to a very good approximation means that angular modes probe 3D scales $\bm{k} \approx \bm{l}/\chi$ at comoving distance $\chi$. The one-halo bispectrum is an integral over the emission profiles of the foregrounds and matter. For the typical masses of the halos that we know contribute, these must be probed at $k<O(10)\,h/\mathrm{Mpc}$ before falling off quickly. Since the structures are at comoving distances of $(3000\text{--}6000)\,h/\mathrm{Mpc}$, this means that there can be significant contributions from a given halo as long as all the modes being probed have wavenumbers of at most $l\sim O(10,000)$. Importantly, also, $ b_{\phi s_1 s_2}$ is always positive.

Let us now consider two limits. First, suppose we are probing lensing modes on scales much smaller than those that dominate the reconstruction, such that $l\gg L$. The two terms in the QE weights above cancel at leading order in this squeezed limit, leaving
\begin{equation}
g(\bm{l},\vL) \rightarrow \frac{L^2}{4}\frac{\tilde{C}_l^{TT}}{(C_l^{TT,\text{obs}})^2} \left[\frac{d}{d\ln l} \ln(l^2 \tilde{C}_l^{TT}) + \cos 2(\psi_{\bm{l}} - \psi_{\vL})\frac{d}{d\ln l} \ln \tilde{C}_l^{TT}\right] \, .
\end{equation}
In the damping tail, both derivative terms are negative. (We note also that the shear term, with the $\cos 2(\psi_{\bm{l}} - \psi_{\vL})$ angular dependence, does not contribute to the one-halo bispectrum in this squeezed limit as the profiles are circularly symmetric.) This, combined with 
$b_{\phi s_1 s_2}>0$, makes the primary bispectrum bias negative. On the other hand, if $L \sim l_{\text{max}}$, the region that we integrate over in $\bm{l}$ --- i.e., the intersection of $|\bm{l}|\leq l_{\text{max}}$ and $|\vL-\bm{l}| \leq l_{\text{max}}$ --- is skewed towards $\bm{l} \cdot \vL > 0$, giving a positive contribution to the bias from the second term in the weights (the first is similarly positive by symmetry).
Somewhere between these two regimes, a sign change must occur and this will shift to larger $L$ as $l_{\text{max}}$ is increased.
It is also no coincidence that the trispectrum bias typically sees a dip at the scale where the primary bispectrum changes sign (cf. figure~\ref{fig:tsz_cib_mixed_biases_to_auto_vs_Mcut}). The one-halo contribution to the trispectrum bias, which dominates on these scales, is simply a sum over halos of terms that are each a product over four profiles and the integrals over $\bm{l}$ and $\bm{l}'$ separate into factors, each of which is just like the integral of the weight function and two foreground profiles, as in the one-halo primary bispectrum bias. Each then vanishes at some $L$ due to cancellations across the $\bm{l}$ integral, but instead of going negative beyond that $L$ as in the bispectrum, here the result is positive because we are dealing with the product of two equal (negative) factors.
Incidentally, this suggests there is typically a regime around these scales where the secondary bispectrum bias --- which is not affected by these considerations --- can become more important relative to the other two terms.

\subsection{Biases to $B$-mode delensing}\label{sec:results_delensing}

Following on from section~\ref{sec:delensing}, we can take predictions for the CMB lensing auto-spectrum biases and use them to model corrections to the power spectrum of $B$-mode polarization after delensing with $TT$ quadratic estimators.

As is clear from equation~\eqref{eqn:full_theory_auto}, this calculation also takes as input the \emph{total} power spectrum of the $E$-mode observations used to construct the template. In realistic applications, these $E$-modes may have been processed through various foreground-cleaning schemes, affecting their power spectrum. But on the intermediate and small scales relevant for delensing, it is the primary CMB anisotropies that dominate over most other components, so we expect a minimum-variance ILC to be apt for most purposes\footnote{Except perhaps if the CIB is included as a matter tracer for delensing, in which case the very largest scales could be affected by a bispectrum of residual dust which could be mitigated by deprojecting dust from the $E$-modes~\cite{ref:cib_delensing_biases}.}. This is therefore the only $E$-mode cleaning scheme that we implement for now. We also assume that the $E$-modes are obtained from the same telescope as the inputs to the quadratic estimator of lensing.

\begin{figure}
    \centering
    \includegraphics[width=\textwidth]{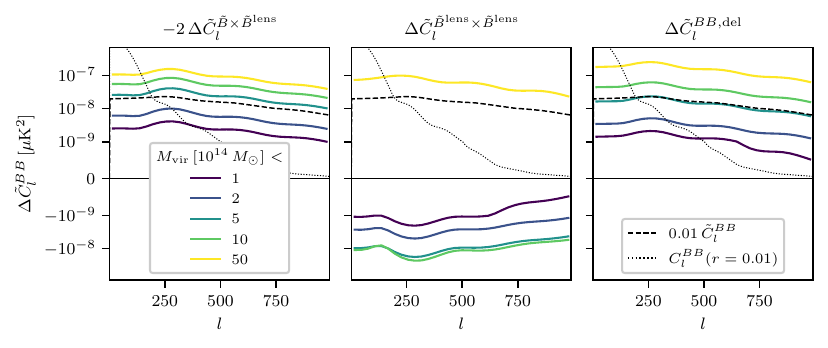}
    \caption{Breakdown of contributions to the absolute bias to the power spectrum of delensed $B$-modes as a function of the maximum halo mass featured in the calculation. The left panel shows the cross-spectrum between the lensing $B$-mode template and the true lensing $B$-modes, which only receives contributions from the primary bispectrum bias. The center panel shows the auto-spectrum of the template, which is sensitive to both primary and secondary bispectra as well as the trispectrum; for low enough mass cuts, bispectra dominate and produce a negative correction that partially cancels the contributions in the leftmost panel. The rightmost panel shows the total bias to the $B$-mode power spectrum. For reference, dashed, black curves show the lensing $B$-mode power spectrum rescaled down to 1\% of its original amplitude, while dotted black curves show a primordial component with $r=0.01$. This plot assumes a minimum-variance combination of SO-like observations with $\ell_{\rm max}=3000$ and captures contributions from both tSZ and CIB.}
\label{fig:delensing_biases_vs_Masscut}
\end{figure}

Figure~\ref{fig:delensing_biases_vs_Masscut} shows absolute biases to the power spectrum of $B$-modes after delensing with $TT$ QE lensing reconstructions obtained from an experiment with characteristics similar to SO, fixing $\ell_{\rm max}=3000$ and using a minimum-variance combination of all six frequency channels. We consider various cases with different cuts on the maximum mass of halos. It is useful to split this delensing bias (the total of which is shown in the rightmost panel of the figure) into contributions from the lensing $B$-mode template auto-spectrum (center panel) or this template's cross-correlation with the true lensing $B$-modes (left panel). Recall that the latter is only sensitive to a primary bispectrum bias, while the former can in principle receive contributions from both primary and secondary couplings of the bispectrum and also from the trispectrum. 

\begin{figure}
    \centering
    \includegraphics[width=\textwidth]{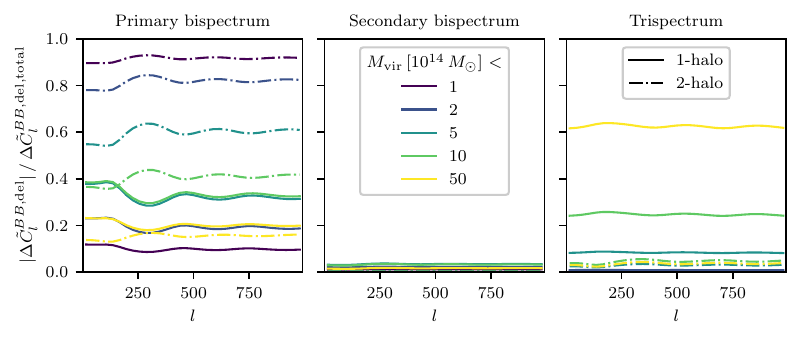}
    \caption{Fractional contribution to the total $B$-mode delensing bias coming from different couplings, as a function of the maximum halo mass threshold. The sum across panels and line styles of all curves of a given color is unity. Higher halo mass thresholds lead to a larger preponderance of one-halo and trispectrum contributions. On the other hand, primary bispectrum contributions eventually dominate as massive halos are masked/removed, with the emphasis also shifting to two-halo contributions. Notably, secondary bispectrum terms are typically negligible. This plot assumes a minimum-variance combination of SO-like observations with $\ell_{\rm max}=3000$ and captures contributions from both tSZ and CIB.}
\label{fig:delensing_biases_structure_vs_Masscut}
\end{figure}

The relevance of the different couplings will depend on what range of lens-scales (i.e., $L$) are most important when delensing large-scale $B$-modes. We can use this to glean some immediate insights. Figure 3 of Ref.~\cite{ref:baleato_and_ferraro_22} shows that the overwhelming majority of the delensing effect comes from lensing modes with $L\lesssim 1500$. Referring to figure~\ref{fig:tsz_cib_mixed_biases_to_auto_vs_Mcut}, we see that the secondary coupling of the bispectrum is typically highly subdominant to other contributions on these scales. Since the primary bispectrum bias is always negative on these scales, it follows that $-2\,\Delta \tilde{C}_{L}^{\tilde{B} \times \tilde{B}^{\rm lens}}$ always induces a positive correction to the $B$-mode power spectrum. This is indeed what we see in the leftmost panel of figure~\ref{fig:delensing_biases_structure_vs_Masscut}. On the other hand, the picture is not so clear-cut for the template auto-spectrum in the central panel. For low enough halo mass limits, the bispectrum biases dominate and the overall bias is negative. As the mass limit is raised, however, the trispectrum grows faster than the bispectra, and the overall contribution can take a positive sign, spoiling cancellations with the template-cross-spectrum term. Though generally the total bias (rightmost panel) grows as the mass cut is raised, growth is faster once these cancellations disappear.

The relevance of different couplings is made more explicit in figure~\ref{fig:delensing_biases_structure_vs_Masscut}, where we plot their fractional contribution to the total bias while we vary the maximum-halo-mass cuts. We also break up contributions into one- and two-halo terms. An interesting trend arises where, as halos of an increasingly lower mass are excluded, the emphasis shifts from one-halo trispectra to a combination of one- and two-halo primary bispectra at first, and finally to the two-halo primary bispectrum alone. This plot also confirms our expectation laid out above that the secondary coupling of the bispectrum has a negligible effect.

\begin{figure}
    \centering
     \includegraphics[width=\textwidth]{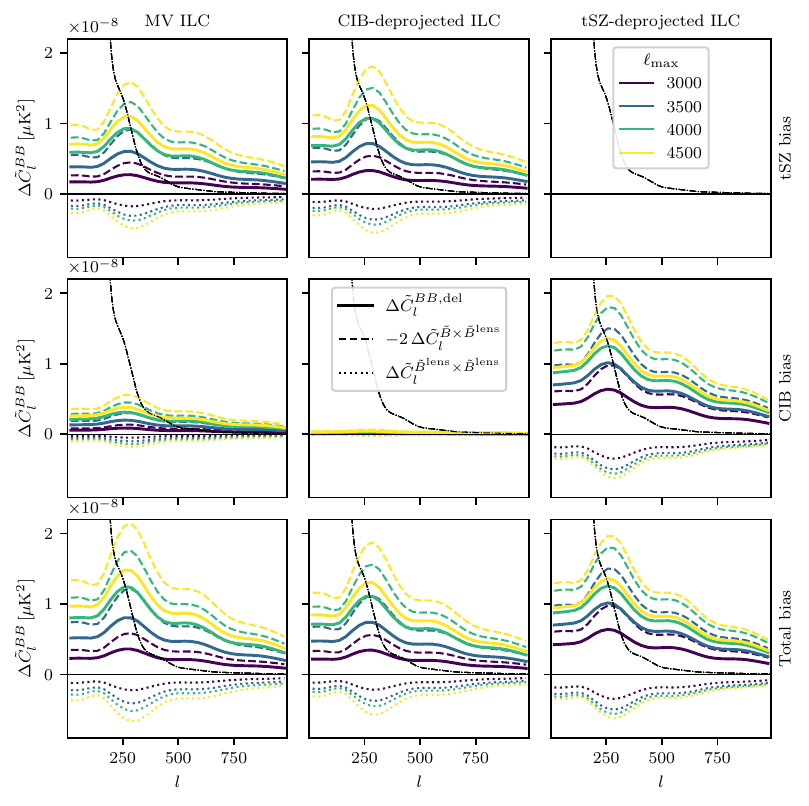}
    \caption{Absolute bias to the power spectrum of $B$-modes after delensing with $TT$ QE reconstructions from an SO-like experiment, shown as a function of the multi-frequency cleaning prescription (different columns), maximum multipole used in the reconstructions ($\ell_{\rm max}$; different colors) and extragalactic foreground (different rows). We assume halos with $M_{\rm vir}>2\times 10^{14}\,\mathrm{M}_{\odot}$ have been masked away. The bias to the total power spectrum is shown as solid lines, while corrections to the lensing $B$-mode template auto- and cross-spectrum are shown as dotted and dashed lines, respectively. For reference, a primordial $B$-mode component with $r=0.01$ is shown via black, dot-dashed lines.}
\label{fig:delensing_biases_vs_fgcleaning}
\end{figure}

Finally, in figure~\ref{fig:delensing_biases_vs_fgcleaning}, we also look into different multi-frequency combinations of the different channels, as well as different $\ell_{\rm max}$, fixing the mass cut to $M_{\rm vir}<2\times 10^{14}\,\mathrm{M}_{\odot}$.\footnote{Any quantitative differences between our results and those in Ref.~\cite{ref:baleato_and_ferraro_22} likely originate from differences in maximum-halo-mass cuts, with their results including radio and kSZ components in addition to tSZ and CIB, and lastly, from the fact that the \texttt{WebSky} CIB maps they used have since been updated due to the discovery of issues (in addition, the mass resolution of \texttt{WebSky} is insufficient to reproduce the CIB non-Gaussianity, as we will soon discuss).} The merit of different foreground cleaning prescriptions is a sensitive function of the extent to which the lensing $B$-mode template auto- and cross-spectrum biases cancel each other out~\cite{ref:baleato_and_ferraro_22}. The overall bias thus results from a detailed balance of halo-masking schemes, $\ell_{\rm max}$ and multi-frequency cleaning that can sometimes give counter-intuitive results~\cite{ref:baleato_and_ferraro_22}. We do not explore this large parameter space any further, though we note that our code and framework can be used to do so efficiently.

\subsection{Sensitivity to redshifts, masses and scales}\label{sec:hyperparam_requirements}
Our analytic formalism and code can also be used to determine the ranges of halo masses, redshifts and scales that contribute to the overall bias signal. This is useful not only to assess the convergence of the calculation, but also to inform requirements on any simulations if they are to characterize these effects faithfully.

\begin{figure}
    \centering
    \includegraphics[width=0.8
    \textwidth]{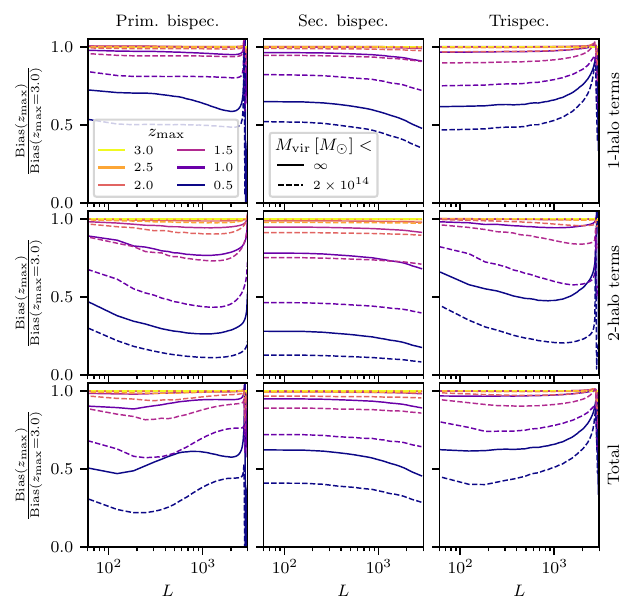}
    \caption{Convergence of tSZ-bias contributions to the CMB lensing auto-spectrum as a function of the maximum redshift $z_{\rm max}$ included in the calculation. We consider contributions from halos with a minimum mass $10^{7}< M_{\rm vir}  [\,\mathrm{M}_{\odot}]$ and a maximum mass of either $  M_{\rm vir}  [\,\mathrm{M}_{\odot}]< \infty$ (solid) or $  M_{\rm vir}  [\,\mathrm{M}_{\odot}]< 2\times10^{14}$ (dashed).  Columns show different couplings of the four-point function, and rows how many halos pair together to produce a given term. The bottom row shows the total contribution (the sum of one- and two-halo terms) for each coupling. We see that the tSZ contribution converges fast with redshift: approximately 90\% (75\%) of contributions to $L<3000$ come from $z<1$ when considering $  M_{\rm vir}  [\,\mathrm{M}_{\odot}]< \infty$ ($  M_{\rm vir}  [\,\mathrm{M}_{\odot}]< 2\times10^{14}$). Note that curves often lie on top of each other at a ratio value of unity, indicating that the calculation has converged. }
\label{fig:tsz_convergence_vs_zmax}
\end{figure}

The natural first thing to look at is what redshifts are more important. Figure~\ref{fig:tsz_convergence_vs_zmax} shows the angular power spectrum of the tSZ bias to the CMB lensing auto-spectrum\footnote{The shape of these fractional plots is to a large extent dictated by the fact that we are plotting the ratio of quantities that can come very close to zero (in the case of the trispectrum) or even change sign (in the case of the primary bispectrum, which cross zero around $L\sim 2800$ for $\ell_{\rm max}=4000$). The ratio becomes formally infinite when the denominator crosses zero, giving rise to kinks in the curves.}, calculated by including structures only up to some redshift $z_{\rm max}$, and expressed as a function of the result for $z_{\rm max}=3.0$, by which point the observable has effectively converged.
For definiteness, we will assume throughout this section that the CMB observations are noiseless and go up to $\ell_{\rm max}=4000$, and carry out the calculation using a minimum redshift of $z_{\rm min}=0.05$ and scales $k<10\,\rm{Mpc}^{-1}$ to be very conservative, though we will optimise these choices later. Let us first consider halos with $10^{7}< M_{\rm \rm vir} [\,\mathrm{M}_{\odot}]< \infty$; this corresponds to the solid lines in the figure. Since the tSZ signal is a strong function of halo mass and the most massive halos on our past lightcone are found preferentially at lower redshifts, the signal is dominated by low-$z$ contributions: the bottom row shows that over 90\% of the bias on scales $L<3000$ comes from $z<1$, with approximately $40\text{--}50\%$ originating from the range $0.5<z<1$. By contrast, contributions above $z>1.5$ provide quickly diminishing returns. When massive halos are masked away (dashed curves in the figure correspond to $M_{\rm vir} [\mathrm{M}_{\odot}]<2\times 10^{14}$) the calculation converges more slowly with redshift, though the qualitative picture remains unchanged.

\begin{figure}
    \centering
    \includegraphics[width=0.8
    \textwidth]{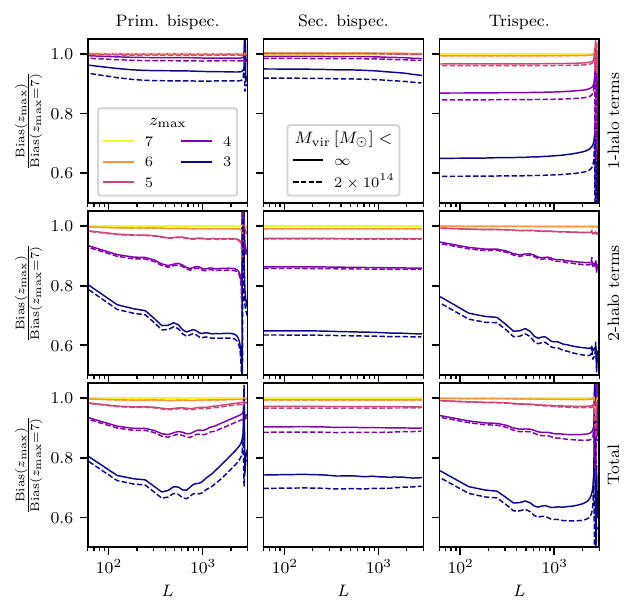}
    \caption{As figure~\ref{fig:cib_convergence_vs_zmax}, but for the CIB. Convergence is now slower, requiring $z_{\rm max}\approx 5$ to get to accuracies of a few percent.}
\label{fig:cib_convergence_vs_zmax}
\end{figure}

Figure~\ref{fig:cib_convergence_vs_zmax} plots the same quantities for the CIB, painting a signficantly different picture where higher redshifts now feature more prominently and convergence is not attained until $z_{\rm max}\approx 7$. Limiting the calculation to $z<3$ misses out on around $30\%$ of contributions. Restricting instead to $z<5$ allows for calculations accurate at the 5\% level; this will be our default choice of maximum redshift going forward.

Both figures~\ref{fig:tsz_convergence_vs_zmax} and~\ref{fig:cib_convergence_vs_zmax} break down convergence by bias coupling (`bispectrum', `trispectrum', etc.) and how many halos are involved. We note that all other things being equal, one-halo contributions converge faster (i.e., at lower redshift) than two-halo ones. This can be explained from the fact that the most massive halos which dominate the signal are found primarily at lower redshifts. We also notice that the trispectrum bias for the CIB converges rather slowly compared to the bispectrum ones.

\begin{figure}
    \centering
    \includegraphics[width=0.8
    \textwidth]{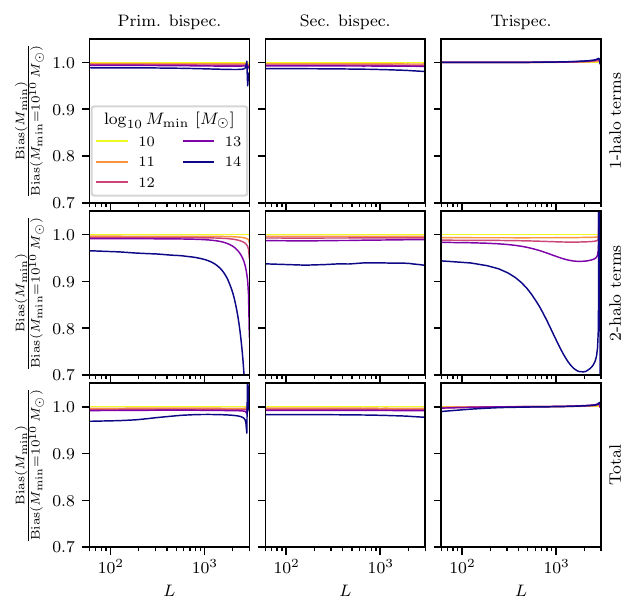}
    \caption{Convergence of tSZ bias contributions to the CMB lensing auto-spectrum as a function of the minimum halo mass $M_{\rm min}$ included in the calculation. Columns show different couplings of the four-point function, and rows how many halos pair together to produce a given term. The bottom row shows the total contribution (the sum of one- and two-halo terms) for each coupling. We see that the tSZ contribution is dominated by the most massive halos. Note that curves often lie on top of each other at a ratio value of unity, indicating that the calculation has converged. }
\label{fig:tsz_convergence_vs_Mmin}
\end{figure}

\begin{figure}
    \centering
    \includegraphics[width=0.8
    \textwidth]{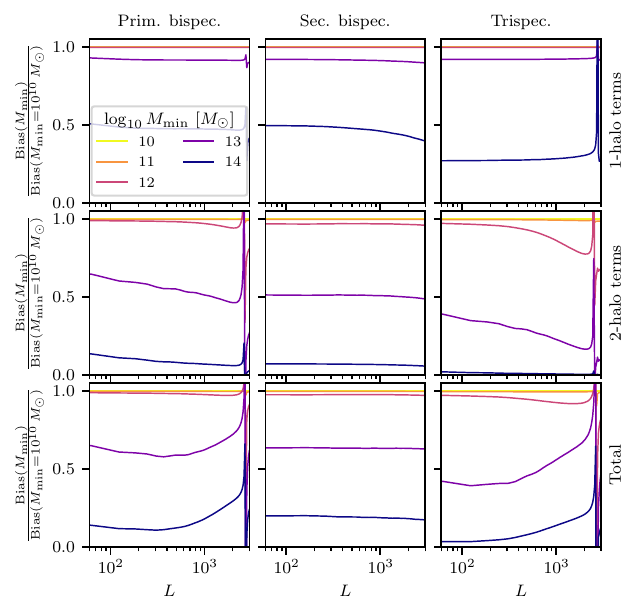}
    \caption{As figure~\ref{fig:tsz_convergence_vs_Mmin}, but for the CIB. Now, lower-mass halos are comparatively more important: halos with mass in the range $M>10^{12}\,M_{\rm \odot}$ need to be taken into account in order to achieve accuracies of the order of a few percent.}
\label{fig:cib_convergence_vs_Mmin}
\end{figure}

We can also investigate the relative importance of different halo mass ranges in producing these CMB lensing biases. Figures~\ref{fig:tsz_convergence_vs_Mmin} and~\ref{fig:cib_convergence_vs_Mmin} explore the convergence of results as a function of the minimum halo mass included in the calculation of the tSZ and CIB biases, respectively. In this case, we impose no upper limit on the mass of halos, so the tSZ calculation is dominated by the most massive structures: over 95\% of the signal originates from $M_{\rm vir}>10^{14}\,M_{\rm \odot}$. This behavior is even more acute for the one-halo contributions (top row), and among those, the trispectrum coupling (rightmost column). Though we do not do it here for conciseness, it is trivial to use our code to quantify how the minimum-halo-mass and redshift requirements vary with $\ell_{\rm max}$ and the ability to mask massive clusters.

It may not come as a surprise that it is high mass and low-redshift halos that dominate the tSZ biases given that only CMB modes up to $\ell_{\rm max}=4000$ are used in this particular example. Reference~\cite{ref:komatsu_seljak_02} reached a similar conclusion for the angular power spectrum of the tSZ. They also showed that lower-mass and higher-redshift structures become increasingly important once smaller-scale CMB modes are considered (cf. their figures 5 and 6). We expect a similar story to apply for CMB lensing biases, though we emphasize that these are sensitive to higher-order functions --- bispectra and trispectra. Once again, our code naturally lends itself to understanding these effects, though a more comprehensive study of this is beyond the scope of this work.

By contrast, the CIB biases, shown in figure~\ref{fig:cib_convergence_vs_Mmin}, are sensitive to much lower-mass halos. In this case, carrying out the calculation with $M>10^{14}\,M_{\rm \odot}$ accounts for only 20--30\% of the power. We find that a minimum mass of $M>10^{12}\,M_{\rm \odot}$ is needed in order to achieve accuracies of the order of a couple percent.

With regards to the relevant physical scales, we find that setting $k_{\rm max}\gtrsim 5\,\rm{Mpc}^{-1} $ suffices to obtain percent-level convergence in the calculation for all the relevant multipoles. We do not explore this aspect further. 

\subsection{Response to astrophysical and cosmological parameters}\label{sec:results_varying_params}
\subsubsection{CMB lensing auto-correlations}
Perhaps the single most important advantage of our new formalism is that it allows us to evaluate lensing biases in different cosmological and astrophysical models, whereas previously these were only measured at a handful of points in parameter space where simulations were run.

\begin{figure}
    \centering
\includegraphics[width=\textwidth]{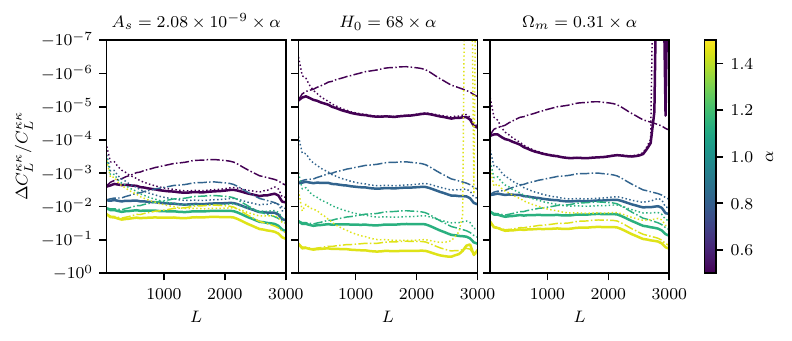}
    \caption{Impact of varying $\Lambda$CDM cosmological parameters on the relative bias to the CMB lensing auto-spectrum measured by an ACT-DR6-like experiment with $\ell_{\rm max}=3000$ and $M_{\rm vir}<2\times 10^{14}\,\mathrm{M}_{\odot}$. Dotted, dashed and solid lines correspond to the one-halo, two-halo and total bias, respectively. Though the relative amplitude of the bias depends on cosmology, the shape is largely insensitive to it as long as variations remain within a few percent of the fiducial parameter values. Note that $H_0$ is in units of $\rm{km}\,\rm{s}^{-1}\,\rm{Mpc}^{-1}$.}
    \label{fig:auto_biases_vs_cosmo}
\end{figure}

For a start, let us see how biases change when a few key parameters in the $\Lambda$CDM cosmological model are varied one at a time. Figure~\ref{fig:auto_biases_vs_cosmo} shows how biases to the total ACT-DR6-like lensing auto-spectrum change as we vary $A_s$, $H_0$ and $\Omega_m$ by up to 40\% of their central values measured by Planck 2018. Of course, the signal is also modified once we alter the cosmological parameters, and we take this into account in the fractional bias ratios we plot\footnote{We do not vary the fiducial cosmology assumed in the QE weights, which in principle degrades the optimality of the reconstructions, albeit by an amount that is negligible for our purposes.}. We find that, although the amplitude of the relative bias can respond quite strongly to changes in parameters (particularly $H_0$), the shape remains rather constant, especially once we restrict to parameter changes of order a few percent compatible with current constraints. The insensitivity to $A_s$, which may be counterintuitive at first given how strongly the halo mass function (and thus tSZ emission) depends on it, is likely due to the fact we are assuming that halos above $M_{\rm vir}<2\times 10^{14}\,\mathrm{M}_{\odot}$ are identified and removed in any of these cosmologies.

\begin{figure}
    \centering
    \includegraphics[width=\textwidth]{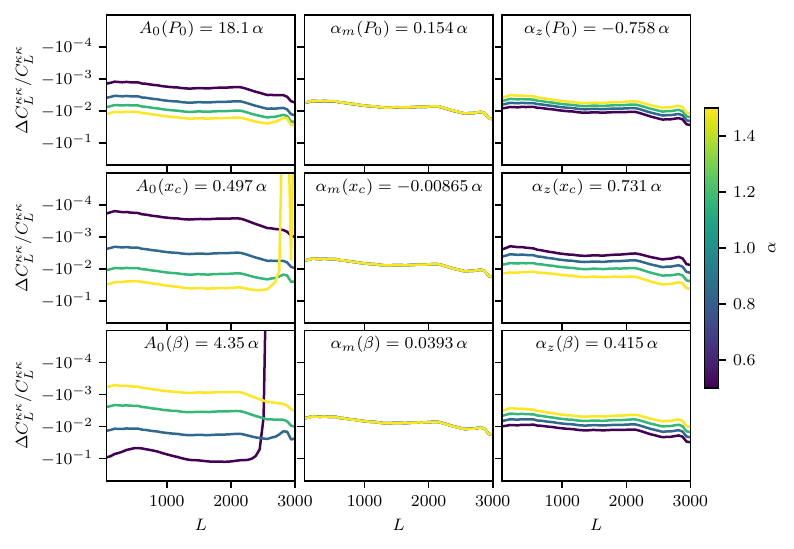}
    \caption{Response of tSZ-induced CMB lensing auto-spectrum biases to variations in the pressure profile parameters. We consider the auto-spectrum of ACT-like lensing reconstructions with $\ell_{\rm max}=3000$ and assuming that halos above $M_{\rm vir}=2\times 10^{14}\,\mathrm{M}_{\odot}$ have all been removed. We vary one model parameter at a time about the central values given in table~\ref{tab:battaglia_fit_params}, and by as much as 40\%. As explained in section~\ref{sec:tsz}, the right-most and central columns parameterize the evolution in redshift and mass, respectively, of the profile parameters in the left-most column. We find that sensitivity is greatest to changes in the latter. Importantly, variations manifest largely as scale-independent rescalings of the relative bias amplitude.}
\label{fig:tsz_auto_biases_vs_tszparams}
\end{figure}

A similar conclusion applies to changes in parameter values within the baseline tSZ and CIB models. Figure~\ref{fig:tsz_auto_biases_vs_tszparams} shows the impact of varying parameters controlling the tSZ profile described in table~\ref{tab:battaglia_fit_params} within 40\% of their fiducial values. We see a stronger dependence on the parameters' overall amplitude (left column) rather than their redshift or mass evolution (central and right columns); shifts to the former at the 20\% level can lead to order-of-magnitude changes to the bias amplitude. But notably, once again, the shape of the relative bias is typically altered only marginally if parameters vary only within 20\% of their fiducial values. Most likely, this insensitivity arises because for $\ell_{\rm max}< 3000$, as considered here, the SZ contribution is dominated by very massive halos ($M_{\rm vir}\gtrsim10^{14}\,\mathrm{M}_{\odot}$). It is well known that contributions from such massive structures come mainly from gas away from the cluster core and are thus relatively insensitive to the details of feedback~\cite{ref:komatsu_seljak_02, efstathiouPowerSpectrumThermal2025}. This situation may well change once smaller-scale CMB modes are used for lensing reconstruction as these probe less-massive halos at higher redshifts.

\begin{figure}
    \centering
    \includegraphics[width=\textwidth]{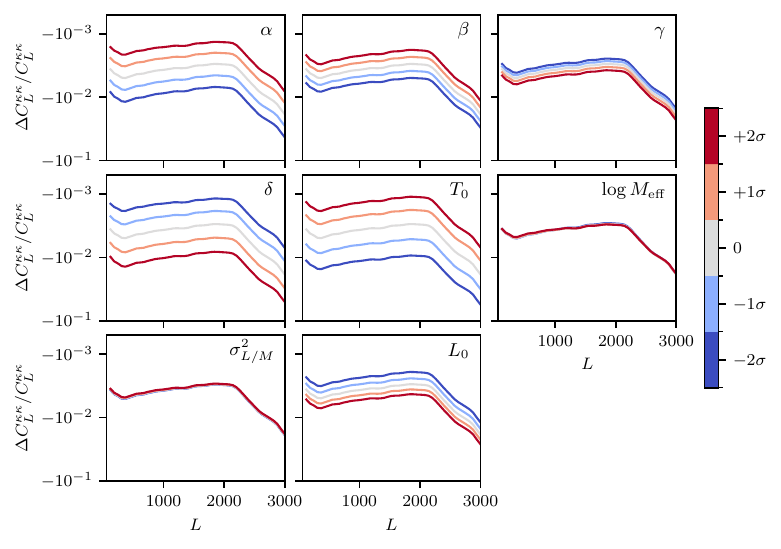}
    \caption{Response of CIB-induced CMB lensing auto-spectrum biases to variations in the CIB halo-model parameters. We consider an ACT-like CMB experiment with $\ell_{\rm max}=3000$ and assume that halos above $M_{\rm vir}=2\times 10^{14}\,\mathrm{M}_{\odot}$ have all been removed. We vary one model parameter at a time about the central values given in table~\ref{tab:planck_cib_model_params}, and by a number of standard deviations given in the same table (for $L_0$ and $\sigma^2_{L/M}$, we take $\sigma$ to be 10\%). These variations manifest largely as scale-independent rescalings of the relative bias amplitude. Note also that uncertainties in $\delta$ and $T_0$  dominate the error budget.}
\label{fig:cib_biases_to_auto_vs_cibparams}
\end{figure}

The equivalent exploration of variations in the CIB halo-model parameters is given in figure~\ref{fig:cib_biases_to_auto_vs_cibparams}. This time, we have access to an uncertainty associated with the model's best-fit parameters (see table~\ref{tab:planck_cib_model_params} in appendix~\ref{appendix:hod_factorial_moments}), so we are able to quantify the size of the variations we explore in terms of the number of $\sigma$ away from the best-fit that that these correspond to. We find that uncertainties are dominated by the $T_0$,  $\alpha$ and $\delta$ parameters. These parametrize, respectively, the CIB monopole temperature, its redshift evolution, and the power-law evolution of the IR luminosity function of dusty galaxies. However, the shape of the relative biases is once again unaffected by parameter shifts.

The remarkable insensitivity of the shape of the relative bias to changes in cosmological or astrophysical parameter values suggests that any residual bias could be dealt with by marginalizing over the amplitude of a bias template determined either from simulations or analytically using the tools developed in this work. The addition of this single extra parameters is unlikely to degrade constraints significantly. We note, however, that this is but a preliminary exploration, and in the future, these conclusions should be validated against different foreground parameterizations, not just changes to parameter values within a single model. Those alternative parameterizations can likely be built into the analytic framework we have developed.

\subsubsection{CMB lensing cross-correlations}
\begin{figure}
    \centering
    \includegraphics[width=\textwidth]{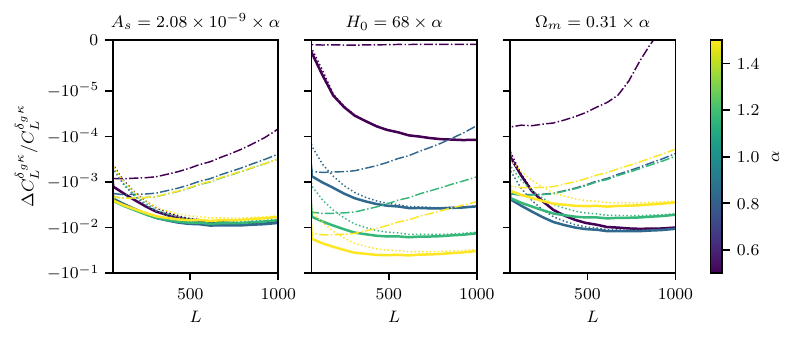}
    \caption{As figure~\ref{fig:auto_biases_vs_cosmo}, but for the cross-correlation of ACT-like lensing with DESI-like LRGs. We now zoom in on $L<1000$ as these are the multipoles that will carry the bulk of the signal-to-noise for the foreseeable future and also the ones for which accurate models exist~\cite{modiModelingCMBLensing2017, modiSimulationsSymmetries2020}. This time, the shape of the bias responds more strongly to changes in $\Omega_m$ and, to a lesser extent, $A_s$. }
    \label{fig:cross_biases_vs_cosmo}
\end{figure}

\begin{figure}
    \centering
    \includegraphics[width=\textwidth]{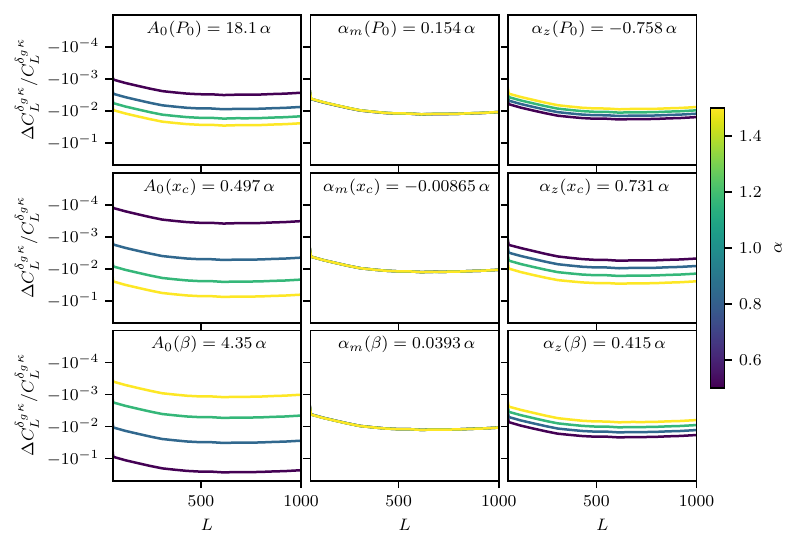}
    \caption{As figure~\ref{fig:tsz_auto_biases_vs_tszparams}, but for the cross-correlation of DESI-like LRGs with ACT-like CMB lensing maps. Once again, variations manifest largely as scale-independent rescalings of the relative bias amplitude.}
\label{fig:tsz_cross_biases_vs_astro}
\end{figure}

\begin{figure}
    \centering
    \includegraphics[width=\textwidth]{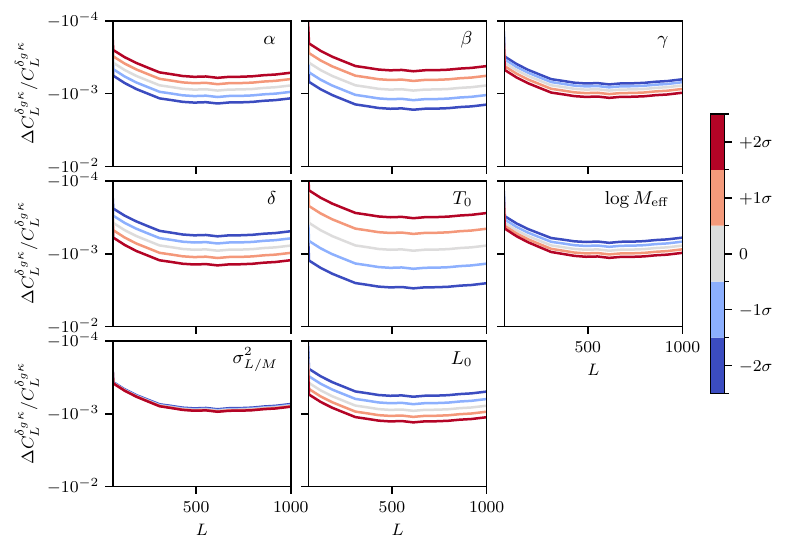}
    \caption{Same as figure~\ref{fig:cib_biases_to_auto_vs_cibparams}, but the cross-correlation of DESI-like LRGs with ACT-like CMB lensing maps. Once again, parameter variations manifest largely as scale-independent rescalings of the relative bias amplitude. Note also that uncertainties in $T_0$  dominate the error budget.}
\label{fig:cib_biases_to_cross_vs_cibparams}
\end{figure}

Analogous tests to those in the previous subsection can be done for cross-correlations. For definiteness, we consider the cross-correlation of ACT-like lensing reconstructions with DESI-like LRGs. Figure~\ref{fig:cross_biases_vs_cosmo} explores how biases to such a cross-correlation measurement are affected by changes to $A_s$, $H_0$ and $\Omega_m$. The situation is now somewhat different to the case of auto-correlations: the shape of the relative bias now \emph{does} seem to respond to changes in $\Omega_m$.\footnote{It is especially important to be aware of this dependence given the current discrepancy between the values of $\Omega_m$ inferred from the primary CMB, uncalibrated supernovae and low-redshift BAO~\cite{loverdeMassiveNeutrinosCosmic2024}.} The good news is that the relative bias shape is still robust to changes in astrophysical model parameters, as evidenced by figure~\ref{fig:tsz_cross_biases_vs_astro} for changes in the tSZ pressure profile and figure~\ref{fig:cib_biases_to_cross_vs_cibparams} for variations in the CIB halo-model parameters.

\section{Comparison to simulations}\label{sec:comparison_with_websky}

\subsection{Simulations of the extragalactic sky at microwave frequencies}
We now set out to validate our calculations against simulations, noting that we can only aspire to attain qualitative agreement
%with the mocks
given the approximate nature of the halo model, the large uncertainties in our understanding of the astrophysical processes driving the lensing biases, and the computational limitations associated with the simulations.

Let us briefly comment on the latter issue. Generating mock observations of the microwave sky over the large sky fractions covered by modern CMB experiments is extremely challenging from a computational standpoint. Huge volumes are needed to account for the phenomena on our past lightcone that impact the CMB lensing signal. Large volumes are also required to sample the high-mass tail of the mass function, which is crucial for effects such as the tSZ. At the other end of the spectrum, we saw in section~\ref{sec:hyperparam_requirements} that halo masses as low as $10^{11}\,\mathrm{M}_{\odot}$ are needed to capture the CIB appropriately. A very large number of particles is therefore required in order to bridge these requirements and attain both sufficient volume and mass resolution.

Unfortunately, it is not possible at present to run hydrodynamical simulations of this size featuring baryonic effects. For this reason, the CMB community relies on mocks based on approximations to the LSS, calibrated to observations and hydrodynamical simulations. The simulations of Refs.~\cite{ref:sehgal_sims_07,ref:sehgal_sims_10, ref:websky, omoriAgoraMultiComponentSimulation2022}, which have proven invaluable for over a decade in the analysis of CMB data, are an example of this. In them, astrophysical prescriptions are used to assign emission to dark matter halos.

Two popular representatives of this lineage of simulations of the microwave sky are the \texttt{WebSky}~\cite{ref:websky} and \texttt{Agora}~\cite{omoriAgoraMultiComponentSimulation2022} simulations. While the large volume of \texttt{WebSky} (around $600\,(\mathrm{Gpc}/h)^3$) is key for many applications, allowing it to reach $z=4.6$ without repetition, it can only resolve halos more massive than $10^{12}\,\mathrm{M}_{\odot}$, which could potentially compromise the accuracy with which it can simulate the CIB. On the other hand, \texttt{Agora} replicates a box of side-length $1\,h^{-1}$\,Gpc many times along the line-of-sight out to $z=3$ for the tSZ and $z=8.6$ for the CIB, with a particle mass resolution of $1.5\times 10^9 h^{-1}\,\mathrm{M}_{\odot}$ (and a few tens of times higher than that for halos). We thus expect it to reproduce CIB emission more accurately.

\texttt{WebSky} calculates the tSZ and CIB components by assigning emission properties to halos based on their mass and redshift. Given the fact that all halos above $M>10^{12}\,\mathrm{M}_{\odot}$ are represented, this is an excellent approximation for the tSZ, but, as explained in section~\ref{sec:hyperparam_requirements}, misses out on a significant fraction of the CIB emission from low-mass halos below the limit. The process of assigning emission to halos is carried out based on prescriptions from astrophysical models, under the assumption that halos are spherically symmetric. The cornerstone of the tSZ model is the `Battaglia' pressure profile of Ref.~\cite{ref:battaglia_et_al_12}; while for the CIB, point-like galaxies are distributed in halos according to the CIB halo model of Ref.~\cite{ref:shang_et_al_12} (explained in detail in section~\ref{sec:halo_model_cib}) with the best-fit parameters of Ref.~\cite{ref:hermes_viero_wang}. The simulations show a level of correlation between tSZ and CIB that is in good agreement with the observations of Ref.~\cite{ref:planck_cib_tsz_correlation}, especially at 143 and 353\,GHz. Since the astrophysical models used by \texttt{Websky} are the same that we use in our analytic calculations, we expect to see rather good agreement between the two approaches.

In \texttt{Agora}, on the other hand, electron pressure profiles inferred from the BAHAMAS hydrodynamical simulations~\cite{mccarthyBahamasProjectCalibrated2017} by Ref.~\cite{ref:mead_et_al_20} are pasted onto $N$-body halos with masses greater than $ 10^{12} h^{-1}\,\mathrm{M}_{\odot}$. Three variants of these maps are available, corresponding to AGN-heating temperatures of $10^{7.6}, 10^{7.8}$, and $10^{8.0}$\,K. Here we consider only the latter one for simplicity. In simulating the CIB, \texttt{Agora} harnesses the \texttt{UniverseMachine} model~\cite{behrooziUniverseMachineCorrelationGalaxy2019} to obtain star-formation rates (SFRs) and $M_*$ values for each halo in the simulation based on the halo's properties and past accretion history. Through several modeling steps informed by state-of-the art observations, SFR is converted to IR luminosity, and through a modified-blackbody SED, to angular power spectra that are then compared to CIB measurements from~\cite{ref:lenz_19} in order to determine the free parameters of the model.

\subsection{Measuring the biases from simulations}
From the discussion above, it is clear that we can use the simulations to calculate the lensing reconstruction biases in equations~\eqref{eqn:auto_biases} and~\eqref{eqn:cross_biases} and compare them to our analytic predictions. Let us now detail how each of the bias terms in those equations can be estimated.

Our simulation-based measurements use the \texttt{QuickLens} implementation of the quadratic estimators. We choose to work with $\ell_{\rm max}=3000$ for concreteness. Though the simulations are noiseless, we apply to them inverse-variance filters appropriate for an experiment with the characteristics of the ACT DR6 150\,GHz channel.

The primary bispectrum bias could, in principle, be calculated as $2\,\langle \hat{\phi}[\tilde{T},\tilde{T}]\, \hat{\phi}[s,s]\rangle$. However, if computed this way, the estimate would be subject to the sample variance of the quadratic estimator --- this would be particularly problematic given that we only have access to a single realization of each of \texttt{WebSky} and \texttt{Agora}\footnote{For this reason, the angle brackets in the equations do not have an effect on the results of this section.}. In order to avoid this issue, we harness the fact that, to leading order, the quadratic estimator is an unbiased estimator of the lensing potential, and we compute instead $2\,\langle \phi\, \hat{\phi}[s,s]\rangle$, where $\phi$ is the input lensing potential of the simulation --- in reality, this would, of course, not be known.

The same concerns regarding sample variance apply to the secondary bispectrum bias if it is calculated as $4\, \langle \hat{\phi}[\tilde{T},s]\,\hat{\phi}[\tilde{T},s]\rangle_{\mathrm{c}}$. We avoid the problem by following Ref.~\cite{ref:schaan_ferraro_18} and computing instead $8\, \langle \hat{\phi}[T,s]\,\hat{\phi}[\tilde{T}^{(1)},s]\rangle$, where $T$ is the unlensed CMB temperature and  $\tilde{T}^{(1)}$ is the first-order part of the lensed CMB temperature field; in appendix~\ref{appendix:curved_sky_T_template}, we explain how $\tilde{T}^{(1)}$ can be calculated in an efficient manner for full-sky maps, and we make our implementation available\footnote{The code to compute $\tilde{T}^{(1)}$ on the full, curved sky can be found at \url{https://github.com/abaleato/curved\_sky\_B\_template}.}.

Finally, to estimate the trispectrum bias, $\langle \hat{\phi}[s,s]\,\hat{\phi}[s,s]\rangle_{\mathrm{c}}$, we first cross-correlate two lensing reconstructions where all input legs are foregrounds. Then, we subtract the Gaussian part of this correlation, which can be calculated analytically by replacing the angular power spectra in equation~\eqref{eqn:QE_norm_TT} with spectra measured from the foreground simulations. (In actual analyses, this would be part of the $N^{(0)}$ noise bias.)

\subsection{Simulated vs. analytic results}

\begin{figure}
    \centering
    \includegraphics[width=\textwidth]{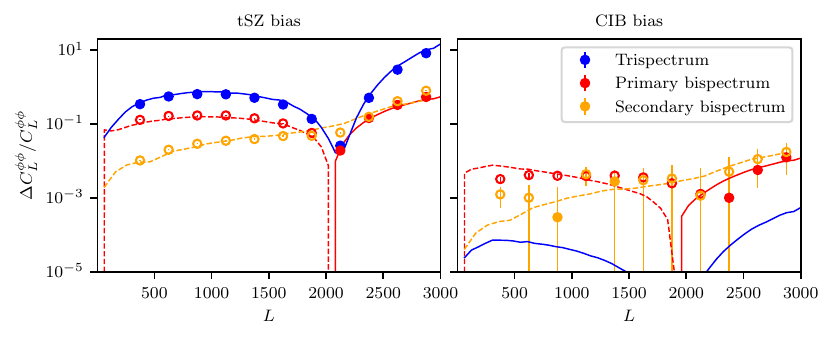}
    \caption{Comparison of analytic predictions for the CMB lensing auto-spectrum biases to measurements from the \texttt{WebSky} simulations for an experiment similar to the 150\,GHz channel of ACT (temperature-only reconstructions up to $\ell_{\rm max}=3000$ and no foreground cleaning or cluster masking). The left and right panels show biases arising from the tSZ and CIB, respectively. In each panel, the primary bispectrum bias is shown in red, the secondary bispectrum bias in orange, and the trispectrum bias in blue. Positive (negative) values of our theory calculation are plotted as solid (dashed) curves, while positive (negative) measurements from the simulation are shown as filled (empty) markers. Error bars denote the standard error on the bandpowers as estimated from the per-multipole measurements, and are too small to be appreciable in the tSZ panel on this log-plot. In the case of the CIB, we remove the 3\% brightest point sources to unveil the clustering contribution, and we do not plot the trispectrum from simulations because it is consistent with zero after subtracting an $N^{(0)}$-like contribution. We emphasize that the analytic predictions are for the fiducial parameter values, not a fit.}
    \label{fig:biases_on_websky}
\end{figure}

\begin{figure}
    \centering
    \includegraphics[width=\textwidth]{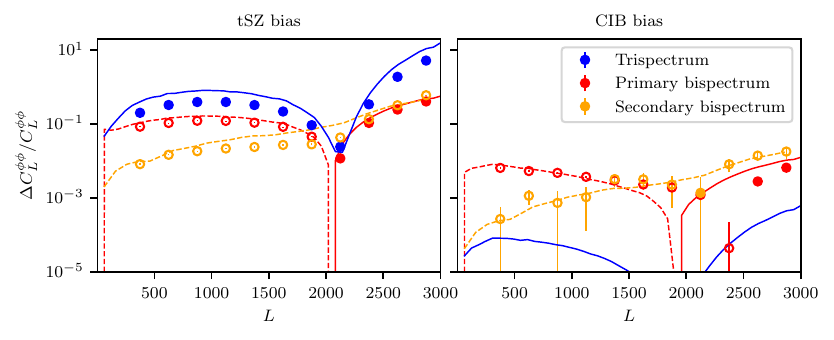}
    \caption{Same as figure~\ref{fig:biases_on_websky} but for the \texttt{Agora} simulations. Note that we have made no attempt to vary the astrophysical model in the simulation or in the analytic prediction to attain a closer match.}
    \label{fig:biases_on_agora}
\end{figure}

In figure~\ref{fig:biases_on_websky}, we show the biases measured by applying a TT quadratic estimator to a single, full-sky \texttt{WebSky} simulation at 150\,GHz, when only temperature modes up to $\ell_{\rm max}=3000$ are used. In the left panel, we isolate the biases coming from the tSZ alone, assuming that no attempt is made to mask massive clusters. The agreement is excellent, stemming from the fact that \texttt{WebSky} covers all the relevant halo masses and redshifts and uses the same tSZ modeling prescription as we do in our analytic approach. A similar conclusion holds when we look at the \texttt{Agora} simulations in figure~\ref{fig:biases_on_agora}. Despite the fact that the underlying astrophysical prescriptions are now different and we have made no attempt to attain a better match by varying model parameters, the bias curves predicted by our analytic method appear to be related to those measured from \texttt{Agora} by a simple amplitude shift that preserves the shape. This is consistent with the findings of section~\ref{sec:results_varying_params}.

\begin{figure}
    \centering
    \includegraphics[width=0.7\textwidth]{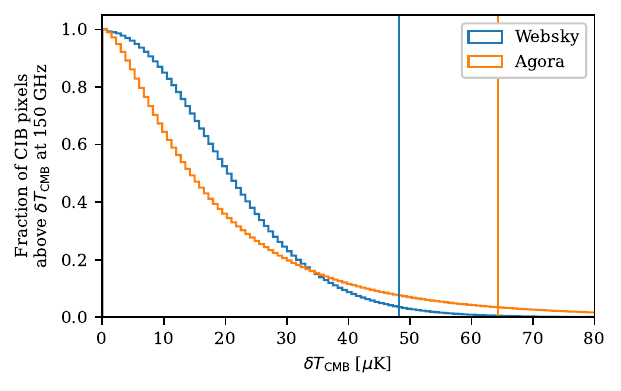}
    \caption{Fraction of pixels in the native \texttt{WebSky} (orange) and \texttt{Agora} (blue) CIB maps with values (in units of differential CMB temperature) above a certain threshold. Vertical lines delimit the region above which we discard point sources: a mere 3.5\% of them.}
    \label{fig:sim_source_cuts}
\end{figure}

Let us also compare results for the CIB. These are shown in the right panels of figures~\ref{fig:biases_on_websky} and~\ref{fig:biases_on_agora} for \texttt{Websky} and \texttt{Agora}, respectively. The biases one measures at face value from the simulations are dominated by shot-noise contributions from sources just below the typical mass cuts. Since these are straightforward to capture in actual analyses given a point-source luminosity function, we choose to remove the brightest 3\% of sources in order to unveil the clustering contributions that we can model within our framework; figure~\ref{fig:sim_source_cuts} shows that these sources live far in the tails of the respective distributions. After subtracting an $N^{(0)}$-like noise bias from estimates of the trispectrum bias, the residual is consistent with zero, so we do not plot it in the figures. Though the agreement is not as good as for the tSZ, we still see decent qualitative agreement, which serves to vindicate our analytic approach as a valid tool to understand the physics behind the lensing biases and their response to cosmology, astrophysics, and experimental configurations.


\section{Conclusions}\label{sec:outlook}

The auto- and cross-correlations of CMB lensing reconstructions constitute powerful cosmological probes, as they are sensitive to physical effects that would be highly degenerate if only unlensed anisotropies could be observed; these include the sum of the neutrino masses, dark energy or modifications of gravity. Moreover, internal lensing reconstructions are key when it comes to delensing  the observed CMB anisotropies in order to probe more precisely primordial $B$-mode polarization associated with inflationary gravitational waves. 

For the foreseeable future, lensing reconstructions from wide-area surveys are going to depend on the temperature anisotropies for a significant fraction of their signal-to-noise. They are in fact always the primary source of information when reconstructing lenses on arcminute scales and smaller. This poses a major challenge, because the measured intensity of the microwave sky receives important contributions from extragalactic foregrounds such as the tSZ and the CIB, which have been shown to bias CMB lensing reconstructions significantly. These contaminants, in fact, constitute the main limitation when it comes to extracting information reliably from the small-scale CMB anisotropies.

In this work, we present an analytic model for the lensing biases sourced by tSZ and CIB as they affect CMB lensing auto- and cross-correlations as well as $B$-mode delensing. Our modelling is based on the halo model of dark matter clustering, with a prescription for assigning tSZ emission to halos that is based on the Battaglia profile of Ref.~\cite{ref:battaglia_et_al_12}, and a mass-dependent-luminosity halo model of the CIB that builds on the work of Ref.~\cite{ref:shang_et_al_12}. We show that the bulk of these effects can be captured by implementing a subset of the one- and two-halo terms --- those that factor the most under the quadratic estimator weights. 

Harnessing the approximate spherical symmetry of the emission profiles, we speed up the evaluation of the quadratic estimators of lensing by up to four orders of magnitude relative to what is achievable by a naïve application of standard codes; they can now be evaluated in $O(1\,\mathrm{ms})$ on a modern laptop. Thanks to this, we are able to evaluate the dominant contributions to the primary bispectrum and trispectrum biases --- the one- and two-halo terms --- very fast, at $O(10\,\mathrm{s})$ per foreground component. Though the secondary couplings of the bispectrum do not lend themselves to the same efficient approach, we can nevertheless evaluate them in order one minute per foreground. In section~\ref{sec:comparison_with_websky}, we show that the resulting predictions are in good qualitative agreement with measurements derived from simulations of the microwave sky.

The key advantage of our approach, however, is its flexibility, enabling fast computation of the dominating bias terms in a wide range of cosmologies, astrophysical foreground models, experimental configurations (including noise levels, beam widths or frequency channels) and analysis choices (varying scale cuts, multi-frequency cleaning prescriptions or source-masking protocols).

In section~\ref{sec:results_auto}, we identify how biases to the CMB lensing auto-spectrum depend on the extent of cluster masking, scale cuts and foreground cleaning prescription. There is a nuanced interplay between the different couplings, including cancellations that can cause rather dramatic changes such as to the sign of the correction. We are able to identify the regimes where terms that are more difficult to bias-harden against can come to dominate; we are referring, primarily, to the secondary bispectrum bias. As a concrete example, we find that there is always an angular scale where the QE weights become highly constrained, leading to a dip in the trispectrum bias and a sign change in the primary bispectrum bias. (In section~\ref{sec:sign_change}, we show that in practice the location of this sign change, which affects both auto and cross-correlations, depends only on $\ell_{\rm max}$.) Around this scale, the secondary coupling of the bispectrum can become the dominant source of bias to the auto-spectrum. More generally, this is always the dominant contribution on small scales to tSZ biases if cluster-masking is sufficiently extensive and for CIB biases in general. We also point out that for upcoming experiments, two-halo contributions will dominate tSZ biases at multipoles smaller than a few hundred, and across all relevant scales for the CIB biases.

The insights we glean regarding the primary bispectrum bias can be used to understand biases to cross-correlations with low-redshift probes of the large-scale structure; we do this in section~\ref{sec:results_cross}. Then, in section~\ref{sec:results_delensing}, we proceed similarly in the context of delensing CMB $B$-mode polarization, which is biased by the same terms that affect the lensing auto-spectrum when internal, temperature-based reconstructions are used as matter tracers. Once again, we emphasize the detailed interplay of terms. We report that secondary couplings of the bispectrum typically have a negligible impact due to the fact that only lenses with $L\lesssim 1500$ factor into the delensing of degree-scale $B$-modes, a regime where said couplings are subdominant. Our tool can be used to explore efficiently the dependence of these biases on multi-frequency cleaning, source masking and scale cuts. This is particularly relevant given that, as argued in Ref.~\cite{ref:baleato_and_ferraro_22}, such delensing biases can likely be forward-modeled from direct measurements of the (biased) two-point functions of CMB lensing, in which case delensing performance just hinges on pursuing whatever choices minimise the residual $B$-mode power spectrum after delensing (including foreground contributions).

We also use our framework to identify the ranges of redshifts and masses that contribute the bulk of the foreground biases in section~\ref{sec:results_varying_params}. This is useful not only to determine what hyperparameters guarantee the convergence of our calculations, but also in order to identify requirements on simulations of the microwave sky if they are to capture the effects of interest faithfully. We find that, if no source masking is applied, the tSZ contribution is dominated by halos with mass $M_{\rm vir}\gtrsim 10^{13}\,\mathrm{M}_{\odot}$ at $z<1.5$, whereas in order to reproduce the CIB biases simulations need to resolve $M_{\rm vir}\gtrsim 10^{11}\,\mathrm{M}_{\odot}$ out to $z\sim6$. The latter is a particularly challenging requirement that is not met by the \texttt{WebSky} simulations. 

Being able to compute the biases efficiently under different cosmological and foreground models enables a leap in our understanding of the uncertainties involved in our modeling of the foregrounds, which at present are derived from a very limited set of simulations, each of which offer a single realization of the microwave sky in a single cosmology and foreground model~\cite{ref:sehgal_sims_10, ref:websky, omoriAgoraMultiComponentSimulation2022}. We study the lensing biases to the CMB lensing power spectrum and find that for current and soon-upcoming experiments, the amplitude of the biases is a strong function of cosmological and astrophysical parameters, but their shape relative to the signal does not change in any significant way for realistic variations in parameters. Pending a more thorough investigation using different foreground parametrizations (not just changing the parameters within a given model), this suggests that residual biases could be effectively dealt with through a simple marginalization over the amplitude of a bias template.

We carried out the same investigations in the context of the cross-correlation of ACT-like lensing reconstructions with DESI-like LRGs and found that, while the relative shape of the bias remains rather insensitive to foreground modeling, it does depend rather strongly on cosmological parameters, mainly $\Omega_m$. Fortunately, cross-correlation biases probe only the primary coupling of the bispectrum, which is rather easy to harden against~\cite{ref:sailer_et_al}. Alternatively, these biases could be investigated further from simulations; our findings suggest that it is more critical for these simulations to feature different cosmologies rather than different astrophysical models.

Although the halo model underlying our calculation is known to be only accurate at the level of tens of percent, it could in principle be used to model these foreground biases jointly along with the cosmological signal as long as they constitute a relatively small correction to the signal. This approach would allow for a more principled propagation of modeling uncertainties than is currently standard. Ideally, if the modeling is accurate enough, one could consider using smaller-scale modes (higher $\ell_{\mathrm{max}}$) than would have been previously advisable while still being able to model the biases incurred. In principle, this analysis would also extract information about the foregrounds and their cosmological dependence, potentially breaking degeneracies and improving constraints beyond a simple increase in the number of modes available.

We make our code, \texttt{CosmoBLENDER}\footnote{Cosmological Biases to LENsing and Delensing due to Extragalactic Radiation: \url{https://github.com/abaleato/CosmoBLENDER}.}, publicly available. We hope it can be of use when optimising experiment designs and filtering choices, as the space of experimental and modeling parameters can be explored very efficiently. It can also be used to complement existing bias-mitigation techniques; for example, by understanding which bias couplings and $n$-halo contributions it is more important to harden against in different regimes, or the extent to which deprojecting certain components in multi-frequency cleaning will boost other contributions in light of masking prescriptions and other factors.

The tool will be useful not only for lensing and delensing analyses internal to the CMB, but also in cross-correlation with lower-redshift probes of the matter distribution. Our code can already calculate biases to cross-correlations with a galaxy number density tracer parameterized via an HOD. Generalizing this to cross-correlations with cosmic shear, $y$-$\kappa^{\text{CMB}}$ correlations, or other low-redshift probes is straightforward and will be done in future work.

Numerous other improvements to our framework are possible. Though the terms we incorporate are believed to be the dominant ones, it should not be difficult to implement several other ones, including three-halo terms. It would also be interesting to extend this methodology to the kSZ: though this contributes a smaller bias, it is a particularly insidious one given that we have no frequency control over this foreground. However, it is unclear how well this effect can be captured in a halo model.

More work on the lensing reconstruction side would also enable interesting extensions. A first, simple, step would be to quantify how these biases are diluted when combining the $TT$ QE with other polarization-based reconstructions. Beyond that, it would be very interesting to modify the estimator weights so as to implement the possibility of bias-hardening or even the `gradient-inversion' estimators that are known to give optimal reconstructions of small-scale lenses~\cite{horowitzReconstructingSmallScale2019a, ref:hadzhiyska_et_al_19}. We reiterate that, in the small-scale limit, temperature-based reconstructions become once again the dominant source of information.

Of course, our method has limitations. There is evidence that realistic effects such as cluster asphericity or substructure increase the tSZ power spectrum by as much as $10\text{--}20 \%$ on small angular scales $l\sim 2000\text{--}8000$~\cite{ref:battaglia_et_al_12}; it is reasonable to expect these effects to impact higher-point functions as well, and for their importance to extend to other foregrounds. It would therefore be desirable to validate our analytic method against hydrodynamical simulations, which properly take these effects into account. Unfortunately, hydrodynamical simulations of the size required for our purposes appear to be unattainable in the near term. In the meantime, carefully-selected zoom-in simulations might prove useful for our purposes.

\acknowledgments
ABL is grateful to Marcelo Álvarez, Boris Bolliet, Fiona McCarthy, Simone Ferraro, Martin White, Mat Madhavacheril, Stephen Bailey and Minas Karamanis for useful discussions during the course of this project. AC acknowledges support from the STFC (grant number ST/W000977/1). This research has made use of NASA's Astrophysics Data System, the arXiv preprint server, the Python programming language and libraries \textsc{JAX, NumPy, Matplotlib, SciPy, AstroPy}, \textsc{HealPy}~\cite{ref:healpy_paper}, \textsc{Hankl}~\cite{karamanisHanklLightweightPython2021}, \texttt{hmvec} and \texttt{BasicILC}.

This work was carried out on the territory of xučyun (Huichin), the ancestral and unceded land of the Chochenyo speaking Ohlone people, the successors of the sovereign Verona Band of Alameda County.
\appendix

\section{The formation of cosmic structures}\label{sec:halo_model}
\subsection{Motivation}
Linear perturbation theory is a powerful framework to understand the Universe at early times (including the CMB) but it breaks down once perturbations grow to a certain level and modes stop evolving independently. The relevance of non-linear evolution is a function of scale and time, but a rough measure is that it cannot be ignored when the dimensionless matter power spectrum approaches unity, $\mathcal{P}_{\mathrm{m}}(k_{\mathrm{nl}},z)\approx1$; for the concordance cosmological model, $k_{\mathrm{nl}}\approx 0.3 h\,\mathrm{Mpc}^{-1}$ today. In this section, we describe ways in which the evolution of matter fluctuations can be followed well into the non-linear regime, and translated to the abundance of observable tracers such as galaxies or clusters. Our goal will be to introduce the halo model of dark matter clustering, which will be an essential ingredient of our work in this paper.

The fundamental assumption underlying the halo model of dark matter clustering (based on the pioneering work of Ref.~\cite{ref:neyman_scott_52}; see, e.g.,~\cite{ref:cooray_sheth_02} for a review) is that all the matter in the Universe is enclosed in bound objects, which have come to be known as halos. The model then posits that the distribution of matter can be described by combining an understanding of the internal structure of the halos, relevant chiefly on small scales, with a characterization of how the halos are clustered, the dominant factor on large scales. This separation of scales is an essential ingredient of the model: the claim is that the clustering can be described using the tools of perturbation theory (see~\cite{ref:bernardeau_et_al_02}), while more sophisticated treatments (invoking possibly hydrodynamical simulations) are needed to comprehend the structure of individual halos.

Ever since the seminal work of Ref.~\cite{ref:white_rees_78}, the idea that galaxy formation is determined by the characteristics of their host halo has become widely accepted. This suggests that the statistical properties of the galaxy distribution can begin to be understood by studying the statistics of the halo population, including prescriptions for how galaxies occupy halos which depend on the galaxy type under scrutiny. We introduce this extension of the halo model to galaxies in section~\ref{sec:halo_model_for_galaxies}, and apply it to the CIB in section~\ref{sec:halo_model_cib}.

\subsection{The halo mass function}\label{sec:hmf}
An essential ingredient of the halo model is a prediction for the number of bound objects of a given mass that are to be found at different redshifts: the halo mass function. In this section, we outline the analytic derivation of the Press--Schechter mass function, a particularly important one that will also shed light on the physics at play. Despite proving surprisingly accurate, analytically-derived mass functions no longer live up to the precision afforded by observational probes. Therefore, we conclude this subsection by reviewing efforts to obtain halo mass functions from simulations.

Before we can study the properties of the ensemble of collapsed objects, we must understand the dynamics of collapse at the level of individual halos. Unfortunately, keeping track of this process once the perturbations have gone non-linear is only possible in a few, highly-idealised scenarios. The simplest case is that of the spherical collapse of a top-hat overdensity, first calculated by Ref.~\cite{ref:gunn_gott_72}. In this model, perturbations behave like closed FLRW universes decoupled from their environment: they initially expand; then they stop, turn around, and collapse. Ultimately, the fate of the collapsed object deviates from the closed FLRW prediction and, rather than ending up in a singularity, it forms a virialised halo with a density approximately 178 times that of the background. A crucial feature of the spherical collapse model is that it provides a relationship between the non-linear overdensity of the object and that which it would have had, had it evolved according to linear theory: a virialised object (a halo) which forms at redshift $z$ must have originated from a region of the primordial density field whose overdensity, $\delta_{\mathrm{i}}(\bm{x})$, if evolved using the linear-theory growth rate ($D(z)$, see e.g.~\cite{ref:dodelson_03}, but normalised to unity at present), reaches the threshold $\delta_{\mathrm{i}}(\bm{x})D(z) = \delta_{\mathrm{c}} \approx  1.69$.\footnote{This value of $\delta_{\mathrm{c}}$ is specific to spherical collapse in an Einstein--de Sitter cosmology (with $\Omega_{\mathrm{m}}=1$, $\Omega_{K}=0$ and $\Omega_{\Lambda}=0$)} Since general relativity --- which governs the dark matter interactions --- is a scale-free theory, this number is independent of mass and redshift\footnote{In principle, the critical threshold for collapse could become a function of mass and redshift if we considered baryons, which introduce a physical scale (the sound horizon); or if collapse was in fact not spherical (see, e.g.,~\cite{ref:sheth_et_al_01}). We ignore these and other realistic effects in this succint exposition. Note, however, that the formation time \emph{is} a function of the initial conditions and the cosmological model (e.g.,~\cite{ref:bryan_norman_98}).}.

Being able to relate the non-linear density of the collapsed object to the density of the perturbations evolved using linear theory is very useful: given the statistics of the primordial density field, one can predict which regions will have collapsed into bound structures at any later time. Moreover, it has the added benefit that a linear transformation preserves the simple, Gaussian character of the primordial fluctuations, greatly simplifying calculations.

The formalism that paved the way for our modern understanding of the mass function was developed by Press and Schechter in~\cite{ref:press_schechter_74}. Their ansatz was that the fraction of mass which is in the form of halos of mass $M$ or greater at any specified time is given by the probability that the primordial density field, smoothed over a volume\footnote{The seminal Press--Schechter paper~\cite{ref:press_schechter_74} studied the case of smoothing with a top-hat filter --- this is, in fact, the only choice that lends itself to an analytic treatment.} $V_{\mathrm{f}}= M/\bar{\rho}$ (where $\bar{\rho}$ is the comoving matter density of the background) and linearly-evolved to the time in question, exceeds the critical threshold for collapse, $\delta_{\mathrm{c}}$. For Gaussian primordial fluctuations, they found that the comoving number density of collapsed objects in the range $M \rightarrow M+\mathrm{d}M$ is
\begin{equation}\label{eqn:press_schechter_mf}
    n(M,t) \mathrm{d}M = \frac{\bar{\rho}}{M^2} f_{\mathrm{PS}}(\nu)   \left|\frac{\mathrm{d}\ln \nu}{\mathrm{d} \ln M}\right| \mathrm{d}M\,,
\end{equation}
where
\begin{equation}\label{eqn:multiplicity_fn}
    f_{\mathrm{PS}}(\nu) \equiv \sqrt{\frac{2}{\pi}} \nu  e^{- \nu^2 / 2}\,
\end{equation}
is known as the `multiplicity function', and we have defined $\nu \equiv \delta_{c}(t) /\sigma(M) $.
\iffalse
\begin{equation}\label{eqn:press_schechter_mf}
    n(M,t) \mathrm{d}M = \sqrt{\frac{2}{\pi}}\frac{\bar{\rho}}{M^2} \frac{\delta_{c}(t)}{\sigma} e^{\left(-2 \frac{\delta_{c}^2(t)}{2\sigma^2} \right)} \left|\frac{\mathrm{d}\ln \sigma}{\mathrm{d} \ln M}\right| \mathrm{d}M\,.
\end{equation}
\fi
Note that, in this formulation, the threshold for collapse, $\delta_{c}(t)=\delta_{c}/D(t)$, is time-dependent; while the density field is linearly extrapolated to the present time, $\delta_{0}(\bm{x})$, and smoothed as
\begin{equation}
    \delta_{\mathrm{s}} (\bm{x},R) \equiv \int \mathrm{d}^3\bm{x}' \delta_{0}(\bm{x}') W(\bm{x}-\bm{x}',R),
\end{equation} 
where $W(\bm{x},R)$ is a window function picking out a comoving volume $V_{\mathrm{f}}(R)= M/\bar{\rho}$. The mass variance of the smoothed density field is, therefore,
\begin{equation}
    \sigma^2(M) = \langle \delta^{2}_{\mathrm{s}} (\bm{x},R) \rangle = \frac{1}{2\pi^2} \int_{0}^{\infty} \mathrm{d}k P(k) \widetilde{W}^2(\bm{k} R) k^2  \, ,
\end{equation}
where $\widetilde{W}(\bm{k}R)$ is the Fourier transform of $W(\bm{x},R)$, and $P(k)$ is the linear-theory matter power spectrum at the present day.

Equation~\eqref{eqn:press_schechter_mf} suggests that halos of mass $M$ cannot become abundant until $\sigma(M)\gtrsim \delta_{c}/D(t)$. At $z=0$, $\sigma(M_*)=\delta_{\mathrm{c}}$ for a characteristic mass scale $M_*=2\times10^{13}\,\mathrm{M}_{\odot}$; halos more massive than this are indeed rare. More generally, this mass scale of non-linearity increases with time as $M_*(t)\propto \left[D(t)\right]^{6/(n+3)}$ for a matter power spectrum of the form $P(k)\propto k^n$~\cite{ref:mo_et_al_textbook}; since CDM spectra generically have $n>-3$ on all relevant scales, structure formation in these models proceeds \emph{hierarchically}, with lower-mass halos typically forming first.

In hierarchical models of structure formation, all matter should collapse into halos at some point in time. This was in fact not guaranteed from the initial derivation in~\cite{ref:press_schechter_74}, which had to be corrected by a `fudge factor' of 2 to ensure it. The inaccuracy arises from an internal inconsistency of the formalism: a mass cell which has $\delta_1>\delta_{\mathrm{c}}$ when smoothed on scale $M_1$ and $\delta_2<\delta_{\mathrm{c}}$ when smoothed on scale $M_2$ will only count towards the fraction of halos with masses greater than $M_1$, even if $M_1>M_2$.

A correct derivation was presented in~\cite{ref:bond_et_al_91} by studying the excursion set of a Gaussian density field, iteratively filtered with sharp band-passes narrowly-separated in $k$-space. The resulting trajectory in $(\sigma(M), \delta_{\mathrm{s}}(R))$-space can be shown to be a Markovian random walk with simple, Gaussian statistics. This approach makes it clear that it is a trajectory's \emph{first} crossing of $\delta_{\mathrm{c}}$ that matters when counting halos, irrespective of whether the overdensity might fluctuate back below the threshold when smaller-scale modes are taken into account. The resulting picture is that the fraction of random walks which first impinge the barrier on scale $R(M)$ corresponds to the fraction of the total mass contained in halos of mass $M$.

The assumption of spherical collapse is engrained in the Press--Schechter mass function --- specifically, through the value and time-independence of the density threshold for collapse. Alternative models of the mass function exist which are valid in the more realistic limit of ellipsoidal collapse; in this case, the critical threshold for collapse depends on the ellipticity and prolateness of the tidal field (the second derivatives of the gravitational potential). The excursion set approach can be extended to incorporate these effects~\cite{ref:sheth_et_al_01, ref:sheth_tormen_02}, producing a mass function that can be written as equation~\eqref{eqn:press_schechter_mf}, but replacing $f_{\mathrm{PS}}$ with
\begin{equation}
    f_{\mathrm{ST}}(\nu)\equiv A \left( 1 + \frac{1}{\tilde{\nu}^{2 q}}\right) f_{\mathrm{PS}}(\tilde{\nu})\,.
\end{equation}
Here, $\tilde{\nu} = 0.84\nu$, $q=0.3$ and $A\approx0.322$. This is the `Sheth--Tormen' mass function, which provides a better fit to simulations than the Press--Schechter prediction; in particular, it alleviates a dearth of high-mass halos and an excess of low-mass ones which started to become apparent in the high-resolution cosmological simulations coming online around the turn of the millenium; see, e.g., ~\cite{ref:jenkins_et_al_01}. Overall, the Sheth--Tormen mass function agrees with state-of-the art simulations to approximately $20\%$ accuracy~\cite{ref:warren_et_al_06}.

Finally, the mass function can also be calibrated to simulations for improved agreement; see, e.g.,~\cite{ref:jenkins_et_al_01, ref:evrard_et_al_02, ref:reed_et_al_03, ref:warren_et_al_06, ref:tinker_et_al_08, ref:bocquet_et_al_16, ref:bocquet_et_al_20}. Even then, care must be taken that one's definition of the mass of virialised objects in the simulations does not bias the calibration. In particular, Ref.~\cite{ref:tinker_et_al_08} showed that the `spherical overdensity' halo finder of Ref.~\cite{ref:lacey_cole_94} is a better choice than the popular `Friends-of-friends' algorithm: the former correlates more tightly with actual observables, and it offers greater self-consistency when used in analytic halo model calculations, which for the most part assume that halos are spherical.

Reference~\cite{ref:tinker_et_al_08} used a suite of $N$-body cosmological simulations of various $\Lambda$CDM cosmologies to calibrate the mass function of halos at redshifts $z\sim 0 \text{--} 2$ and with masses in the range $ \left(10^{11}\text{--}10^{15}\right) \,h^{-1}\,\mathrm{M}_{\odot}$. Their fitting form of choice was
\begin{equation}\label{eqn:tinker_mf}
    f_{\mathrm{T}}(\nu) \equiv A \left[ 1 + \left(\beta \nu\right)^{-2\phi}\right] \nu^{2\nu} e^{-\gamma \nu^2 /2}\,,
\end{equation}
where $\beta,\phi, \eta$ and $\gamma$ are parameters determined from simulations, and $A$ is set by the normalisation constraint that the mass function integrate to unity at $z=0$ (that is, that all the matter in the Universe resides in halos at present). They found that the mass function deviated significantly from `universality'\footnote{In this context, universality refers to the shape of $f(\nu)$ being approximately independent of cosmology and redshift.}, as halos of a fixed smoothing scale became less abundant at higher redshifts\footnote{This is likely due to the evolution of $\Omega_{\mathrm{m}}$ with redshift, and the impact this has on the halo concentration, which in turn affects any observables defined over a fixed aperture~\cite{ref:tinker_et_al_08}. See~\cite{ref:castorina_et_al_14}, and references therein, for further discussions on the universality of halo abundances and halo clustering.}. This redshift evolution is captured rather well by $X(z)=X_0 (1+z)^{\alpha_X}$, where $X_0$ is the value of parameter $X$ at redshift $z=0$, and $\alpha_X$ is a new parameter to be fit for; values are shown in table~\ref{tab:tinker_mf_params}. The combination of fitting formula and best-fit parameters has come to be known as the `Tinker' mass function. It agrees with simulations at the level of $5\%$ up to $z\lesssim 2.5$.
\begin{table}
    \centering
    \begin{tabular}{l l l}
        \toprule
        $X$ & $X_0$ & $\alpha_X$\\ \midrule
        A & 0.368 & --\\ 
        $\beta$ &0.589 & 0.20\\
        $\gamma$ & 0.864 & -0.01\\
        $\phi$ & -0.729 & -0.08 \\
        $\eta$ & -0.243 & 0.27\\
        \bottomrule
    \end{tabular}
    \caption[Best-fit parameters for the halo mass function]{Best-fit parameters for the Tinker halo mass function parametrisation~\cite{ref:tinker_et_al_08}, equation~\eqref{eqn:tinker_mf}, in the case where $r_{\mathrm{h}} = r_{200}$. The redshift evolution of a given parameter, $X$, is modelled as $X(z)=X_0 (1+z)^{\alpha_X}$, where $X_0$ is the value of the parameter at $z=0$. Values obtained from Ref.~\cite{ref:tinker_et_al_10}.}\label{tab:tinker_mf_params}
\end{table}

All in all, simulations constitute the most promising route to more accurate characterizations of the mass function. However, the work ahead will be arduous, owing to the large dynamic range that is required: from large volumes to track the highest-mass objects, all the way down to fine resolutions to reproduce the insides of halos and their substructure. Furthermore, $N$-body simulations are unlikely to be able to reproduce the true halo mass function with percent-level accuracy until they can properly take into account baryonic effects such as AGN and supernova feedback, which significantly alter the distribution of matter within halos relative to the dark-matter-only scenario~\cite{ref:kravtsov_and_borgani_12}.

\subsection{The bias of dark matter halos}\label{sec:halo_bias}
The halo mass function we introduced in the previous section lets us predict the mean number density of halos of a certain mass that we should expect to find at some redshift in a given cosmology --- one might think that, so far, we have learnt nothing about how these halos will be arranged in space. However, note that these halos form once the uncollapsed regions from which they originate have become more overdense than some critical value, and that this is more likely to happen in denser-than-average environments. This implies that the spatial distribution of halos ought to be correlated with their mass: higher-mass halos preferentially form in denser regions, so they are more clustered; on the other hand, halos with low masses can only form and avoid significant accretion or mergers if they form in under-dense regions, so their clustering is penalised relative to that of the linear field (or that of the dark matter that is not yet bound in halos). Halos are therefore said to be \emph{biased} tracers of the underlying dark matter distribution~\cite{ref:kaiser_84}. In this section, we briefly review the formalism for relating the statistics of the halo population to those of the dark matter, in some highly idealised scenarios.

Consider a spherical region of comoving radius $R_{0}$ and comoving volume $V_0$, containing mass $M_{0} = V_0\bar{\rho}(z_0)$ at $z_0$. Denote the overdensity of this region, linearly extrapolated to $z_0$, as $\delta_{0}$. We are interested in the overdensity of halos of mass $M_1$, which were formed at $z_1>z_0$, and are found in this region; that is,
\begin{equation}\label{eqn:halo_bias}
    \delta_{\mathrm{h}}\left(M_1, z_1 | M_0, R_0, z_0\right) = \frac{N\left(M_1, z_1 | M_0, R_0, z_0\right)}{n\left(M_1,z_1\right) V_0}\,\left[1+\delta(z)\right] - 1\,.
\end{equation}
Here, $n\left(M_1,z_1\right)$ is the usual halo mass function, $N\left(M_1, z_1 | M_0, R_0, z_0\right)$ is the average number of halos of mass $M_1$ which formed at $z_1$ and are found in the patch, and $\delta(z)$ is the non-linear matter overdensity within our region at some redshift $z<z_1$ --- the redshift at which we will later ask questions about halo bias. The factor of $[1+\delta(z)]$ is there to account for the change in volume of the region as it contracts or expands by virtue of being over- or under-dense\footnote{Equation \eqref{eqn:halo_bias} thus leads to the halo bias in the \emph{Eulerian} picture. For its \emph{Lagrangian} equivalent, one must drop the factor of $[1+\delta(z)]$.}.

One of the advantages of the excursion-set approach to calculating mass functions~\cite{ref:bond_et_al_91} is that it can be used to calculate quantities such as $N\left(M_1, z_1 | M_0, R_0, z_0\right)$: this is just the average number of halos with mass $M_1$, identified at $z_1$, that merge to form a more massive halo with mass $M_0$ at $z_0$. The formalism can be applied irrespective of whether the region characterized by $R_0$ has actually collapsed into a halo by $z_0$ (only the volume of the region and its impact on the background density matter). It gives~\cite{ref:mo_white_96}
\begin{equation}\label{eqn:mean_num_in_cell}
    N\left(M_1, z_1 | M_0, R_0, z_0\right) \mathrm{d} M_{1} = \frac{M_{0}}{M_{1}} f(1|0) \left|\frac{\mathrm{d}\sigma^2(M_1)}{\mathrm{d} M_1}\right| \mathrm{d} M_{1} \,,
\end{equation}
where
\begin{equation}\label{eqn:progenitor_mf}
    f(1|0)\, \mathrm{d}\sigma^2(M_1) = \frac{1}{\sqrt{2\pi}} \frac{\delta_{\mathrm{c}}(z_1) - \delta_0}{\left[\sigma^2(M_1) - \sigma^2(M_0)\right]^{3/2}} \exp \left[- \frac{\left(\delta_{\mathrm{c}}(z_1) - \delta_0\right)^2}{2\left[\sigma^2(M_1) - \sigma^2(M_0)\right]}\right]\, \mathrm{d}\sigma^2(M_1)\,.
\end{equation}

Consider the limit where $R_0$ is large, so that $M_0\gg M_1$ and $\sigma(M_1) \gg \sigma(M_0)$. Equation~\eqref{eqn:progenitor_mf} then tells us that the number of halos in our region of choice grows with $\delta_0$. Heuristically, $\delta_{0}$ can be intepreted as a long-wavelength background modulating shorter-wavelength fluctuations\footnote{This idea of decomposing the density field into short and long wavelength modes was first introduced by Ref.~\cite{ref:kaiser_84}, and it has been referred to as the `peak-background' split since~\cite{ref:bardeen_et_al_86}; see~\cite{ref:desjacques_et_al_18} for a review of the uses of this powerful tool.}, and causing the critical density for collapse at $z_1$ to be replaced with $\delta_{\mathrm{c}}(z_1) \rightarrow \delta_{\mathrm{c}}(z_1) - \delta_0$; hence, collapse is easier in denser regions.

In the limit of large $R_0$, the halo overdensity of equation~\eqref{eqn:halo_bias} is
\begin{align}\label{eqn:non_linear_bias}
    \delta_{\mathrm{h}}\left(M_1, z_1 | M_0, R_0, z_0\right) & \approx \left(1-\frac{\delta_{0}}{\delta_{\mathrm{c}}(z_1)}\right) \exp \left[\frac{\delta_{0}}{2\sigma^2(M_1)}(2\delta_{0}-\delta_{\mathrm{c}}(z_1))\right] \left[1+\delta(z)\right]- 1 \nonumber \\
    & = \delta(z) + \delta_{0} \frac{\nu_1^2 - 1}{\delta_{\mathrm{c}}(z_1)} +   \delta(z) \delta_{0} \frac{\nu_1^2 - 1}{\delta_{\mathrm{c}}(z_1)} + O(\delta_{0}^2)\,,
\end{align}
where $\nu_1 \equiv \delta_{\mathrm{c}}(z_1) / \sigma(M_1)$. In the linear regime, $\delta(z) \approx \delta_{0} D(z) \ll 1$, so we can drop the third term in the last line and replace $\delta_{0} \approx \delta(z)/D(z)$. Thus, in this limit, fluctuations in the halo abundance can be related to those of the dark matter rather accurately as
\begin{equation}
    \delta_{\mathrm{h}}\left(M_1, z_1 | M_0, R_0, z_0\right) = b(M_1,  z_1 , z) \delta(z)\,,
\end{equation}
where
\begin{equation}\label{eqn:linear_bias}
    b(M_1,  z_1 , z)  \equiv 1 + \frac{1}{D(z)} \left(\frac{\nu_1^2 -1}{\delta_{\mathrm{c}}(z_1)}\right)\,.
\end{equation}
is the `linear bias'\footnote{As was the case with the mass function, more accurate results can be obtained analytically if the derivation of the bias relation is carried out based on the physics of ellipsoidal collapse, as opposed to spherical~\cite{ref:jing_98, ref:sheth_tormen_99}.}, at redshift $z$, of halos of mass $M_1$, formed at $z_1$. These expressions reflect the tendency for massive halos to cluster more extensively than the matter field, since $\nu_1 >1 $ implies $b>1$; while low-mass halos are anti-biased, because $\nu_1 < 1 $ means that $ b<1$. The high-mass behaviour might remind the reader of the results in Ref.~\cite{ref:kaiser_84}: massive halos form at high peaks of the density field, which have a higher probability of existing in dense, compact environments. Low-mass halos, on the other hand, can only escape mergers and accrete only moderately if they formed in underdense, sparse environments. Note also that halos of a given mass were more strongly biased at higher redshifts.

In deriving the linear bias relation of~\eqref{eqn:linear_bias} from equation~\eqref{eqn:non_linear_bias}, we ignored terms beyond first order in $\delta(z)$, arguing that they were small when considering perturbations in the linear regime. Though this is a good approximation on large scales, halo bias is, in general, non-linear --- particularly on small scales. If we Taylor-expand $\delta_{\mathrm{h}}(\delta)$ as
\begin{equation}
    \delta_{\mathrm{h}}\left(M_1, z_1 | M_0, R_0, z_0\right) = \sum_{k>0} b_{k}(M_1, z_1, z)\delta^{k}\,,
\end{equation}
we can then define $b_{k}$, the $k$-th order bias coefficients. In certain analytic models --- such as spherical or ellipsoidal collapse --- $\delta(z)$ is fully determined by $\delta_0$ and $R_0$, so higher-order bias terms can be calculated analytically~\cite{ref:mo_et_al_97b, ref:scoccimarro_et_al_01, ref:catelan_et_al_98}.

Halo bias must satisfy the `consistency relation' that
\begin{equation}\label{eqn:consistency_relation}
    \int \mathrm{d}M \frac{M}{\bar{\rho}(z)} n(M,z) b(M,z) = 1\,,
\end{equation}
at all redshifts and to all orders. As will become clearer from the discussion around equation~\eqref{eqn:large_scale_lim_2h}, the consistency relation is needed for the halo power spectrum to equal the matter power spectrum on large scales --- and hence, for dark matter to be unbiased against itself, once we assume that all of it resides in halos.

If the spread around the mean number of halos predicted by equation~\eqref{eqn:mean_num_in_cell} were negligible, then the (possibly non-perturbative) halo bias extracted from equation~\eqref{eqn:non_linear_bias} would be all that is needed to relate the halo population to the matter field. In reality, however, halo formation is affected by peculiarities of the environment other than its mass and redshift, so the relationship is stochastic rather than deterministic~\cite{ref:mo_white_96, ref:sheth_lemson_99, ref:casas-miranda_et_al_02}. Since the physical processes that source this stochasticity tend to be rather local, the bias can be treated as being deterministic on large scales.

In addition, the halo bias can also depend on the smoothing scale, $R_0$, in which case it is known as `scale-dependent' bias; this effect is small on large scales\footnote{Note that we are not referring to the effect identified in~\cite{ref:dalal_08}, which is associated with primordial non-Gaussianity.}, so we ignore it here.

Improved characterizations of the halo bias can be obtained from simulations. In this work, we will use the description put forth by~\cite{ref:tinker_et_al_10}, who studied the large-scale bias of halos ranging from masses of $10^{10}\,h^{-1}\,\mathrm{M}_{\odot}$, to the most massive clusters found in their simulated, cosmological volumes. They introduced the functional form
\begin{equation}\label{eqn:tinker_bias}
    b(\nu) = 1- \mathrm{A} \frac{\nu^\mathrm{a}}{\nu^\mathrm{a} + \delta^\mathrm{a}_c} + \mathrm{B}\nu^\mathrm{b} + \mathrm{C} \nu^\mathrm{d}\,,
\end{equation}
and fit for the parameters A,a,b,B,C and d, subject to the consistency relation of equation~\eqref{eqn:consistency_relation}. The best-fit parameters for the case where the mass function is that from Ref.~\cite{ref:tinker_et_al_08} are shown in table~\ref{tab:tinker_bias_params}. These fit the bias measured from individual simulations with a typical scatter of approximately $6\%$. Interestingly, Ref.~\cite{ref:tinker_et_al_10} find that the form of $b(\nu)$ changes by less than $5\%$ (if at all) in the redshift range $0<z<2.5$. 
\begin{table}
    \centering
    \begin{tabular}{l l}
        \toprule
        Parameter & Value \\ \midrule
        A & $1.0 + 0.24 y\exp \left[-(4/\epsilon)^4\right]$ \\ 
        a & $0.44 \epsilon -0.88$ \\
        B & $0.183$ \\
        b & $1.5$ \\
        C & $0.019 + 0.107 \epsilon + 0.19 \exp\left[-(4/\epsilon)^4\right]$ \\
        d & $2.4$ \\
        \bottomrule
    \end{tabular}
    \caption[Best-fit parameters for the large-scale halo bias]{Best-fit parameters for the large-scale halo bias parametrisation of equation~\eqref{eqn:tinker_bias}, in the case where $r_{\mathrm{h}} = r_{200}$. Here, $\epsilon \equiv \log_{10}200 $. Values were obtained from Ref.~\cite{ref:tinker_et_al_10}. These parameter choices ensure that the constraint of equation~\eqref{eqn:consistency_relation} is satisfied to better than percent-level precision when used in conjunction with the mass function of equation~\eqref{eqn:tinker_mf} (and its associated best-fit parameter values given in table~\ref{tab:tinker_mf_params}).}\label{tab:tinker_bias_params}
\end{table}

To conclude this section, we emphasise that, in the limit where the halo bias is deterministic, the halo mass function is all that is needed to relate the halo population to the matter field. The bias can then be calculated (to all orders) from it. 

\subsection{The density profile of dark matter halos}
To lowest order, we can treat dark matter halos as being spherical\footnote{It has been known since the earliest simulations of structure formation in CDM cosmologies~\cite{ref:davis_et_al_85} that this unlikely to be strictly true in nature; see~\cite{ref:allgood_et_al_06} and references therein for details.}, in which case they are characterized by their density profile, $\rho(r)$. Further analytic insights into the internal structure of these halos can only be derived in the context of highly-idealised scenarios of gravitational collapse. For example, the self-similar solution for the spherical collapse of dark matter perturbations with a spectrum compatible with current constraints predicts a density run scaling as $\rho(r)\propto r^{-2}$ --- similar to that of an isothermal sphere (see, e.g.,~\cite{ref:mo_et_al_textbook}).

Early explorations using $N$-body simulations yielded dark matter density profiles that indeed resembled this isothermal solution~\cite{ref:frenk_et_al_88}. However, with the advent of higher-resolution simulations in the late 1990s, it became clear that the slopes of simulated profiles were actually shallower than in the isothermal case near the centre, $\rho(r)\propto r^{-1}$, but steeper in the outskirts, $\rho(r)\propto r^{-3}$. The NFW profile --- after Navarro, Frenk and White --- of Ref.~\cite{ref:nfw_96} proved to be a particularly good fit to halos of a wide range of masses in CDM cosmologies~\cite{ref:navarro_et_al_97}, so it became widely adopted. It will also be our parametrisation of choice in this work to model the distribution of dark matter inside halos. It has the form
\begin{equation}\label{eqn:nfw_profile}
    \rho(r) = \frac{\rho_{\mathrm{crit}} \delta_{\mathrm{char}}}{ (r/r_{\mathrm{s}}) \left(1+r/r_{\mathrm{s}}\right)^2}\,,
\end{equation}
where $\rho_{\mathrm{crit}}$ is the critical density to make the Universe flat at the redshift of interest (though we omit this dependence for notational compactness), $\delta_{\mathrm{char}}$ is a characteristic overdensity of the halos, and $r_{\mathrm{s}}$ is a scale radius parametrising the transition from the $\rho(r)\propto r^{-1}$ to the $\rho(r)\propto r^{-3}$ scaling, where the profile agrees with the isothermal prediction of $\rho(r)\propto r^{-2}$ (in the sense that $d \ln \rho / d \ln r = -2$).

The mass enclosed in the halo, out to radius $r$, is therefore
\begin{equation}\label{eqn:nfw_mass}
    M(r) = 4\pi \rho_{\mathrm{crit}} \delta_{\mathrm{char}} r^3_{\mathrm{s}} \left[ \ln (1+r/r_{\mathrm{s}}) - \frac{r}{r_{\mathrm{s}} + r}\right]\,.
\end{equation}
This expression does not converge when integrated out to an infinite radius, so a halo radius, $r_{\mathrm{h}}$, must be defined. Conventionally, it is chosen to be the radius within which the mean matter density is some number, $\Delta_{\mathrm{h}}$, times the background matter density or the critical density at the redshift in question. In this work, we adopt the former definition, and use $\rho_{\mathrm{h}}= \Delta_{\mathrm{h}} \rho_{\mathrm{m}} = \Delta_{\mathrm{h}} \rho_{\mathrm{crit}} \Omega_{\mathrm{m}}(z)$, where $\Omega_{\mathrm{m}}(z)\equiv \rho_{\mathrm{m}} / \rho_{\mathrm{crit}}$ is redshift-dependent. Two of the most popular choices are $ \Delta_{\mathrm{h}}=200$ and $ \Delta_{\mathrm{h}}= \Delta_{\mathrm{vir}}\approx176$. Note that the latter choice\footnote{See the discussion of spherical collapse of dark matter into virialised halos in section~\ref{sec:hmf}.} is more consistent with the definition of halos used to derive the halo mass function, so it will be the one we preferentially use in our halo model calculations.

Notice, from equation~\eqref{eqn:nfw_mass}, that $\delta_{\mathrm{char}}$ can be directly calculated from the mass of the halo once $r_{\mathrm{h}}$ is defined and $r_{\mathrm{s}}$ is specified\footnote{From the definition of the halo radius, $M(r_{\mathrm{h}}) \equiv (\Delta_{\mathrm{h}} \rho_{\mathrm{crit}} \Omega_{\mathrm{m}}(z))(4\pi r_{\mathrm{h}}^3/3)$, we learn that
    \begin{equation}
        \delta_{\mathrm{char}} = \frac{\Delta_{\mathrm{h}} \Omega_{\mathrm{m}}(z)}{3} \frac{c^3}{\ln (1+c) - c/(1+c)}\,,
\end{equation} where $c$  is the concentration parameter, to be introduced shortly.}. This means that, in a given cosmology, the NFW profile is fully determined by the mass and the scale radius of the halo; or, equivalently, by the mass and the concentration, defined as $c\equiv r_{\mathrm{h}} / r_{\mathrm{s}}$.

The concentration parameter is useful because it highlights interesting trends seen in the simulations. Typical concentration values range between 4--40, with the Milky way halo sitting at roughly $c\approx 10$--15 (see, e.g., Ref.~\cite{ref:belokurov_13} and references therein). In general, it is closely related to the formation history of a halo~\cite{ref:navarro_et_al_97}; as such, it depends on halo mass, redshift and cosmology. Much effort has been devoted to modelling the evolution of the mean concentration of NFW halos, $\bar{c}(M,z)$, in simulations of $\Lambda$CDM cosmologies; see, e.g.~\cite{ref:navarro_et_al_97, ref:bullock_et_al_01b, ref:duffy_et_al_08, ref:diemer_kravtsov_15}. This evolution is captured rather well by power laws in mass and redshift\footnote{Note that power-law parametrisations of $\bar{c}(M,z) $ do not extrapolate well to redshifts, masses and cosmologies different from those on which the model was calibrated --- for applications requiring this, a calibration based on $\nu$ might be more appropriate, as there is evidence that $\bar{c}(\nu)$ is more universal than $\bar{c}(M,z)$~\cite{ref:zhao_et_al_09, ref:prada_et_al_12, ref:bhattacharya_et_al_13, ref:diemer_kravtsov_15}.},
\begin{equation}\label{eqn:duffy_c}
    \bar{c}(M,z) = \mathrm{A}  \left(\frac{M}{M_{\mathrm{pivot}}}\right)^{\mathrm{B}} (1+z)^{\mathrm{C}}\,,
\end{equation}
where $\mathrm{A}$, $\mathrm{B}$ and $\mathrm{C}$ are parameters to be fit for. In this work, we will use the parameter values shown in table~\ref{tab:duffy_params}, which were obtained by Ref.~\cite{ref:duffy_et_al_08} (though note that more recent constraints exist; see, e.g.,~\cite{ref:diemer_joyce_19}). These reflect a trend of decreasing concentration with halo mass and redshift: higher-mass halos typically form later, so they are more likely to have experienced recent mergers and have an abundance of material orbiting far from their central regions. At a fixed halo mass, there is significant scatter around the mean concentration owing to the diversity of halo formation histories; the relation can be approximated as a log-normal distribution,
\begin{equation}\label{eqn:concentration_spread}
    p(c|M,z) \mathrm{d}c = \frac{\mathrm{d \ln c}}{\sqrt{2\pi\sigma^2_{\ln c}}} \exp \left[ - \frac{\ln^2 \left[c/\bar{c}(M,z)\right]}{2 \sigma^2_{\ln c}}\right] \, ,
\end{equation}
the width of which, $\sigma_{\ln c} \approx 0.11$--$0.27$ (for relaxed halos, using $r_{\mathrm{h}} = r_{\mathrm{vir}}$), is largely independent of halo mass; see, e.g.,~\cite{ref:jing_00, ref:bullock_et_al_01b, ref:wechsler_et_al_02, ref:neto_et_al_07}.
\begin{table}
    \centering
    \begin{tabular}{l l l l l}
        \toprule
        Convention & Redshift & A & B & C \\ \midrule
        $r_{\mathrm{h}} = r_{\mathrm{200}}$ & 0$\text{--}$2& $5.71 \pm 0.12$& $-0.084\pm0.006$&$-0.47\pm0.04$\\
        $r_{\mathrm{h}} = r_{\mathrm{vir}}$ & 0$\text{--}$2& $7.85^{\pm0.17}_{\pm0.18}$& $-0.081\pm0.006$ &$-0.71\pm0.04$\\  
        \bottomrule
    \end{tabular}
    \caption[Best-fit parameters for the median halo concentration of NFW halos in the model of Ref.~\cite{ref:duffy_et_al_08}]{Best-fit parameters for the median halo concentration of NFW halos in the model of Ref.~\cite{ref:duffy_et_al_08}. The parameters we quote are calibrated to the full sample of simulated halos on redshifts $z<2$, with a background cosmology as in~\cite{ref:wmap_5yr}. In all cases, $M_{\mathrm{pivot}}=2\times10^{12}\,h^{-1}\,\mathrm{M}_{\odot}$.}\label{tab:duffy_params}
\end{table}

In this work, we will frequently invoke the normalised Fourier transform of the dark matter distribution within halos\footnote{For consistency with the literature, we use the asymmetric Fourier convention in this section. In three dimensions, this is: 
\begin{equation}
    f(\bm{x}) = \int \frac{d^3 \bm{l}}{(2\pi)^{3}} f(\bm{l}) e^{i \bm{l} \cdot \bm{x}} \quad \text{and} \quad f(\bm{l}) = \int d^3 \bm{x} f(\bm{x}) e^{-i \bm{l} \cdot \bm{x}}\,.
\end{equation}}. For a halo of mass $M$, this is
\begin{equation}\label{eqn:profile_normalisation}
    u(\bm{k} | M ) \equiv \frac{\int \mathrm{d}^3 \bm{x} \,\rho(\bm{x} | M ) e^{-i\bm{k}\cdot \bm{x}}}{\int \mathrm{d}^3 \bm{x} \,\rho(\bm{x} | M )} \,,
\end{equation}
which, for a spherically-symmetric profile, simplifies to
\begin{equation}\label{eqn:normalised_radial_nfw}
    u(\bm{k} | M ) = \int_{0}^{r_{\mathrm{h}}} \mathrm{d}r \, 4\pi r^2 \frac{\sin \left(k r\right)}{k r} \frac{\rho(r | M )}{M}\,.
\end{equation}
Mathematically, this is just the statement that the Fourier transform of a spherically-symmetric function is a Hankel transform. Smaller-mass halos have Fourier transforms extending to smaller scales (see, e.g., Ref.~\cite{ref:cooray_sheth_02}). Note also that, on scales much larger than their typical spatial extent, halos can be regarded as point masses; i.e., $ u(|\bm{k}| \rightarrow 0\, | M ) \rightarrow 1$. 

The reason why density profiles of dark matter halos are nearly universal is not yet understood. However, this is a feature that emerges for a very broad range of initial conditions, so it is likely a product of the relaxation processes driving the halo to equilibrium after the messy stages of formation through mergers\footnote{In dynamics, a relaxed system is one which is found in an equilibrium state. There are several pathways for a system to achieve this from a non-relaxed configuration; but for collisionless systems like dark matter halos, the only one available is violent relaxation due to a changing gravitational potential~\cite{ref:mo_et_al_textbook}. One plausible explanation for the universality of dark matter halo profiles is that violent relaxation might have erased virtually all dependence on the initial conditions.}.

Though in this work we will only be using the NFW profile, it is worth mentioning that other profiles exist in the literature that provide a better fit to modern simulations. One notable contender is the Einasto profile~\cite{ref:einasto_65, ref:navarro_et_al_04}, which outperforms NFW even when one of its three parameters is fixed and both fitting forms have the same number of degrees of freedom~\cite{ref:mo_et_al_textbook}.

\subsection{The halo model}\label{sec:hm_calcs}
There is compelling evidence that the dark matter component of our Universe is `cold'. As discussed in section~\ref{sec:hmf}, cold dark matter is expected to all end up eventually bound in virialised halos. In this section, we combine the infrastructure developed in previous sections --- the halo mass function, bias and internal structure --- into a formalism describing the clustering properties of the matter: the `halo model'(e.g.,~\cite{ref:seljak_00, ref:ma_fry_00, ref:cooray_sheth_02}). This will be the main tool we use in section~\ref{sec:analytic_framework} to model extragalactic CMB foregrounds. For this reason, we will go through the derivation in detail.

The total matter density at a point can be calculated by adding the contributions from all individual halos,
\begin{equation}
    \rho (\bm{x}) = \sum_i M_i u\left( \bm{x} - \bm{x}_i | M_i \right)\,,
\end{equation}
where each halo is located at position $\bm{x}_i$, and has mass $M_i$. In a statistical sense, the normalised density profile of halos is fully determined by their mass and redshift: the halo concentration can be obtained statistically from these two using, for example, equations~\eqref{eqn:duffy_c} and~\eqref{eqn:concentration_spread}. For the purpose of this introduction, we consider halos at a fixed redshift, so that the normalised density profile, $u\left( \bm{x} | M \right)$, depends only on mass.

The mean matter density at this redshift, $\bar{\rho}$, can then be calculated as
\begin{align}
    \bar{\rho} &= \langle \rho (\bm{x}) \rangle = \langle \sum_i M_i u\left( \bm{x} - \bm{x}_i | M_i \right) \rangle \nonumber \\
    & = \int \mathrm{d}M n(M) M \underbrace{\int \mathrm{d}^3\bm{x}'  u\left( \bm{x} - \bm{x}' | M \right)}_{=1}\,.
\end{align}
In the second line, we have used the definition of the mass function to replace the ensemble average of sums over halo masses and positions with mass-function-weigthed integrals over mass and volume.

This formalism also enables calculation of the clustering properties of the matter. The two-point function of the density is given by
\begin{align}
    \langle \rho (\bm{x}_1) \rho (\bm{x}_2)\rangle & = \langle \sum_{i,j} M_i u\left( \bm{x}_1 - \bm{x}_i | M_i \right) M_j u\left( \bm{x}_2 - \bm{x}_j | M_j \right)\rangle \,.%\nonumber \\
    %& = \langle \sum_{i=j} M_i^2 u\left( \bm{x}_1 - \bm{x}_i | M_i \right) u\left( \bm{x}_2 - \bm{x}_i | M_i \right)\rangle  \nonumber \\
    %& \quad + \langle \sum_{i \neq j} M_i u\left( \bm{x}_1 - \bm{x}_i | M_i \right) M_j u\left( \bm{x}_2 - \bm{x}_j | M_j \right)\rangle
\end{align}
This received two types of contributions: those where both legs of the correlator come from the same halo ($i=j$), and those stemming from two different halos ($i\neq j$). The one-halo contribution is
\begin{align}
    \langle \sum_{i=j} M_i^2 u\left( \bm{x}_1 - \bm{x}_i | M_i \right) u\left( \bm{x}_2 - \bm{x}_i | M_i \right)\rangle  =  \int& \mathrm{d}M M^2 n(M) \nonumber \\
    & \times \int \mathrm{d}^3\bm{x}' u\left( \bm{x}_1 - \bm{x}' | M \right) u\left( \bm{x}_2 - \bm{x}' | M \right) \, ,
\end{align}
while the two-halo terms are
\begin{align}
    & \langle \sum_{i \neq j} M_i u\left( \bm{x}_1 - \bm{x}_i | M_i \right) M_j u\left( \bm{x}_2 - \bm{x}_j | M_j \right)\rangle  \nonumber \\ & =  \int \mathrm{d}M_1 M_1 n(M_1) \int \mathrm{d}M_2 M_2 n(M_2) \nonumber \\
    &  \quad \times \int \mathrm{d}^3\bm{x}' \mathrm{d}^3\bm{x}'' \left[1 + \xi_{\mathrm{hh}}\left(\bm{x}' - \bm{x}'' | M_1,M_2\right)\right] u\left( \bm{x}_1 - \bm{x}' | M_1 \right) u\left( \bm{x}_2 - \bm{x}'' | M_2 \right)  \nonumber \\
    & =  \left(\bar{\rho}\right)^2 + \int \mathrm{d}M_1 M_1 n(M_1) \int \mathrm{d}M_2 M_2 n(M_2) \nonumber \\
    & \hphantom{ \left(\bar{\rho}\right)^2 + \int} \times \int \mathrm{d}^3\bm{x}' \mathrm{d}^3\bm{x}'' \xi_{\mathrm{hh}}\left(\bm{x}' - \bm{x}'' | M_1,M_2\right) u\left( \bm{x}_1 - \bm{x}' | M_1 \right) u\left( \bm{x}_2 - \bm{x}'' | M_2 \right)   \,.
\end{align}
In the first equality, we have defined the halo correlation function, $\xi_{\mathrm{hh}}\left(r | M_1,M_2\right)$, which describes the excess clustering probability, over a random distribution in space, of a pair of halos of masses $M_1$ and $M_2$ separated by a distance $r$. 

We are now in a position to compute the covariance of the dark matter overdensity,
\begin{align}
    \langle \delta (\bm{x}_1) \delta (\bm{x}_2) \rangle & = \frac{1}{\bar{\rho}^2}\langle \rho (\bm{x}_1) \rho (\bm{x}_2)\rangle -1\nonumber \\
    & \equiv \xi^{\mathrm{1h}}\left(|\bm{x}_1 - \bm{x}_2\right|) + \xi^{\mathrm{2h}}\left(|\bm{x}_1 - \bm{x}_2\right|)  \,,
\end{align}
where we have defined a one-halo contribution,
\begin{align}
    \xi^{\mathrm{1h}}\left(|\bm{x}_1 - \bm{x}_2\right|)  & = \frac{1}{\bar{\rho}^2} \int \mathrm{d}M M^2 n(M) \int \mathrm{d}^3\bm{x}' u\left( \bm{x}_1 - \bm{x}' | M \right) u\left( \bm{x}_2 - \bm{x}' | M \right) \,,
\end{align}
and a two-halo term,
\begin{align}
    \xi^{\mathrm{2h}}\left(|\bm{x}_1 - \bm{x}_2\right|)  = & \frac{1}{\bar{\rho}^2} \int  \mathrm{d}M_1 M_1 n(M_1) \int \mathrm{d}M_2 M_2 n(M_2) \nonumber \\
    &  \times \int \mathrm{d}^3\bm{x}' \mathrm{d}^3\bm{x}'' \xi_{\mathrm{hh}} \left(\bm{x}' - \bm{x}'' | M_1,M_2\right) u\left( \bm{x}_1 - \bm{x}'' | M_1 \right) u\left( \bm{x}_2 - \bm{x}'' | M_2 \right) \,.
\end{align}
We have assumed that the halo distribution is a stationary point process, so that the correlation function depends on the distance between halos but not on their global location or orientation.

Notice that these expressions involve convolutions of density profiles; they are ubiquitous in halo model calculations. Since convolutions in real space become multiplications in Fourier space, it shall prove convenient to work in the latter. The Fourier space equivalent of the correlation function is the power spectrum,
\begin{equation}
    P(k) = P^{1\mathrm{h}}(k) + P^{2\mathrm{h}}(k).
\end{equation}
The first term on the right-hand side is the one-halo contribution,
\begin{align}\label{eqn:1h_dm_power}
    P^{1\mathrm{h}}(k)  & =  \int \mathrm{d}M \left(\frac{M}{\bar{\rho}}\right)^2 n(M) |u\left(k | M \right)|^2 \,,
\end{align}
which only depends on the halo profile and mass function. On the other hand, the two-halo term,
\begin{align}\label{eqn:general_2h_ps_term}
    P^{2\mathrm{h}}(k) =  \int & \mathrm{d}M_1 \left(\frac{M_1}{\bar{\rho}}\right) n(M_1) u\left( k | M_1 \right) \int \mathrm{d}M_2 \left(\frac{M_2}{\bar{\rho}}\right) n(M_2) u\left( k | M_2 \right) P_{\mathrm{hh}}(k|M_1,M_2)\,,
\end{align}
depends also on $P_{\mathrm{hh}}(k|M_1,M_2)$, the cross-power spectrum of halos with masses $M_1$ and $M_2$; in fact, on scales much larger than the size of individual halos, the two-halo term is insensitive to the the shape of the individual halos.

On large scales, it is a good approximation to assume that the halo overdensity is related to that of the matter via deterministic, linear halo bias\footnote{See  Refs.~\cite{ref:mead_and_verde_21, ref:asgari_et_al_23} for generalizations to non-linear halo bias.} (section~\ref{sec:halo_bias}); hence,
\begin{equation}\label{eqn:halo_ps}
    P_{\mathrm{hh}}\left(k_{\mathrm{long}} | M_1,M_2\right) \approx b(M_1) b(M_2) P(k_{\mathrm{long}})\,,
\end{equation}
where $ P(k)$ is the matter power spectrum (for now, we ignore all redshift dependence). Notice that, on scales much larger than the typical size of a halo (call them $k^{-1}_{\mathrm{long}}$), $ u\left(k_{\mathrm{long}} | M \right)\rightarrow 1$; this, together with the consistency relations of equation~\eqref{eqn:consistency_relation}, implies that
\begin{equation}\label{eqn:large_scale_lim_2h}
    P^{2\mathrm{h}}(k_{\mathrm{long}}) \approx P(k_{\mathrm{long}}) \approx P_{\mathrm{lin}}(k_{\mathrm{long}})\,.
\end{equation}
The last equality follows from the fact that, on large scales, the matter power spectrum is very well approximated by the linear-theory calculation, $P_{\mathrm{lin}}$. Hence, in the regime where the two-halo term is important,
\begin{align}
    P^{2\mathrm{h}}(k) \approx  P_{\mathrm{lin}}(k) & \int \mathrm{d}M_1 \left(\frac{M_1}{\bar{\rho}}\right) b(M_1) n(M_1) u\left( k | M_1 \right) \nonumber \\
    & \times \int \mathrm{d}M_2 \left(\frac{M_2}{\bar{\rho}}\right) b(M_2) n(M_2) u\left( k | M_2 \right)\,.
\end{align}
Rigorously speaking, this expression only holds insofar as we are probing scales where both the assumption of deterministic halo bias and our neglect of halo exclusion effects hold. In practice, we shall not need to refine the expression any further, for the one-halo term dominates on scales small enough that these assumptions break.

To conclude, note that, as $k \rightarrow 0$, the two-halo term goes as $P^{2\mathrm{h}}(k_{\mathrm{long}}) \propto  P_{\mathrm{lin}}(k_{\mathrm{long}}) \propto k$ (for CDM spectra), while the one-halo term is a constant (shot noise). At large enough scales, this leads to an unphysical result: the one-halo term dominates the power spectrum. One way to avoid this is to consider `compensated' halo profiles built from combinations of positive and negative overdensities~\cite{ref:cooray_sheth_02, ref:chen_afshordi_20}, but these have problems of their own, such as a lack of power on large scales. Instead, in the halo model calculations of the power spectrum in this paper, we introduce an \emph{ad hoc} factor in the style proposed by Ref.~\cite{ref:mead_et_al_15} that damps the density profiles on large scales when calculating the one-halo contribution:
\begin{equation}\label{eqn:profile_softening}
    u(k|M) \rightarrow u(k|M) \left(1 - e^{-k/k_*}\right)\,,
\end{equation}
where $k_*$ is a damping scale for which we choose the value $k_*\equiv 0.01\,\mathrm{Mpc}^{-1}$. Physically, this is motivated by the fact that halos cannot exist within each other, by definition~\cite{ref:smith_et_al_07}.

The framework can be extended to calculate the higher-point correlations of the matter fluctuations. We quote now some results that will be useful later; and refer the reader to~\cite{ref:cooray_sheth_02} for details. The matter bispectrum is
\begin{equation}\label{eqn:total_matter_bispectrum}
    B(\bm{k}_1,\bm{k}_2,\bm{k}_3|M_1,M_2,M_3) = B^{1\mathrm{h}}+B^{2\mathrm{h}}+B^{3\mathrm{h}}\,,
\end{equation}
where,
\begin{align}
    &B^{1\mathrm{h}} = \int dM\, n(M) \left(\frac{M}{\bar{\rho}}\right)^3\
    \prod_{i=1}^3 u(k_i|M) \,,  \nonumber\\
    &B^{2\mathrm{h}} = \int dM_1\,n(M_1) 
    \left(\frac{M_1}{\bar{\rho}}\right)\,u(k_1|M_1)
    \int dM_2\,n(M_2)\left(\frac{M_2}{\bar{\rho}}\right)^2\, 
    u(k_2|M_2) u(k_3|M_2)  \nonumber \\
    & \quad \quad \times P_{\mathrm{hh}}(k_1|M_1,M_2) +  {\rm perms.} \,,\nonumber\\
    & B^{3\mathrm{h}} = \left[ \prod_{i=1}^3 \int dM_i\,u(k_i|M_i) n(M_i) 
    \left(\frac{M_i}{\bar{\rho}}\right)\right]
    B_{\mathrm{hhh}}(\bm{k}_1,\bm{k}_2,\bm{k}_3|M_1,M_2,M_3) \, . \nonumber \\
    \label{eqn:bi}
\end{align}
The three-halo term depends on the halo bispectrum,
\begin{align}\label{eqn:halo_bispec}
    B_{\mathrm{hhh}}(\veck_1,\veck_2,\veck_3;M_1,M_2,M_3) &= \prod_{i=1}^{3}b_i(M_i)
    \bigg[B^\lin(\veck_1,\veck_2,\veck_3)  + \frac{b_2(M_3)}{b_1(M_3)}P^\lin(k_1)P^\lin(k_2)\bigg]\, ,
\end{align}
where $b_i$ is the $i$th order halo bias (for convenience, we have denoted the linear halo bias as $b_1$, though in earlier discussions we used $b$), and $B^\lin$ is obtained from tree-level perturbation theory; see, e.g.~\cite{ref:cooray_sheth_02}.

The same ideas can be used to calculate the matter trispectrum,
\begin{equation}\label{eqn:total_matter_trispectrum}
    T(\bm{k}_1,\bm{k}_2,\bm{k}_3,\bm{k}_4|M_1,M_2,M_3, M_4) = T^{1\mathrm{h}}+T^{2\mathrm{h}}+T^{3\mathrm{h}}+T^{4\mathrm{h}} \, ,
\end{equation}
where 
\begin{align}
    &T^{1\mathrm{h}} =\int dM\,n(M) \left(\frac{M}{\bar{\rho}}\right)^4\
    \prod_{i=1}^4 u(k_i|M)\,, \\
    & T^{2\mathrm{h}} = \Big[ \int dM_1\,n(M_1)\left( 
    \frac{M_1}{\bar{\rho}}\right)\,u(k_1|M_1)
    \int dM_2\,n(M_2)\left(\frac{M_2}{\bar{\rho}}\right)^3
    \nonumber \\
    & \quad \quad \times u(k_2|M_2) u(k_3|M_2) u(k_4|M_2)  
    P_{\mathrm{hh}}(k_1|M_1,M_2) + {\rm perms.} \Big] \label{eqn:T_2h_factorizableinells} \\
    &+ \Big[ \int dM_1\,n(M_1)\left( 
    \frac{M_1}{\bar{\rho}}\right)^2\,u(k_1|M_1)u(k_2|M_2)
    \int dM_2\,n(M_2)\left(\frac{M_2}{\bar{\rho}}\right)^2\, 
    u(k_3|M_2) u(k_4|M_2)  \nonumber \\
    & \quad \quad \times P_{\mathrm{hh}}(|\veck_1+\veck_2||M_1,M_2) + {\rm perms.} \Big] \label{eqn:T_2h_not_factorizableinells}\, ,
     \\
    &T^{3\mathrm{h}} =  \int dM_1\,n(M_1) 
    \left(\frac{M_1}{\bar{\rho}}\right)\,u(k_1|M_1)
    \int dM_2\,n(M_2) \left(\frac{M_2}{\bar{\rho}}\right)\,u(k_2|M_2)
     \\
    & \quad \quad \times
    \int dM_3\,n(M_3)\left(\frac{M_3}{\bar{\rho}}\right)^2\, 
    u(k_3|M_3) u(k_4|M_3)  
    B_{\mathrm{hhh}}(\veck_1,\veck_2,\veck_3+\veck_4|M_1,M_2,M_3) \, ,\nonumber \\
    & T^{4\mathrm{h}} = \left[ \prod_{i=1}^4 \int dM_i\,u(k_i|M_i) n(M_i) 
    \left(\frac{M_i}{\bar{\rho}}\right)\right] \nonumber \\
    & \hphantom{T^{4\mathrm{h}} = }\times T_{\mathrm{hhhh}}(\veck_1,\veck_2,\veck_3,\veck_4;M_1,M_2,M_3,M_4) .
\end{align}
Notice that the four-halo contribution depends on the halo trispectrum,
\begin{align}\label{eqn:halo_trispec}
    T_{\mathrm{hhhh}}(\veck_1,\veck_2,\veck_3,\veck_4|M_1,M_2,M_3,M_4) &=
    \prod_{i=1}^{4}b_i(M_i) \bigg[T^\lin(\veck_1,\veck_2,\veck_3,\veck_4)
    \nonumber \\
    & \hphantom{\prod_{i=1}^{4}b_i(m_i) \bigg[T^\lin}+ \frac{b_2(M_4)}{b_1(M_4)}P^\lin(k_1)P^\lin(k_2)P^\lin(k_3)\bigg]\, ,
\end{align}
which is a function of the tree-level matter trispectrum, $T^\lin$.

\subsection{Populating dark matter halos with galaxies}\label{sec:halo_model_for_galaxies}
In a seminal paper from the late 70s, White and Rees~\cite{ref:white_rees_78} realised that, for gas to cool efficiently and form stars, it must inhabit potential wells such as those provided by dark matter halos. In the decades since, increasingly sophisticated hydrodynamical simulations have consolidated a picture of galaxy formation within which dark matter halos have retained their early importance: whilst some halos are devoid of galaxies, every galaxy forms in a halo --- the most massive halos can, in fact, contain many galaxies; each one forming in a subhalo.

This understanding might, at first, appear to be in conflict with observations. The galaxy correlation function, $w(\theta)$ --- which measures the clustering of galaxies as a function of the angular separation on the sky, $\theta$ --- has been determined empirically to follow a power-law with approximately $w(\theta)\propto \theta^{-0.7} $ (e.g., figure~1 of~\cite{ref:maddox_et_al_90}; see also~\cite{ref:peebles_80, ref:zehavi_et_al_04}). This implies, in turn, that the three-dimensional correlation function of galaxies must also be a power law~\cite{ref:peebles_80}. Contrast this, however, with the shape of the dark matter power spectrum as measured from simulations (e.g., figure~4 of~\cite{ref:efstathiou_et_al_88}) or predicted by the halo model: galaxies appear to cluster very differently from the dark matter on small scales ($\leq 1$--2\,Mpc).

In this section, we shall see how our understanding of the clustering of dark matter can be extended to model the clustering of galaxies. In doing so, we will make contact with the complex `gastrophysics' at play, which ultimately determines the way in which galaxies populate halos (see, e.g.~\cite{ref:white_frenk_91}). Two extra pieces need to be added to the halo model in order to capture this richness~\cite{ref:peacock_smith_00, ref:scoccimarro_et_al_01}: the probability distribution, $P(N|M)$, that a halo of mass $M$ contains $N$ galaxies; and some knowledge of how those galaxies are distributed in the halos. These two elements constitute the `halo occupation distribution' (HOD)~\cite{ref:berlind_weinberg_02}.

In recent years, $P(N|M)$ has been characterized with increasing accuracy both from simulations (e.g.,~\cite{ref:berlind_et_al_03, ref:kravtsov_et_al_04}) and observations (e.g.,~\cite{ref:leauthaud_et_al_12}). The first moment of the distribution --- the mean number of galaxies per halo, $\langle\Ngal\rangle$ --- is particularly important for our purposes, for if the distribution is Poissonian, the mean specifies all the higher-order factorial moments, $\langle \Ngal(\Ngal -1)\dots (\Ngal - j)\rangle = \langle\Ngal\rangle^{1+j}$. As can be seen from e.g. figure~4 of~\cite{ref:kravtsov_et_al_04}, $\langle\Ngal\rangle$ scales as a power law at high halo masses ($M\geq 10^{13}\,h^{-1}\,\mathrm{M}_{\odot}$), plateaus to values of order unity at intermediate masses, and drops to zero for masses below some threshold. This behaviour is also consistent with theoretical predictions (e.g.,~\cite{ref:berlind_et_al_03}) and fits to observations (e.g.,~\cite{ref:scoccimarro_et_al_01}). No single probability distribution --- be it Poissonian, binomial, or a nearest-integer distribution --- can adequately describe the population statistics across the entire range of halo masses that host galaxies~\cite{ref:berlind_et_al_03, ref:kravtsov_et_al_04}.

However, progress can be made by splitting the galaxy population into two classes~\cite{ref:guzik_and_seljak_02, ref:kravtsov_et_al_04}: centrals, which occupy the centre of their parent halos; and satellites, which form in the substructure of those host halos. There is reason to expect that galaxy formation will not proceed below a certain mass threshold --- supernova explosions of stars formed early on are expected to readily disperse baryons away from low-mass halos with not enough gravitational pull to bind them. The number of central galaxies within a halo, $\Ncen$, can therefore be described as a step function
\begin{equation}\label{eqn:ncen_parametrisation}
    \Ncen = \begin{cases}
        0, & \text{if $M\leq M_{\mathrm{min}}$; \quad and}\\
        1, & \text{if $M\geq M_{\mathrm{min}}$.}
    \end{cases}
\end{equation}

If the host halos are massive enough, they will comprise a number of subhalos large enough for `satellite' galaxies to form within them. To a first approximation, we expect the number of satellites per halo, $\Nsat$, to scale approximately as $\Nsat \propto M$ --- galaxies are made up of baryons, so their number should be proportional to the baryon abundance, which in turn is a fixed fraction of the dark matter abundance. By studying the subhalo mass function in $N$-body simulations (and assuming that it corresponds to the abundance of satellite galaxies\footnote{This assumption --- often referred to as `abundance matching' --- is generally in good agreement with semi-analytic techniques (see, e.g.,~\cite{ref:ghigna_et_al_00, ref:cooray_sheth_02, ref:kravtsov_et_al_04}), although the exact choice of which property of the dark matter halo best correlates with galaxy abundance is a topic of active research~\cite{ref:wechsler_tinker_18}. Note, however, that abundance matching is not guaranteed to capture potentially important baryonic effects, such as the fact that the cooling efficiency of the gas --- and hence, the efficiency of galaxy formation --- diminishes with the mass of the host halo, as its virial temperature grows.}), Ref.~\cite{ref:kravtsov_et_al_04} found that, at sufficiently high masses, the population of satellite galaxies is well described\footnote{The satellite HOD only begins to deviate significantly from the Poissonian approximation in the mass range where the HOD of centrals dominates the overall behaviour; see figure 4 of~\cite{ref:kravtsov_et_al_04}.} by a Poissonian distribution whose mean is given by a power law in mass,
\begin{equation}\label{eqn:hod_satellites}
    \langle \Nsat \rangle_{\Ncen=1} = \left(\frac{M}{M_{\mathrm{sat}}}\right)^{\alpha}\,,
\end{equation}
where, indeed, $\alpha\approx 0.87 \pm0.01$, and $M_{\mathrm{sat}}$ is defined to be the mass of halos which (on average) host a single satellite galaxy. These parametrisations of satellites and centrals can be extended to accommodate non-zero scatter in the relation between host halo mass and number of galaxies; see, e.g.,~\cite{ref:zheng_et_al_05}. These are the essential ingredients of the galaxy HOD we will use in later sections of this paper; in particular, we will use the model parameters determined empirically by~\cite{ref:leauthaud_et_al_12} (the `SIG MOD1’ model for the median redshift bin), as implemented in the publicly-available code \texttt{hmvec}\footnote{The code \texttt{hmvec}, which is based on the work of Ref.~\cite{ref:smith_et_al_18}, can be found at \url{https://github.com/simonsobs/hmvec} .}. In order to reproduce the characteristics of various surveys, we follow~\cite{ref:smith_et_al_18} (see, in particular, their appendix 3) and start from their fiducial galaxy surface density as a function of redshift, convert this to a comoving number density and from that obtain a minimum mass threshold at each redshift bin.

Notice that, in the equation above, we have made a distinction between the mean number of satellites across halos of any type, $\langle \Nsat \rangle$, and the mean number in halos that already have a central galaxy, $\langle \Nsat \rangle_{\Ncen=1}$. Rigorously speaking, it is only in the latter set of halos that the distribution of satellites can be considered Poissonian. The two definitions can be related as
\begin{align}
    \langle \Nsat \rangle &= \sum_{\Ncen} \sum_{\Nsat} \Nsat P(\Nsat|\Ncen,M) P(\Ncen|M)\nonumber \\
    & = P(\Ncen=1|M) \sum_{\Nsat} \Nsat P(\Nsat|\Ncen=1,M)\nonumber\\
    &= \langle \Ncen \rangle \langle \Nsat \rangle_{\Ncen=1}\,.
\end{align}
In going to the second line, we have assumed that satellite galaxies only form once there is a central galaxy in the host halo; and in going to the last, we have used the fact that $P(\Ncen=1|M) = \langle \Ncen \rangle$, since $\Ncen\in\{0,1\}$. Note also that $\Ngal=\Ncen + \Nsat$.

In equations~\eqref{eqn:1h_dm_power} and~\eqref{eqn:general_2h_ps_term}, we weighted the normalised halo profiles by mass in order to draw conclusions about the density distribution. How should this weighting be modified if we were instead interested in the statistics of galaxy number density? We will now address this question in detail; the understanding we develop along the way will be useful to our discussions in section~\ref{sec:analytic_framework} of extensions of the halo model to higher-point functions of the CIB.

Consider, first, the one-halo term. In particular, let us calculate the contribution to the galaxy power spectrum coming from pairs of galaxies\footnote{Contributions from individual galaxies are also possible, but these are part of the `shot-noise' budget.} within a halo of mass $M$, a quantity we will call $Q$. One important ingredient we will need is the spatial distribution of galaxies across the halo, $u_{\mathrm{gal}}\left(r | M \right)$; this is radial, by assumption, and has Fourier transform $u_{\mathrm{gal}}\left(k | M \right)$. In general, it is a very good approximation to replace $ u_{\mathrm{gal}}\left(k | M \right)$ with an NFW profile~\cite{ref:cooray_sheth_02}; however, we will keep the $u_{\mathrm{gal}}$ notation here to retain generality. We will use the fact that central galaxies sit at the centre of halos, where $ u_{\mathrm{gal}}\left(k | M \right) \rightarrow 1$, though this assumption should be relaxed if aiming to incorporate mis-centring. Hence, we can sum over all possible pairs of galaxies in the halo as
\begin{align}
    \mathcal{Q} &= \sum_{\Ngal=2} P(\Ngal | M) \left[\Nsat u_{\mathrm{gal}}\left(k | M \right) + \frac{1}{2} \Nsat (\Nsat -1) u_{\mathrm{gal}}\left(k | M \right)^2  \right] \nonumber \\
    & = \langle\Ncen\rangle \sum_{\Nsat=0} P(\Nsat | \Ncen=1, M)  \left[\Nsat u_{\mathrm{gal}}\left(k | M \right) + \frac{1}{2} \Nsat (\Nsat -1) u_{\mathrm{gal}}\left(k | M \right)^2  \right] \nonumber \\
    & = \frac{1}{2} \langle\Ncen\rangle  \left[2\langle\Nsat\rangle_{\Ncen=1} u_{\mathrm{gal}}\left(k | M \right) + \langle\Nsat\rangle^2_{\Ncen=1} u_{\mathrm{gal}}\left(k | M \right)^2  \right]\nonumber\\
    &= \frac{1}{2\langle\Ncen \rangle} \left[2\langle \Ncen \rangle \langle \Nsat \rangle u_{\mathrm{gal}}\left(k | M \right)  +  \langle \Nsat \rangle^2 u_{\mathrm{gal}}\left(k | M \right)^2 \right]\,.
\end{align}
In the third line, we have used the fact that $\Nsat$ follows a Poisson distribution when conditioned to $\Ncen=1$, so  $\langle\Nsat(\Nsat-1)\rangle_{\Ncen=1} = \langle\Nsat\rangle^2_{\Ncen=1}$. Finally, in the last line, we have restored the unconditional averages, expressed in terms of the entire halo population. Note that $M_{\mathrm{sat}}$ is significantly larger than $M_{\mathrm{min}}$ (see, e.g., Ref.~\cite{ref:kravtsov_et_al_04}), so $\langle \Ncen \rangle\approx1$ for the mass of halos that can host at least a pair of galaxies; for this reason the $\langle \Ncen \rangle$ factor in the denominator is usually dropped. It can be shown (e.g.,~\cite{ref:seljak_00}) that $\mathcal{Q}\approx \langle \Ngal (\Ngal-1) \rangle u_{\mathrm{gal}}^{p}$ in both the high- and low-halo-occupancy regimes (with $p=2$ in the former, and $p=1$ in the latter), and indeed this formulation is more common in the literature. We will encounter it shortly, when discussing generalisations to higher-point functions of the galaxy field. Though this formulation is approximate, the relative error incurred is typically small.

Inspired by this weighting, we can modify equation~\eqref{eqn:1h_dm_power} to obtain the one-halo contribution to the galaxy power spectrum~\cite{ref:cooray_sheth_02}:
\begin{align}\label{eqn:1h_gal_power}
    P^{1\mathrm{h}}_{\mathrm{gal}}(k)  & =  \int \mathrm{d}M  n(M) \frac{1}{\bar{n}^2_{\mathrm{gal}}} \left[2\langle\Ncen\rangle \langle\Nsat\rangle u_{\mathrm{gal}}\left(k | M \right) + \langle\Nsat\rangle^2 u_{\mathrm{gal}}\left(k | M \right)^2\right]\,,
\end{align}
where
\begin{equation}\label{eqn:ngal}
    \bar{n}_{\mathrm{gal}} = \int \mathrm{d}M n(M) \left(\langle\Ncen\rangle +\langle\Nsat\rangle \right) \,
\end{equation}
and the averages are to be evaluated at mass $M$. Note also that, in a magnitude-limited sample, the integration would be restricted to the mass range accessible given the depth of the survey (once a relationship between halo mass and galaxy luminosity has been asserted). This is taken care of by the redshift-dependent minimum-mass thresholds in equations~\eqref{eqn:ncen_parametrisation} and~\eqref{eqn:hod_satellites}.

The two-halo term depends only on the mean number of galaxies per halo; it is given by
\begin{align}\label{eqn:2h_gal_power}
    P^{2\mathrm{h}}_{\mathrm{gal}}(k)  & = P_{\mathrm{hh}}(k) \int \mathrm{d}M_1 \left(\frac{\langle\Ngal\rangle(M_1)}{\bar{n}_{\mathrm{gal}}}\right) b(M_1) n(M_1) u_{\mathrm{gal}}\left( k | M_1 \right) \nonumber \\
    & \hphantom{P_{\mathrm{hh}}(k) \int} \times  \int \mathrm{d}M_2 \left(\frac{\langle\Ngal\rangle(M_2)}{\bar{n}_{\mathrm{gal}}}\right) b(M_2) n(M_2) u_{\mathrm{gal}}\left( k | M_2 \right)\,.
\end{align}
Finally, the total power in galaxy clustering is
\begin{equation}
    P_{\mathrm{gal}}(k) = P^{1\mathrm{h}}_{\mathrm{gal}}(k) + P^{2\mathrm{h}}_{\mathrm{gal}}(k)  + P^{\mathrm{shot}}_{\mathrm{gal}}(k),
\end{equation}
where $P^{\mathrm{shot}}_{\mathrm{gal}}(k)$ is a shot-noise contribution where both legs come from the same galaxy. We will not consider shot-noise terms here; instead, we refer the reader to, e.g.,~\cite{ref:shang_et_al_12}, for details.

If we compare equations~\eqref{eqn:1h_gal_power} and~\eqref{eqn:2h_gal_power} to their dark-matter-only counterparts, equations~\eqref{eqn:1h_dm_power} and~\eqref{eqn:general_2h_ps_term}, we can immediately see that the clustering of galaxies will differ from that of the dark matter to the extent that $\langle\Ngal\rangle$ and $2\langle\Nsat\rangle\langle\Ncen\rangle + \langle\Nsat\rangle^2$ are not exactly proportional to $M$ and $M^2$, respectively. Thanks to this flexibility, the combination of halo model for dark matter clustering and a HOD has been used to successfully fit the two-point function of galaxies (e.g.~\cite{ref:tinker_et_al_10b, ref:coupon_et_al_12}), which has helped constrain possible models of galaxy formation.

Higher-order statistics of the galaxy population can be obtained by generalising the ideas in this section to higher-point functions of the matter field, such as equations~\eqref{eqn:total_matter_bispectrum} or~\eqref{eqn:total_matter_trispectrum}. In the literature, one can find extensions of this kind to calculate, for example, the three-point function of galaxies~\cite{ref:wang_et_al_04}, the redshift-space bispectrum~\cite{ref:smith_et_al_08} or the CIB bispectrum\footnote{We will see more on CIB halo models in section~\ref{sec:halo_model_cib}.}~\cite{ref:lacasa_halo_model}.

This last reference provides a convenient set of rules to build higher-point functions of the 3D galaxy field. The $n$th-order polyspectrum of galaxies (that is, the Fourier transform of the galaxy $n$-point function) will be a sum over all possible ways of arranging $n$ galaxies in halos (including shot-noise-type terms where the same galaxy is featured several times). Each of these terms can be written down as follows:
\begin{itemize}
    \item Start with an overall factor of $\left(1/\bar{n}_{\mathrm{gal}}\right)^n$.
    \item For each one of $p$ halos (with $p\leq n$) labelled by $\alpha_i$, where $i \in \{1,\dots,p\}$, put in a factor of $\int \mathrm{d}M_{\alpha_i}$; a halo mass function $n(M_{\alpha_i})$; and a factorial moment $\langle \Ngal(\Ngal-1)\dots (\Ngal-j) \rangle (M_{\alpha_i})$, where $j+1$ is the number of different galaxies in that halo.
    \item For each galaxy in halo $i$, include a density profile $u_{\mathrm{gal}}\left( k_i | M_{\alpha_i} \right)$. If the term at hand has the same galaxy contracted several times (hence contributing to the shot-noise), $k_i$ is the magnitude of the sum over the momenta of all the legs associated with that galaxy.
    \item Include the $p$-th order halo polyspectrum\footnote{See~\cite{ref:lacasa_halo_model} for a rigorous definition.} evaluated at the masses (and redshifts) of the halos, $\mathcal{P}^{(p)}_{\mathrm{h}}\left(\sum_{r\in\alpha_1}\bm{k}_r,\dots,\sum_{r\in\alpha_p}\bm{k}_r | M_{i},\dots,M_{p}\right)$, where the sum $\sum_{r\in\alpha_i}\bm{k}_r$ is over the wavevectors associated with galaxies in halo $\alpha_i$. The first few halo polyspectra --- with $p=2$, $p=3$ and $p=4$ --- are the halo power spectrum, bispectrum and trispectrum of equations~\eqref{eqn:halo_ps},~\eqref{eqn:total_matter_bispectrum} and~\eqref{eqn:total_matter_trispectrum}, respectively.
\end{itemize} 
The total contribution is then the sum over all possible arrangements of the $n$ galaxies in the halos which produce different diagrams/expressions, weighted by the appropriate permutation factors.

The factorial moment in the second point can be cast in a form that makes more transparent the connection with equation~\eqref{eqn:1h_gal_power}, and with the mass-dependent-luminosity models we will introduce in section~\ref{sec:halo_model_cib}. In appendix~\ref{appendix:hod_factorial_moments}, we show that
\begin{align}\label{eqn:factorial_moment_relation}
    \langle \Ngal(\Ngal-1)\dots (\Ngal-j) \rangle = \langle \Nsat \rangle^{j} \left[ (j+1)\langle \Ncen \rangle + \langle \Nsat \rangle \right ]\,,
\end{align}
where we have once again omitted the mass and redshift dependences of the averages. This relation only holds insofar as the distribution of satellites can be approximated as Poissonian, and that galaxy conformity holds. Though this is a very good approximation, higher-order factorial moments in principle require knowledge of higher-order moments of the full probability distribution, $P(\Ngal|M)$.

\subsection{A halo model of the CIB}\label{sec:halo_model_cib}

As described in section~\ref{sec:intro_to_cib}, the CIB is helping further our understanding of key areas of astrophysics, such as star formation or CMB lensing. These promising use-cases have prompted numerous efforts to model the CIB. Given the existing evidence for clustering contributions to its power spectrum~\cite{ref:hall_CIB, ref:dunkley_et_al_11, ref:planck_cib_11} and bispectrum~\cite{ref:crawford_et_al_14, ref:planck_13_cib}, a framework that can accommodate clustering is evidently preferred over simpler treatments that assume the sources are uncorrelated. The halo model is a natural candidate for this; here, we briefly review how it can be extended to understand the CIB.

Let $I_{\nu}$ denote the specific intensity of the CIB at frequency $\nu$. This is given by
\begin{align}
    I_{\nu} & = \int \mathrm{d}z \frac{\mathrm{d}\chi}{\mathrm{d}z} a(z) j_{\nu}(z) \nonumber \\
    & = \int \mathrm{d}z \frac{\mathrm{d}\chi}{\mathrm{d}z} a(z) \bar{j}_{\nu}(z)\left[1 + \frac{ \delta j_{\nu}(z)}{ \bar{j}_{\nu}(z)}\right] \,,
\end{align}
where $a$ is the scale factor of the Universe, and $\chi$ is the comoving distance to redshift $z$; note also that we have assumed a spatially-flat Universe. The expression above is an integral along the line of sight of the comoving infrared emissivity, $j_{\nu}$, which is sourced by the luminosity of galaxies:
\begin{equation}\label{eqn:emissivity_definition}
    j_{\mathrm{\nu}} (z) = \frac{1}{4\pi}\int \mathrm{d} L\, n(L, z) L_{(1+z)\nu} \,.
\end{equation}
Here, $ n(L, z)$ is the IR galaxy luminosity function.

Early attempts to apply the halo model to the CIB (e.g.,~\cite{ref:planck_cib_11, ref:amblard_et_al_11}) assumed that the emissivity traces the galaxy number density, so that $\delta  j_{\nu} / \bar{ j}_{\nu} = \delta  n_{\mathrm{gal}} / \bar{ n}_{\mathrm{gal}} $. This is equivalent to assuming that the luminosity function above is flat; that is, that all galaxies have the same luminosity. Though conveniently simple, these models failed to fit simultaneously the CIB power spectra across the range of frequencies observed by Planck~\cite{ref:planck_cib_11}.

An improved model was introduced by Ref.~\cite{ref:shang_et_al_12}, who let galaxy luminosity be a function of halo mass. This fix is important: the assumption that all galaxies have the same luminosity leads to over-estimation of the number of satellites in massive halos when fitting the CIB power spectrum on small scales. It is for this reason that previous studies, such as~\cite{ref:amblard_et_al_11}, returned values of $\alpha$ --- the power law exponent in equation~\eqref{eqn:hod_satellites}, which governs the high-mass scaling of the mean number of galaxies per halo --- sigificantly higher than the theoretical prediction of $\alpha \leq 1$. The model of Ref.~\cite{ref:shang_et_al_12} resolves this by having fewer, but more luminous, satellite galaxies in massive halos. Slightly modified versions of this model were subsequently used to fit successfully Planck~\cite{ref:planck_13_cib} and Herschel~\cite{ref:hermes_viero_wang} measurements of the power spectra of CIB anisotropies. This is also the model that we will use later on to describe the CIB. We explain it, in detail, below.

If we allow galaxy luminosity to depend on the mass of the host halo, we can rewrite equation~\eqref{eqn:emissivity_definition} as
\begin{equation}\label{eqn:emissivity_ito_centrals_and_sats}
    j_{\mathrm{\nu}} (z) = \int \mathrm{d} M\, n(M, z) \left[f^{\mathrm{cen}}_{\nu} (M,z) + f^{\mathrm{sat}}_{\nu} (M,z) \right]  \,,
\end{equation}
where we have separated the contributions from centrals,
\begin{equation}
    f^{\mathrm{cen}}_{\nu} (M,z)= \frac{1}{4\pi} \Ncen L_{\mathrm{cen},\,(1+z)\nu}(M,z)  \,,
\end{equation}
and from satellite galaxies,
\begin{equation}
    f^{\mathrm{sat}}_{\nu}(M,z)= \frac{1}{4\pi}\int \mathrm{d} m\, n_{\mathrm{sub}}(m, z|M) L_{\mathrm{sat},\,(1+z)\nu}(m,z)  \,.
\end{equation}
In these expressions, $n(M, z)$ and $n_{\mathrm{sub}}(m, z)$ are the halo and subhalo mass functions, respectively. The latter element can be obtained from the fitting functions of Ref.~\cite{ref:tinker_wetzel_10}, for example.

Note now the similarity between equation~\eqref{eqn:emissivity_ito_centrals_and_sats} and the galaxy number density of equation~\eqref{eqn:ngal}; this suggests that the power spectrum of the emissivity can be obtained by simply replacing $\bar{n}_{\mathrm{gal}} \rightarrow \bar{j}$, $\Ncen \rightarrow f^{\mathrm{cen}}$ and $\Nsat \rightarrow f^{\mathrm{sat}}$ in the expressions for the galaxy power spectrum, equations~\eqref{eqn:1h_gal_power} and~\eqref{eqn:2h_gal_power}. (This can also be shown rigorously by following a derivation analogous to the one in section~\ref{sec:halo_model_for_galaxies}.) The same will be true for higher-point functions of the emissivity.

The key ingredient in this model is a description of how luminosity depends on halo mass, frequency and redshift. Neglecting any scatter in the relation, we can write
\begin{equation}
    L_{(1+z)\nu}(M,z) = L_0 \Phi(z) \Sigma (M) \Theta \left((1+z)\nu, z\right)\,. 
\end{equation}
Here, $L_0$ is a normalisation constant, and the other functions will be defined shortly.

The function $\Phi(z)$ attempts to capture how the IR luminosity of the galaxies comprising the CIB evolves with redshift. For this class of galaxies, luminosity is known to correlate with star formation rate (SFR)~\cite{ref:kennicutt_98}. The SFR grows with redshift because of higher accretion and merger rates, higher gas fractions and objects being more compact at earlier times (see, e.g.,~\cite{ref:wang_et_al_13} and references therein). Reference~\cite{ref:shang_et_al_12} argues that, theoretically, this evolution should be well approximated as a power law,
\begin{equation}
    \Phi(z)=(1+z)^{\delta}\,,
\end{equation}
though they also warn that the data, albeit still uncertain, seem to favour more flexible models with a plateau at high redshifts $z>2$; e.g.,~\cite{ref:tasca_et_al_15}.

The luminosity--mass, or $L$\,--\,$M$ relation is modelled by the function $\Sigma (M)$. It is well known that the stars driving this luminosity only form efficiently in a range of halo masses, since at lower and higher masses their formation is suppresed by feedback processes associated with supernovae and AGN activity, photoionisation heating, or virial shocks; see, e.g.,~\cite{ref:benson_et_al_03, ref:croton_et_al_06}. Denoting as $M_{\mathrm{eff}}$ the mass at which star formation peaks, the dependence of galaxy luminosity on halo mass can be modelled as a log-normal distribution,
\begin{equation}
    \Sigma (M) = M \frac{1}{\left(2\pi \sigma^2_{L/M}\right)^{1/2}} e^{- \left(\log_{10} M-\log_{10} M_{\mathrm{eff}}\right)^2/ (2\sigma^2_{L/M})}\,,
\end{equation}
where $\sigma_{L/M}$ parametrises the characteristic width of the distribution. In order to account for the quenching of gas accretion at low masses~\cite{ref:bouche_et_al_10}, the distribution is truncated by setting $L=0$ below $M<M_{\mathrm{min}}$, where $M_{\mathrm{min}}$ is an additional parameter to be fit for in the range $10^{10}$\,--\,$10^{11}\,\mathrm{M}_{\odot}$. 

Finally, $\Theta (\nu,z)$ reflects the frequency-dependence of the IR luminosity of dusty, star-forming galaxies. Inspired by the CIB SED of Ref.~\cite{ref:hall_CIB}, Ref.~\cite{ref:shang_et_al_12} models it as
\begin{equation}\label{eqn:cib_sed}
    \Theta (\nu,z) = \begin{cases}
        \nu^{\beta} B_{\nu}(T_{\mathrm{d}}), & \text{if $\nu < \nu_0$; \quad and}\\
        \nu^{-\gamma}, & \text{if $\nu > \nu_0$\,,}
    \end{cases}
\end{equation}
where $B_{\nu}$ is the Planck function; and $\beta, \gamma$ and $T_{\mathrm{d}}(z)$ are parameters to be fit for. The temperature of the dust within the galaxies sourcing the CIB, $T_{\mathrm{d}}(z)$, is presumed to be a function of redshift and parametrised as~\cite{ref:planck_13_cib}
\begin{equation}
    T_{\mathrm{d}}(z) = T_0 (1+z)^{\alpha}\,.
\end{equation}
See Ref.~\cite{ref:planck_13_cib} for a discussion of the physical interpretation and appropriate prior ranges for these parameters. To ensure a smooth transition between the two regimes in equation~\eqref{eqn:cib_sed}, the free parameters are fit for subject to the constraint that $\mathrm{d}\ln \left[\nu^{\beta} B_{\nu}(T_{\mathrm{d}})\right] / \mathrm{d}\ln \nu = -\gamma$ at $\nu_0$.

In order to keep the complexity of the model to a minimum, it is standard to assume that both satellites and centrals have the same luminosity if they inhabit halos of equal mass, at the same redshift\footnote{This appears to be a rather good approximation when the mass of the subhalo is measured at the time of accretion~\cite{ref:planck_13_cib}.}; hence, we drop the subscripts and set $L_{\mathrm{sat},\,\nu}(m,z)= L_{\mathrm{cen},\,\nu}(m,z) = L_{\nu}(m,z)$.

Reference~\cite{ref:planck_13_cib} used the framework described above to model the 15 possible combinations of auto- and cross-spectra measured by Planck at 217, 353, 545, 857\,GHz, and by IRAS\footnote{The measurements were made on the reprocessed IRAS maps of~\cite{ref:miville-deschenes_lagache_05}, which go by the name of IRIS.} at 3000\,GHz, on angular scales $187\leq \ell \leq 2649$. They constrained eight halo model parameters, $\{\alpha,T_0, \beta,\gamma,\delta,M_{\mathrm{eff}},M_{\mathrm{min}}, L_0\}$ (the width of the $L$\,--\,$M$ distribution was held fixed at a value of $\sigma^2_{L/M}= 0.5$), along with 15 other ones corresponding to the shot-noise in each channel. Best-fit values for the former set are shown in table~\ref{tab:planck_cib_model_params}; we refer the reader to Ref.~\cite{ref:planck_13_cib} for a discussion of the physical implications of these constraints, as well as for the best-fit values of the shot-noise parameters. The model not only provides a very good fit to the data, but it also predicts with notable accuracy the cross-correlation between CMB lensing reconstructions and CIB observations made by Planck at various frequencies. Reference~\cite{ref:hermes_viero_wang} fit a slightly-modified version of the model to Herschel data, obtaining results that were qualitatively similar.
\begin{table}
    \centering
    \begin{tabular}{l l}
        \toprule
        Parameter & Mean value  \\ \midrule
        $\alpha$ & $0.36\pm0.05$\\ 
        $T_0$ [K] & $24.4\pm 1.9$\\
        $\beta$ &$ 1.75 \pm 0.06$\\
        $\gamma$ & $1.7 \pm 0.2$\\
        $\delta$ & $3.6 \pm 0.2$\\
        $\log (M_{\mathrm{eff}}/\mathrm{M}_{\odot})$ & $12.6 \pm 0.1$\\
        $M_{\mathrm{min}}$ [$\mathrm{M}_{\odot}$] & unconstrained \\
        \bottomrule
    \end{tabular}
    \caption[Best-fit CIB halo model parameters from Planck data]{Mean values and marginalised $68\%$ confidence intervals for the CIB halo model parameters constrained by Ref.~\cite{ref:planck_13_cib}. They belong to a model, based on that of Ref.~\cite{ref:shang_et_al_12}, which allows for mass-dependent galaxy luminosities. These parameters were fit (jointly with parameterizations of the shot-noise amplitude) to all possible auto- and cross- spectra of Planck data at frequencies of 217, 353, 545, 857\,GHz and IRAS data at 3000\,GHz for angular scales of $187\leq \ell \leq 2649$. The normalisation parameter, $L_0$, is then obtained by fitting the model CIB power spectrum calculated from the parameter values above to the Planck data~\cite{mccarthy_2021_ImprovingModelsCosmic}. From table~9 of Ref.~\cite{ref:planck_13_cib}.}\label{tab:planck_cib_model_params}
\end{table}

\section[Factorial moments of the halo occupation distribution]{Factorial moments of the halo occupation distribution}\label{appendix:hod_factorial_moments}
Consider the proposition
\begin{align}\label{eqn:inductive_prop}
    \langle \Ngal(\Ngal-1)\dots (\Ngal-j) \rangle =  (j+1)\langle \Ncen \rangle & \langle \Nsat (\Nsat-1)\dots (\Nsat-j+1) \rangle \nonumber \\ + & \langle \Nsat (\Nsat-1)\dots (\Nsat-j) \rangle\,.
\end{align}
If the population of satellite galaxies obeys Poisson statistics,
\begin{equation}
    \langle \Nsat (\Nsat-1)\dots (\Nsat-j) \rangle = \langle \Nsat \rangle^{j}\,,
\end{equation}
and the expression above reduces to
\begin{equation}
     \langle \Ngal(\Ngal-1)\dots (\Ngal-j) \rangle = \langle \Nsat \rangle^{j} \left[ (j+1)\langle \Ncen \rangle + \langle \Nsat \rangle \right ]
    \,.
\end{equation}
To prove equation~\eqref{eqn:inductive_prop}, suppose it is true for $j=k$. We will now show that it is then necessarily true for $j=k+1$. We will be working under the assumption of galaxy conformity, whereby expectation values over $\Ncen$ and $\Nsat$ can be treated independently. It follows that
\begin{align}
      \langle \Ngal(\Ngal-1)\dots   (\Ngal-(k+1) \rangle &= \langle \big[ (k+1) \Ncen \Nsat (\Nsat-1)\dots (\Nsat-k+1)  \nonumber \\
    & \quad  +  \Nsat (\Nsat-1)\dots (\Nsat-k) \big]\big[\Ncen + \Nsat - (k+1)\big] \rangle \nonumber \\
    &= (k+1)\langle \Ncen (\Ncen-1) \Nsat (\Nsat-1) \dots (\Nsat -k+1)\rangle \nonumber \\
      & \quad +  (k+1) \langle \Ncen \Nsat (\Nsat-1) \dots (\Nsat -k)\rangle \nonumber \\
      & \quad + \langle \Ncen \Nsat (\Nsat-1) \dots (\Nsat -k)\rangle \nonumber \\ 
      & \quad + \langle \Nsat (\Nsat-1) \dots (\Nsat -k -1)\rangle\, . \label{eqn:term_that_cancels}
\end{align}
Under the assumption of conformity, the first term in %line~\eqref{eqn:term_that_cancels} cancels
the final equality is zero
because it is proportional to $\langle \Ncen(\Ncen-1)\rangle$ and $\Ncen$ can only take values $\{0,1\}$. Combining the remaining terms, we have
\begin{align}
      \langle \Ngal(\Ngal-1)\dots   (\Ngal-(k+1) \rangle = \,&(k+2)\langle \Ncen \rangle \langle \Nsat (\Nsat-1) \dots (\Nsat -k)\rangle \nonumber \\
      & + \langle \Nsat (\Nsat-1) \dots \left[\Nsat -(k +1)\right]\rangle \,,
\end{align}
so equation ~\eqref{eqn:inductive_prop} is true for $j=k+1$ whenever it is true for $j=k$.  We can easily show that it is true for $j=k=1$,
\begin{align}
      \langle \Ngal(\Ngal-1) \rangle = &  \langle (\Nsat + \Ncen)(\Nsat + \Ncen-1) \rangle \nonumber \\
     = &  \langle \Ncen(\Ncen-1)\rangle + \langle \Nsat(\Nsat-1)\rangle + 2 \langle \Ncen \Nsat \rangle \nonumber \\
     = & \langle \Nsat(\Nsat-1)\rangle + 2 \langle \Ncen \rangle \langle \Nsat \rangle\,,
\end{align}
so it follows that the proposition is true for all integers $k>0$.

\section[Fast lensing reconstructions using Hankel transforms]{Fast lensing reconstructions using Hankel transforms}\label{appendix:qe_w_fftlog}
The $TT$ quadratic estimator can be written in real space as
\begin{equation}\label{eqn:real_space_tt_qe}
    \hat{\phi}(\vL) = - A^{TT}_{L} \int \frac{d^2\vx}{2\pi}\, e^{-i \vL\cdot\vx} \mathbf{\nabla} \cdot [ F_1(\vx) \mathbf{\nabla} F_2(\vx)],
\end{equation}
where
\begin{equation}
    F_1(\vL) \equiv \frac{\hat{T} (\vL)}{\tilde{C}_l^{\mathrm{tot}}} \quad \text{and} \quad F_2(\vL) \equiv \frac{\tilde{C}_l^{TT}\,\hat{T} (\vL)}{\tilde{C}_l^{\mathrm{tot}}}\,.
\end{equation}
Here, $\hat{T}$ denotes the observed temperature anisotropies of the CMB, and we assume that the filtering is diagonal in harmonic space. Intuitively, the QE harnesses the fact that, to leading order, lensing introduces a dependence on its unlensed gradient into the lensed CMB. Hence, $\phi$ can be extracted by correlating (filtered) observations of small-scale, lensed anisotropies with their (filtered) gradient.

Thanks to the approximate azimuthal symmetry of the projected profiles and the diagonal-independence of all the terms we implement, we can always evaluate the lensing reconstructions letting $\hat{T}(\vl) \approx \hat{T}(l)$, such that $F_1(\bm{l}) \approx F_1(l)$ and $F_2(\bm{l}) \approx F_2(l)$. This will enable us to carry out the angular integrals analytically. As expected, the two-dimensional Fourier transform of an isotropic profile reduces to its Hankel transform (see, e.g., Ref.~\cite{ref:faris_08}),
\begin{align}
    F(\vx) & = \int \frac{d^2\vl}{2\pi}\, F(\vl) e^{- i \vl \cdot \vx} \nonumber \\
    & = \int \frac{dl d\theta}{2\pi}\, l F(l) e^{- i l r \cos \theta} \nonumber \\
    & = \int dl \, l F(l) J_0(lr),
\end{align}
where $r=|\bm{x}|$, and $J_n(x)$ is the $n$-th order Bessel function of the first kind. In the last line, we have used the fact that
\begin{equation}
    \int_0^{2\pi} d\theta e^{- i l r \cos \theta} = 2\pi J_0(lr).
\end{equation}
Consequently, the quadratic estimator can be rewritten as
\begin{align}\label{eqn:isotropic_qe_v0}
    \hat{\phi}(\vL) &= - A^{TT}_{L} \int \frac{dr d\Psi}{2\pi}\,r e^{- i L r \cos\left(\Psi_{\vL} - \Psi\right)} \bm{\nabla} \cdot \left[ F_1(r) \bm{\nabla}F_2(r)\right] \nonumber \\
    & = - A^{TT}_{L} \int dr \,r J_0(L r)  \bm{\nabla} \cdot \left[ F_1(r) \bm{\nabla}F_2(r)\right] \nonumber \\
    & = - A^{TT}_{L} \int dl' \, l' F_1(l') \int dl'' \, l'' F_2(l'') \int dr \, J_0(Lr) \partial_r \left[ r J_0(l'r) \partial_r J_0(l''r)\right],
\end{align}
where, in the first line, we defined $\Psi_{\vL}$ as the angle between the vector $\vL$ and the $x$-axis. Calculating the product of three $J_0$ functions analytically, the expression reduces to
\begin{align}\label{eqn:analytic_qe}
    \hat{\phi}(\vL) = - 2A^{TT}_{L} \int \frac{dl\,dl'}{2\pi} & lF_1(l) l'F_2(l') \Delta(l,l',L) \nonumber \\
    & \times \left(\frac{L^2 + l'^2 - l^2}{2Ll'} \right)\left[1 - \left( \frac{L^2 +l'^2 -l^2}{2Ll'}\right)\right]^{-\frac{1}{2}},
\end{align}
where $\Delta(l,l',L)=1$ if the triangle inequality is satisfied and zero otherwise (note that there is an integrable singularity at the limits of the inequality)\footnote{Note that the integral in equation~\ref{eqn:analytic_qe} is proportional to the square of the $3j$ symbol with all $m$s zero, since it is the flat-sky version of
\begin{equation}
    \int d\cos\theta\, P_L(\cos\theta)  \bm{\nabla} \cdot\left[ P_{l'}(\cos\theta) \bm{\nabla}P_{l''}(\cos\theta)\right] = - \int d\cos\theta\, P_{l'}(\cos\theta)  \bm{\nabla}P_{L}(\cos\theta) \cdot \bm{\nabla}P_{l''}(\cos\theta) \, .
\end{equation}
This is basically ${}_2 I_{l' L l''}^{000}$ of equation~\eqref{eq:harmonic_firstorder}.
Integrating by parts several times reduces to something proportional to the integral of the product of three Legendre polynomials (with no derivatives).}.
In practice, these integrals are computed between some limits $[l_{\mathrm{min}}, l_{\mathrm{max}}]$. Although this approach scales as $O(N^2)$, it can be evaluated very fast using Gaussian quadratures, which require relatively few integration points, implemented in the form of matrix multiplication.

\subsection{Algorithm for fast evaluation using Gaussian quadratures}\label{appendix:gaussian_quad}
Quadrature rules for numerical integration work by approximating the integral of a function as a weighted sum of the values of that function at $n$ key locations, called nodes:
\begin{equation}
    \int_{-1}^{1}dx\, f(x) \approx \sum_{i=1}^{n}w_i f(x_i)\,.
\end{equation}
Such a rule is said to be of order $n$. Though there are myriad quadrature rules, the `Gaussian rule' is particularly important due to being exact for polynomials of degree $2n-1$ or less. In this case, node $x_i$ is the $i$-th root of the Legendre polynomial of order $n$, and the weights are equally well understood and tabulated in any numerical integration package. The method can be applied to a different integration domain through a trivial change of variables:
\begin{equation}
    \int_{a}^{b}dx\, f(x) \approx \frac{b-a}{2}\sum_{i=1}^{n}w_i f\left(\frac{b-a}{2}x_i + \frac{a+b}{2}\right)\,.
\end{equation}

Our calculation entails carrying out a large number of double integrals of the form of~\eqref{eqn:analytic_qe}: one per step in mass, redshift and $L$, for each of the many terms. These operations thus become our bottleneck and call for careful optimization. Let us therefore revisit equation~\eqref{eqn:analytic_qe} while introducing some key definitions:
\begin{align}\label{eqn:key_integ_qe}
    \hat{\phi}(\vL) / A^{TT}_{L} =  \int_{l_{\mathrm{min}}}^{l_{\mathrm{max}}} dl\,dl' & F_1(l) F_2(l') W(L, l, l') \,,
\end{align}
with
\begin{align}
    W(L, l, l') \equiv -   \Delta(l,l',L) l l' \left(\frac{L^2 + l'^2 - l^2}{2\pi Ll'} \right)\left[1 - \left( \frac{L^2 +l'^2 -l^2}{2Ll'}\right)\right]^{-\frac{1}{2}}\,.
\end{align}
The integral on the RHS of~\eqref{eqn:key_integ_qe} can be carried out numerically as
\begin{align}
    \int_{l_{\mathrm{min}}}^{l_{\mathrm{max}}} dl\,dl' & F_1(l) F_2(l') W(L, l, l') \approx \sum_i \sum_j \omega_i \omega_j F_1(l_i) F_2(l_j) W(L, l_i, l_j)\,,
\end{align}
where the nodes and weights are adapted to our integration domain as
\begin{align}
    l_i \equiv \frac{l_{\mathrm{max}}-l_{\mathrm{min}}}{2}x_i + \frac{l_{\mathrm{max}}+l_{\mathrm{min}}}{2} \quad \text{and}\quad \omega_i \equiv \frac{l_{\mathrm{max}} - l_{\mathrm{min}}}{2}\, w_i \,.
\end{align}
In fact, defining a vector $\bm{F}_\eta \equiv  F_{\eta}(l_i)$ and a matrix $ \mathcal{W}(L) \equiv \omega_i \omega_j W(L, l_i, l_j) $, we can cast the integration as matrix multiplication:
\begin{align}
    \int_{l_{\mathrm{min}}}^{l_{\mathrm{max}}} dl\,dl' & F_1(l) F_2(l') W(L, l, l') \approx \bm{F}_1^{T}\mathcal{W}(L) \bm{F}_2 \,.
\end{align}

The matrices $\mathcal{W}(L) = W(L, l_i, l_j)$ can be tabulated at an initial stage, once the quadrature order and output $L$'s are fixed, and used for every lensing reconstruction we need to perform. Similarly, the $F_{\eta}(l_i)$ only need to be computed once for each reconstruction (as in, it need not be recomputed for different $L$). These two caching steps significantly reduce the computational cost. 

In addition, the use of matrix multiplication in the double integrals (and in the iteration over $L$) also enables extensive computational gains owing to the use of efficient libraries, such as \texttt{BLAS}, or, better yet, the use of GPUs. We use 
\texttt{JAX} to harness GPU acceleration when available while retaining the ability to run on CPUs if needed. On a GPU, we can calculate the lensing reconstruction at 50 output $L$'s in approximately 1\,ms, an order of magnitude faster than using \texttt{NumPy} to run the same operation on a CPU\footnote{We run on an Apple M2 Max chip, using just-in-time compilation, single-point precision and a Gaussian quadrature rule of order 1000.}.

\subsection{Alternative evaluation using Hankel tranforms}\label{appendix:fftlog}
The unnormalized quadratic estimator of equation~\eqref{eqn:isotropic_qe_v0} can also be approximated as a nested series of Hankel transforms. We emphasize that this is only an approximation, for the Hankel transform is defined on the interval $[0,\infty]$, whereas our integrals are truncated by the scale cuts $[l_{\mathrm{min}}, l_{\mathrm{max}}]$. Nevertheless, let us quote the approach for reference. 

First, we express the derivatives of $J_0$ in terms of higher-order Bessel functions~\cite{ref:handbook_math_functions}. The integral over three Bessel functions becomes
\begin{align}
    \int dr \, J_0(Lr) \partial_r \left[ r J_0(l'r) \partial_r J_0(l''r)\right] = &\int dr \, r J_0(Lr) \nonumber \\
    \,& \times \bigg[ -\frac{l''}{r} J_0(l'r) J_1(l''r) + l'l'' J_1(l'r)J_1(l''r) \nonumber\\
    \,&\qquad - \frac{1}{2}(l'')^2 J_0(l'r) J_0(l''r) + \frac{1}{2} (l'')^2 J_0(l'r) J_2(l''r)\bigg].
\end{align}
Therefore, if we denote as $\mathcal{H}_n[f]$ the $n$-th order Hankel tranform of function $f$, the quadratic estimator applied to azimuthally-symmetric profiles can be written as
\begin{align}\label{eqn:isotropic_qe_hankel}
    \hat{\phi}(\vL) = - A^{TT}_{L} \mathcal{H}_0\bigg[& -\frac{1}{r}\mathcal{H}_0[F_1(l)]\mathcal{H}_1[l F_2(l)] + \mathcal{H}_1[l F_1(l)]\mathcal{H}_1[l F_2(l)] \nonumber \\
    & - \frac{1}{2}\mathcal{H}_0[F_1(l)]\mathcal{H}_0[l^2 F_2(l)] + \frac{1}{2}\mathcal{H}_0[F_1(l)]\mathcal{H}_2[l^2 F_2(l)]   \bigg].
\end{align}
This can be further simplified to
\begin{align}\label{eqn:isotropic_qe_hankel_v2}
    \hat{\phi}(\vL) = - A^{TT}_{L} \mathcal{H}_0\bigg[& \mathcal{H}_1[l F_2(l)] \left( -\frac{1}{r}\mathcal{H}_0[F_1(l)] + \mathcal{H}_1[l F_1(l)]\right) \nonumber \\
    & + \frac{1}{2}\mathcal{H}_0[F_1(l)] \left(- \mathcal{H}_0[l^2 F_2(l)] + \mathcal{H}_2[l^2 F_2(l)] \right) \bigg]\,.
\end{align}
Hence, we have reduced the problem to the evaluation of seven continuous Hankel transforms.

These transforms can be approximated in a very computationally-efficient way using the FFTlog algorithm for computing discrete Hankel transforms with logarithmically-spaced samples. The computational complexity is thus reduced to $O(4N\log N)$. However, this approach is hindered by the sharp scale-cuts that we impose on the transform inputs, which translate to ringing in the conjugate space. Though these effects can be fine-tuned away for fixed inputs, the parameters of choice do not extrapolate well to different inputs, to the point that we are unable to reach the desired level of accuracy in our results. Our baseline approach is therefore that of appendix~\ref{appendix:gaussian_quad}.

\section{Evaluating the secondary bispectrum bias}\label{appendix:evaluating_secondary_bispec_bias}
Harnessing statistical isotropy, which imposes $\vL = - \vL'$, we can evaluate equation~\eqref{eqn:secondary_bispec_bias} at a fixed $L$ as a QE reconstruction
\begin{align}\label{eqn:outer_rec}
     \langle \hat{\phi}(\vL)\hat{\phi}(\vL') \rangle \supset - \frac{8}{2\pi} \, A^{TT}_{L} \int &  \frac{\mathrm{d}^2\bm{l}'}{2\pi} g(\bm{l}',\vL) \langle G(\vl')_{L} T^{s}(\vL+\bm{l}') \rangle \,,
\end{align}
where one of the input legs is itself a QE reconstruction,
\begin{align}\label{eqn:inner_rec}
     G(\vl')_{L} & \equiv A^{TT}_{L} \int  \frac{\mathrm{d}^2\bm{l}''}{2\pi} g(\bm{l}'',\vL) \left[C^{TT}_{l''} T^{s}(\vL-\vl'') \vl''\right] \cdot \left[  (\vl'+\vl'') \phi(\vl'+\vl'')\right].
\end{align}
The outer reconstruction of equation~\eqref{eqn:outer_rec} has to be carried out by brute force, but the inner one can be computed fast using the convolution theorem, given that $\vL$ is a constant for the purpose of each reconstruction.

\section{Halo model of projected higher-point functions}\label{appendix:projected_higherpnt_fns}
It is evident from equations~\eqref{eqn:prim_bispec_bias} and~\eqref{eqn:trispec_bias} that calculating these lensing biases requires being able to compute angular bispectra and trispectra involving the foreground contaminants. In general, projected observables take the form
\begin{equation}
    q(\hat{\bm{n}}) = \int \mathrm{d}z \frac{\mathrm{d}\chi}{\mathrm{d}z} W(z) Q(\hat{\bm{n}}\chi(z),z)\,,
\end{equation}
where $\chi(z)$ is the comoving distance to redshift $z$, $W(z)$ is some weight function, and $Q(\bm{x},z)$ is a three-dimensional quantity whose projection onto the celestial sphere we observe in the direction of $\hat{\bm{n}}$ as $q(\hat{\bm{n}})$. The following are some relations of this form that are relevant for our purposes\footnote{Note that these relations only hold in flat universes.}. First, the CMB lensing convergence,
\begin{equation}
    \kappa^{\mathrm{CMB}} (\hat{\bm{n}}) = \int \mathrm{d}z \frac{\mathrm{d}\chi}{\mathrm{d}z} W_{\kappa}(z)  \delta_{\mathrm{m}}(\hat{\bm{n}}\chi(z),z) \quad \mathrm{with} \quad  W_{\kappa}(z)  = \frac{3\Omega_{m}H_0^2 \chi }{2ac^2}\frac{\chi_* - \chi}{\chi_*}\,,\label{eqn:convergence}
\end{equation}
where $\chi_*$ is the comoving distance to the surface of last scattering. The convergence is related to the CMB lensing potential, the quantity we use in the main text, by $\kappa=-\nabla^2 \phi /2$.

It will also be useful to consider an external tracer, $\mathcal{G}$, to help us gauge the impact of extragalactic-foreground biases on CMB lensing cross-correlations. Suppose this tracer has redshfit distribution $dN/dz$. Then (e.g.,~\cite{ref:baleato_white_23}),
\begin{equation}
    \mathcal{G} (\hat{\bm{n}}) = \int \mathrm{d}z \frac{\mathrm{d}\chi}{\mathrm{d}z} W_{\mathcal{G}}(z)  \delta_{\mathcal{G}}(\hat{\bm{n}}\chi(z),z) \quad \mathrm{with} \quad  W_{\mathcal{G}}(z)  = \frac{H }{c}\frac{dN}{dz}\,.
\end{equation}

As per the foregrounds themselves, the CIB fluctuation at frequency $\nu$ can be written as
\begin{equation}
    \delta I_{\nu} (\hat{\bm{n}}) = \int \mathrm{d}z \frac{\mathrm{d}\chi}{\mathrm{d}z} W_{\mathrm{CIB}} \delta j_{\nu}(\hat{\bm{n}}\chi(z),z) \quad \mathrm{with} \quad W_{\mathrm{CIB}}(z)  = a(z) \,.
\end{equation}
In appendix~\ref{appendix:fg_cleaning}, we explain how to generalize our calculations when multi-frequency observations are used to mitigate the impact of foregrounds.

On the other hand, the 2D Compton-$y$ map is
\begin{equation}
    y (\hat{\bm{n}}) = \int \mathrm{d}z \frac{\mathrm{d}\chi}{\mathrm{d}z} W_{\mathrm{tSZ}}(z) y_{\mathrm{3D}}(\hat{\bm{n}}\chi(z),z) \quad \mathrm{with} \quad  W_{\mathrm{tSZ}}(z)  = a(z) \,.\label{eqn:3d_y}
\end{equation}
When observing at a single frequency, this can be converted to a CMB temperature increment by $\Delta T/T = g_{\nu} y(\hat{\bm{n}})$, as given by equation~\eqref{eqn:T_change_due_to_tsz}. Alternatively, when dealing with multi-frequency-cleaned maps, the frequency scaling $g_{\nu}$ needs to be replaced by the frequency-weighted harmonic weights defined in equation~\eqref{eqn:freq_indep_tsz_scaling}. 

Note that, in equation~\eqref{eqn:3d_y}, we have introduced a 3D Compton-$y$ field. Following Ref.~\cite{ref:hill_pajer_13}, this is a rescaled version of the electron pressure profile we encountered in section~\ref{sec:tsz},
\begin{equation}
    y_{\mathrm{3D}}(\bm{x},z) \equiv \frac{\sigma_{\mathrm{T}}}{m_{e}c^2} P_{e}(\bm{x},z)\,,
\end{equation}
with dimensions of inverse length. In our calculations, we use the fitting form of Ref.~\cite{ref:battaglia_et_al_12} for the electron pressure profile and its evolution with mass and redshift; the details are described in section~\ref{sec:tsz}.

In order to calculate the `bispectrum' biases --- those arising from the correlation between the matter or galaxy density and the extragalactic foregrounds --- we need a model for the dark matter overdensity, $\delta_{\mathrm{m}}$, which produces the lensing effect. Following the halo model ansatz,  we model the dark matter overdensity as an NFW profile; we let the halo concentration vary with mass and redshift according to the fitting functions of Ref.~\cite{ref:duffy_et_al_08} for the mean concentration seen in simulations. Note that, in assuming that the matter overdensity is entirely associated with halos, we are ignoring possible correlations between material in the field (which actually plays an important role in producing the lensing deflections) and the material that produced the CIB and tSZ emission, which lives mostly in halos.

In the Limber approximation~\cite{ref:limber_53, ref:loverde_afshordi_08, ref:lemos_et_al_17}, angular polyspectra of these projected quantities are related to the equal-redshift polyspectra of the 3D quantities by (e.g.,~\cite{ref:lacasa_halo_model})

\begin{align}\label{eqn:limber_polyspectra}
    \mathcal{P}_{q_1 \dots q_n}^{(n)}(\bm{l}_1, \dots, \bm{l}_n) = \int \mathrm{d}z \frac{\mathrm{d}\chi}{\mathrm{d}z} \chi^{2(1-n)}(z) W_1(z)\dots W_n(z)\mathcal{P}_{Q_1 \dots Q_n}^{(n)}(\bm{k}^*_1, \dots, \bm{k}^*_n;z) \,,
\end{align}
where $\bm{k}_i^* = \bm{l}_i/\chi(z)$.

The expression above depends on the Fourier transform of $Q$. In this work, we assume that all quantities of the form of $Q$ are spherically-symmetric. It is straightforward to show that the Fourier transform of such spherically-symmetric functions simplifies to\footnote{We give equation~\eqref{eqn:sym_ft_of_gen_prof} in the asymmetric Fourier convention to make contact with earlier discussions involving the normalised NFW profile (cf. equation~\eqref{eqn:normalised_radial_nfw}); to convert to the symmetric convention $Q(\bm{k},M,z)$ must be divided by $(2\pi)^{3/2}$.}
\begin{equation}\label{eqn:sym_ft_of_gen_prof}
    Q(\bm{k},M,z) = 4 \pi \int \mathrm{d}r\,r^2\,\frac{\sin (k r)}{k r}\,Q(r,M,z)\,,
\end{equation}
where $k=|\bm{k}|$. This is merely the three-dimensional instance of a more general theorem that relates the Fourier transform of a radial function to its Hankel transform; see, e.g., Ref.~\cite{ref:faris_08}. In appendix~\ref{appendix:qe_w_fftlog}, we use its two-dimensional version to speed up the evaluation of the quadratic estimators of lensing.

Finally, we can use the halo model to calculate the equal-redshift bispectra and trispectra needed to evaluate equation~\eqref{eqn:limber_polyspectra}. At present, we consider only one- and two- halo contributions to equations~\eqref{eqn:total_matter_bispectrum} and~\eqref{eqn:total_matter_trispectrum}, because these are the ones we expect to dominate. Moreover, three- and and four-halo trispectra are typically diagonal-dependent (in the language of~\cite{ref:lacasa_halo_model}) and thus more difficult to project\footnote{Note that the three-halo bispectrum contains diagonal-independent terms that would be rather straight-forward to implement. We defer doing this to future work.}.

Starting with the bispectrum, the terms we consider have the form
\begin{align}\label{eqn:B_1halo}
    B_{Q_1 Q_2 Q_3}^{\mathrm{1h}}(\bm{k}_1,\bm{k}_2,\bm{k}_3;z) &= \int \mathrm{d}M\, n(M,z)Q_1(k_1,M,z)Q_2(k_2,M,z)Q_3(k_3,M,z)\,, 
\end{align}
and
\begin{align}\label{eqn:general_prim_bispec_2h}
    B_{Q_1 Q_2 Q_3}^{\mathrm{2h}}(\bm{k}_1,\bm{k}_2,\bm{k}_3;z) &= P_{\mathrm{lin}}\left(k_1,z\right)\int \mathrm{d}M'\, n(M',z)b(M',z)Q_1(k_1,M',z ) \nonumber \\
    & \quad \times \int \mathrm{d}M\, n(M,z) b(M,z)Q_2(k_2,M,z)Q_3(k_3,M,z)\, +\,\mathrm{perms.}
\end{align}
Here and throughout, we have used $P_{\mathrm{hh}}(k_1|M_1,M_2) \approx b(M_1) b(M_2) P_{\mathrm{lin}}(k_1)$, which is a good approximation when dealing with the two-halo term, as discussed above.

In the second equation, we have left implicit all possible permutations of $\{k_1, k_2, k_3\}$. Terms like these will appear in the bispectrum biases to CMB lensing auto- and cross-correlations. Assuming $Q_1$ takes the role of $\kappa$ or $\mathcal{G}$ in those cases, the two integrals in the term we show explicitly factor out under the action of the QE, and the two-halo term can be evaluated by performing a lensing reconstruction on profiles $Q_1$ and $Q_2$ at every $M$. On the other hand, permutations of the two-halo bispectrum where $\kappa$ or $\mathcal{G}$ live in the same halo as one of the foreground profiles can be evaluated by first performing the $M'$-integral over the third profile (the one that lives alone in its halo), feeding this as input to the QE along with the other foreground profile, and finally doing the integral over M.

In these expressions, $n(M,z)$ is the halo mass function, and $b(M,z)$ is the large-scale halo bias; these are explained in more detail in sections~\ref{sec:hmf} and~\ref{sec:halo_bias}, respectively. We evaluate them from functional forms and parameters fit to simulations: the mass function of Ref.~\cite{ref:tinker_et_al_08}, and the bias of Ref.~\cite{ref:tinker_et_al_10}, as implemented in the \texttt{hmvec}\footnote{\url{https://github.com/simonsobs/hmvec}} code.

On the other hand, the subset of trispectra we study includes the one-halo term
\begin{align}
    T_{Q_1 Q_2 Q_3 Q_4}^{\mathrm{1h}}&(\bm{k}_1,\bm{k}_2,\bm{k}_3, \bm{k}_4;z) \nonumber \\
    &=\int \mathrm{d}M\, n(M,z)Q_1(k_1,M,z)Q_2(k_2,M,z)Q_3(k_3,M,z)Q_4(k_4,M,z)\,,
\end{align}
as well as two types of two-halo terms. First, we consider a 2-2 contribution
\begin{align}\label{eqn:T_2halo}
    &T_{Q_1 Q_2 Q_3 Q_4}^{\mathrm{2h},\, 2\text{-}2}(\bm{k}_1,\bm{k}_2,\bm{k}_3, \bm{k}_4;z) \nonumber \\
    &\qquad =P_{\mathrm{lin}}\left(|\bm{k}_1+\bm{k}_2|,z\right)\int \mathrm{d} M' \, n(M',z)b(M',z)Q_1(k_1,M',z )Q_2(k_2,M',z) \nonumber \\
    & \hphantom{\qquad =P_{\mathrm{lin}}(|\bm{k}_1+\bm{k}_2|z }\times \int \mathrm{d}M \,n(M,z)b(M,z)Q_3(k_3,M,z)Q_4(k_4,M,z)\, + \mathrm{perm}\,,
\end{align}
where the permutation is one where $|\bm{k}_1+\bm{k}_2| \leftrightarrow |\bm{k}_3+\bm{k}_4|$ -- the reason we can calculate these terms at all despite them being diagonal-dependent is that, upon applying the QEs, $|\bm{k}_1+\bm{k}_2| = |\bm{k}_3+\bm{k}_4|\rightarrow L$. We also implement a 1-3 contribution of the form
\begin{align}\label{eqn:general_trispec_2h_13}
    &T_{Q_1 Q_2 Q_3 Q_4}^{\mathrm{2h},\, 1\text{-}3}(\bm{k}_1,\bm{k}_2,\bm{k}_3, \bm{k}_4;z) \nonumber \\
    & \qquad =  P_{\mathrm{lin}}(k_1) \int dM\,n(M) b(M_ z) Q_2 (k_2, M, z)  Q_3 (k_3, M, z)  Q_4 (k_4, M, z) \nonumber \\
    & \qquad \qquad  \times
    \int  dM'\,n(M') b(M', z) Q_1 (k_1, M', z)  + {\rm perms.}
\end{align}
To evaluate these contributions, we follow a similar approach to that delineated above for some of the two-halo bispectra, pre-computing the integral over $M'$ before feeding it to one of the QEs at every $M$.

Be it because we want to investigate contributions from halos in a certain mass range, or for reasons of computational feasibility, our mass integrals will be carried out within some limits, $M_{\mathrm{min}}\leq M \leq M_{\mathrm{max}}$, rather than $0\leq M \leq \infty$, effectively ignoring contributions from halos lighter than the cut. Though this will not significantly impact one-halo terms (e.g.,~\cite{ref:mead_et_al_20}), it can lead to problems with many-halo ones: for example, at the two-halo level, the halo model matter power spectrum might no longer approach $P(k\rightarrow0)\approx P_{\mathrm{lin}}(k\rightarrow0)$ on large scales because the consistency relation in equation~\eqref{eqn:consistency_relation} will not be satisfied strictly. A way to get around this issue is to alter the mass function so that the signal from halos below the lower mass limit of the integral, $M_{\mathrm{min}}$, is assumed to come instead from halos with mass exactly $M_{\mathrm{min}}$~\cite{ref:schmidt_16} (see also~\cite{ref:mead_et_al_20}). In effect, this entails replacing, in the two-halo expressions above,
\begin{align}\label{eqn:consistency}
    \mathcal{I}(M_{\mathrm{min}}, M_{\mathrm{max}}; k) \rightarrow \mathcal{I}(M_{\mathrm{min}}, M_{\mathrm{max}}; k) + \left[1 - \mathcal{I}_{m}(M_{\mathrm{min}}, M_{\mathrm{max}}; k\rightarrow0) \right] \frac{Q(k,M,z)}{M_{\mathrm{min}}/\bar{\rho}}\,,
\end{align}
where $\mathcal{I}$ is the integral over a general profile,
\begin{equation}
    \mathcal{I}(M_{\mathrm{lower}}, M_{\mathrm{upper}}; k) \equiv \int_{M_{\mathrm{lower}}}^{M_{\mathrm{upper}}} \mathrm{d}M\, n(M,z)b(M,z)Q(k,M,z)\,,
\end{equation}
and $\mathcal{I}_{m}$ is that  over matter,
\begin{equation}
    \mathcal{I}_{m}(M_{\mathrm{lower}}, M_{\mathrm{upper}}; k) \equiv \int_{M_{\mathrm{lower}}}^{M_{\mathrm{upper}}} \mathrm{d}M\, n(M,z)b(M,z) \frac{M}{\bar{\rho}}\,.
\end{equation}
This correction is unlikely to be significant for the foregrounds we model here, because only halos above a certain minimum mass can form stars or produce a significant tSZ effect; computationally, it should not be a problem to set an $M_{\mathrm{lower}}$ as low as this threshold. However, the correction could be much more important when calculating the two-halo contribution to the bispectrum bias or a CMB lensing cross-correlation with cosmic shear, because in those cases low-mass dark matter halos can play an significant role. This means that it is particularly important to apply the consistency condition to the integral over $M'$ in equation~\ref{eqn:general_prim_bispec_2h} (assuming that $Q_1\rightarrow \kappa$).

Finally, in order to ensure that one-halo terms do not contribute unphysically to the large-scale power, we `damp' the emission profiles at low $k$ when calculating the one-halo contribution; see equation~\eqref{eqn:profile_softening} and the discussion around it. We find this to have a very small effect.

In appendices~\ref{appendix:hm_bispectra} and~\ref{appendix:hm_trispectra}, explicit expressions are provided for the equal-redshift bispectra and trispectra we consider in this work. But first, let us explain how foreground cleaning can be incorporated into the formalism.

\section{Multi-frequency cleaning}\label{appendix:fg_cleaning}
To begin with, we write down the spherical harmonics of the observed temperature field at frequency $\nu$ as 
\begin{equation}
    d^{\nu}_{lm} = s_{lm} + t^{\nu}_{lm} + I^{\nu}_{lm} + n^{\nu}_{lm} \,,
\end{equation}
where $s$ is the primary CMB signal, $t$ is the tSZ contribution, $I$ is the CIB emission, and $n$ is a noise component where we lump every other contribution including instrument noise and foregrounds different from tSZ and CIB. We work in CMB temperature units, so the primary signal is unchanged across frequencies.

Let us denote the ILC weights at frequency $\nu$ and multipole $l$ as $w^{\nu}_l$. We refer the reader to the many good references that exist (e.g.,~\cite{ref:remazeilles_contrained_ilc}) for details on how these are calculated -- suffice it to say that what is needed is the angular auto- and cross- spectra between all the observed channels. With the weights in hand, a foreground-cleaned map can be obtained as
\begin{equation}
    d^{\mathrm{ILC}}_{lm} = \sum_{\nu} w^{\nu}_l d^{\nu}_{lm} \,.
\end{equation}
By construction, $\sum_{\nu} w^{\nu}_l=1$ so as to retain unit response to the primary CMB signal. In contrast, the other components propagate in non-trivial ways through the process.

Consider, the tSZ emission. This component can be factored as the product of a spatial template $y_{lm}$ times a frequency-dependence $g_{\nu}$. After forming the ILC, the tSZ contributes
\begin{equation}\label{eqn:freq_indep_tsz_scaling}
    \sum_{\nu} w^{\nu}_l t^{\nu}_{lm} =  \sum_{\nu} w^{\nu}_l g_{\nu} y_{lm} \equiv  g^{\mathrm{ILC}}_l y_{lm}\,,
\end{equation}
where we have defined the frequency-independent scaling $g^{\mathrm{ILC}}_l$. This means that the contribution from tSZ to the $n$-th order projected polyspectrum after ILC cleaning can be written as
\begin{align}
    \mathcal{P}_{d^{\mathrm{ILC}}_1 \dots d^{\mathrm{ILC}}_n}^{(n)}(\bm{l}_1, \dots, \bm{l}_n) \supset  \sum_{\nu_1 \dots \nu_n} w^{\nu_1}_{l_1} \dots w^{\nu_n}_{l_n}   \mathcal{P}_{t_1 \dots t_n}^{(n)}(\bm{l}_1, \dots, \bm{l}_n) =  g^{\mathrm{ILC}}_{l_1} \dots g^{\mathrm{ILC}}_{l_n} \,\mathcal{P}_{y_1 \dots y_n}^{(n)}(\bm{l}_1, \dots, \bm{l}_n)\,.
\end{align}
This simplification will allow us to calculate lensing biases from tSZ after foreground cleanining with no additional cost once the ILC weights have been generated.

The case of the CIB is more complicated because, unlike the tSZ, its spatial emission is a function of frequency, so it is in general not possible to proceed as above and factor out the frequency dependence. In other words, the CIB contribution to the foreground-cleaned polyspectrum is
\begin{align}
    \mathcal{P}_{d^{\mathrm{ILC}}_1 \dots d^{\mathrm{ILC}}_n}^{(n)}(\bm{l}_1, \dots, \bm{l}_n) \supset  \sum_{\nu_1 \dots \nu_n} w^{\nu_1}_{l_1} \dots w^{\nu_n}_{l_n}   \mathcal{P}_{I^{\nu_1}_1 \dots I^{\nu_n}_n}^{(n)}(\bm{l}_1, \dots, \bm{l}_n) \,.
\end{align}
Nevertheless, we will see large gains in computational efficiency by defining
\begin{equation}
    \sum_{\nu} w^{\nu}_l f_{\nu}^{\mathrm{sat}}(M,z) \equiv w^{\mathrm{sat}}_l (M,z)\,,
\end{equation}
and
\begin{equation}
    \sum_{\nu} w^{\nu}_l f_{\nu}^{\mathrm{cen}}(M,z) \equiv w^{\mathrm{cen}}_l (M,z)
\end{equation}
and casting our calculations in terms of these. To motivate why, consider as an example the 1-halo CIB trispectrum,
\begin{align}
    \mathcal{T}^{1\mathrm{h}}_{I_1 \dots I_4}(\bm{l}_1, \dots, \bm{l}_4) =  \sum_{\nu_1 \dots \nu_n} & w^{\nu_1}_{l_1} \dots w^{\nu_n}_{l_n}   \int  \mathrm{d}z \frac{\mathrm{d}\chi}{\mathrm{d}z} \chi^{-6}(z) W^{4}_{\mathrm{CIB}}(z) \nonumber \\
    & \times \int \mathrm{d}M\, n(M,z) u(k_1^*,M,z)\dots u(k^*_4,M,z) \big[ f_{\nu_1}^{\mathrm{sat}}(M,z)\dots f_{\nu_4}^{\mathrm{sat}}(M,z) \nonumber \\
    & \qquad  + \frac{f_{\nu_1}^{\mathrm{cen}}(M,z)}{u(k_1^*,M,z)}f_{\nu_2}^{\mathrm{sat}}(M,z)\dots f_{\nu_4}^{\mathrm{sat}}(M,z)+ 3\,\mathrm{perms.}\big] \nonumber \\
    =  &   \int  \mathrm{d}z \frac{\mathrm{d}\chi}{\mathrm{d}z} \chi^{-6}(z) W^{4}_{\mathrm{CIB}}(z) \nonumber \\
    & \times \int \mathrm{d}M\, n(M,z) u(k_1^*,M,z)\dots u(k^*_4,M,z) \big[ w^{\mathrm{sat}}_{l_1} (M,z)\dots w^{\mathrm{sat}}_{l_4} (M,z) \nonumber \\
    & \qquad  + \frac{w^{\mathrm{cen}}_{l_1} (M,z)}{u(k_1^*,M,z)}w^{\mathrm{sat}}_{l_2} (M,z)\dots w^{\mathrm{sat}}_{l_4} (M,z)+ 3\,\mathrm{perms.}\big] 
    \,.
\end{align}
where $k_i^* = l_i/\chi(z)$, and the three permutations correspond to switching $l_1$ in the second term with $l_2$ through $l_4$. 

At this point, it is worth remembering that calculating polyspectra is not our end goal. Rather, we are ultimately interested in applying to them quadratic estimators of lensing in the way of equations~\eqref{eqn:prim_bispec_bias} or~\eqref{eqn:trispec_bias}. This means that the polyspectrum configurations of interest are symmetric under exchange of the two legs going into each QE, and also under swapping the two QEs when considering the lensing power spectrum. This allows for simplifications that make the evalution of polyspectra involving the CIB more efficient. In the trispectrum example above, these symmetries imply an equivalence of terms related by a replacement of $(\bm{l}_1, \bm{l}_2)$ with $(\bm{l}_3, \bm{l}_4)$, as well as possibly an exchange of wavevectors within each pair. Consequently, under the action of the QEs, the expression above is equivalent to
\begin{align}
    \mathcal{T}^{1\mathrm{h}}_{I_1 \dots I_4}(\bm{l}_1, \dots, \bm{l}_4) \rightarrow &   \int  \mathrm{d}z \frac{\mathrm{d}\chi}{\mathrm{d}z} \chi^{-6}(z) W^{4}_{\mathrm{CIB}}(z) \int \mathrm{d}M\, n(M,z) u(k_1^*,M,z)\dots u(k^*_4,M,z) \nonumber \\
    & \quad \times \big[ w^{\mathrm{sat}}_{l_1} (M,z)\dots w^{\mathrm{sat}}_{l_4} (M,z)  + 4\frac{w^{\mathrm{cen}}_{l_1} (M,z)}{u(k_1^*,M,z)}w^{\mathrm{sat}}_{l_2} (M,z)\dots w^{\mathrm{sat}}_{l_4} (M,z)\big] 
    \,.
\end{align}
Note that these simplified integrands can also be obtained easily from equation~\eqref{eqn:factorial_moment_relation} (proved in appendix~\ref{appendix:hod_factorial_moments}), after replacing $ \langle \Ncen \rangle \rightarrow w^{\mathrm{cen}}_{l_i} / u(k_i^*,M,z)$ and $ \langle \Nsat \rangle \rightarrow w^{\mathrm{sat}}_{l_i}$.

All in all, the prescription for writing down polyspectra involving the CIB can be summarized as follows. First, use equation~\eqref{eqn:factorial_moment_relation} to relate the factorial moments of the total number of galaxies in a halo to those of satellites and centrals. Then, trivially recast these into single-frequency expressions in terms of $f_{\nu_i}^{\mathrm{cen}}$ and $f_{\nu_i}^{\mathrm{sat}}$. Finally, replace $f_{\nu_i}^{\mathrm{cen}}\rightarrow w^{\mathrm{cen}}_{l_i} $ and $f_{\nu_i}^{\mathrm{sat}}\rightarrow w^{\mathrm{sat}}_{l_i} $ to obtain the foreground-cleaned version.

\section[Halo model for bispectrum biases]{Halo-model for bispectrum biases in auto- and cross-correlation}\label{appendix:hm_bispectra}

We now present the halo model formalism for calculating the bispectra that produce bias to CMB lensing power spectra; at the end of the section, we explain how to extend this to cross-correlations with matter tracers.

We can calculate the relevant matter-tSZ-tSZ bispectrum by following the prescriptions of appendix~\ref{sec:hm_calcs}, but replacing dark matter density profiles with their 3D Compton-$y$ equivalent where needed --- that is, $ u M / \bar{\rho} \rightarrow y_{\mathrm{3D}}$. We find 
\begin{align}
    B_{\kappa y y}^{\mathrm{1h}}(\bm{k}_1,\bm{k}_2,\bm{k}_3;z) &= \int \mathrm{d}M\, n(M,z)\frac{M}{\bar{\rho}}u(k_1,M,z)y_{\mathrm{3D}}(k_2,M,z)y_{\mathrm{3D}}(k_3,M,z)\,,
\end{align}
and
\begin{align}
    B_{\kappa y y}^{\mathrm{2h}}(\bm{k}_1,\bm{k}_2,\bm{k}_3;z) &= P_{\mathrm{lin}}\left(k_1,z\right)\int \mathrm{d}M'\, n(M',z)b(M',z)\frac{M'}{\bar{\rho}}u(k_1,M',z) \nonumber \\
    & \quad \times \int \mathrm{d}M\, n(M,z) b(M,z)y_{\mathrm{3D}}(k_2,M,z)y_{\mathrm{3D}}(k_3,M,z) \nonumber \\
    & + 2 P_{\mathrm{lin}}\left(k_3,z\right)\int \mathrm{d}M'\, n(M',z)b(M',z) y_{\mathrm{3D}}(k_3,M',z)\nonumber \\
    & \quad \times \int \mathrm{d}M\, n(M,z) b(M,z) \frac{M}{\bar{\rho}}u(k_1,M,z)  y_{\mathrm{3D}}(k_2,M,z)\,,
\end{align}
where we have defined $\bar{\rho}\equiv \rho_{\mathrm{m}}(z)$. Everywhere in this work, $u(k,M,z)$ refers to the Fourier-transformed NFW profile, normalised as prescribed by equation~\eqref{eqn:profile_normalisation}, and with concentration $c(M,z)$ obtained from the fitting function of Ref.~\cite{ref:duffy_et_al_08}; see equation~\eqref{eqn:duffy_c} and the discussion around it for details. The subscript notation we have adopted is intended to denote the role of the higher-point function upon projection.

When writing down bispectra and trispectra of the CIB, we will be guided by the prescriptions of sections~\ref{sec:halo_model_cib} and~\ref{sec:halo_model_for_galaxies}. In particular, we will use equation~\eqref{eqn:factorial_moment_relation} (proved in appendix~\ref{appendix:hod_factorial_moments}) to relate the factorial moments of the total number of galaxies in a halo to those of satellites and centrals, which can trivially be translated to expressions in terms of $f^{\mathrm{cen}}$ and $f^{\mathrm{sat}}$. Furthermore, we will follow appendix~\ref{appendix:fg_cleaning} to incorporate also the effects of multi-frequency cleaning, in which case $f^{\mathrm{cen}}$ and $f^{\mathrm{sat}}$ are replaced with the frequency-and-multipole-weighted emissivities $w^{\mathrm{cen}}_l$ and $w^{\mathrm{sat}}_l$.

These tools suggest a one-halo contribution of the form\footnote{Note that the expressions we provide in this and subsequent appendices are not strictly true for a given 3D polyspectrum because they do not include the correct permutations, but they do give the right result under the action of the quadratic estimators and are faster to evaluate.}
\begin{align}
    B_{\kappa I I}^{\mathrm{1h}}(\bm{k}_1,\bm{k}_2,\bm{k}_3;z) &= \int \mathrm{d}M\, n(M,z) \frac{M}{\bar{\rho}} u(k_1,M,z) u(k_2,M,z) u(k_3,M,z) \nonumber \\
    & \hphantom{= \int } \times w^{\mathrm{sat}}_{l^*_{2}}(M,z)  \left[2 \frac{w_{l^*_{3}}^{\mathrm{cen}}(M,z)}{u(k_3,M,z)} + w^{\mathrm{sat}}_{l^*_{3}}(M,z) \right] \,, 
\end{align}
where $l^*_i = k_i  \chi(z)$, and two-halo terms
\begin{align}
    B_{\kappa I I}^{\mathrm{2h}}(\bm{k}_1,\bm{k}_2,\bm{k}_3;z) &= P_{\mathrm{lin}}\left(k_1,z\right)\int \mathrm{d}M'\, n(M',z)b(M',z)\frac{M'}{\bar{\rho}}u(k_1,M',z) \nonumber \\
    & \quad \times \int \mathrm{d}M\, n(M,z) b(M,z)u(k_2,M,z)u(k_3,M,z)\nonumber\\
    & \qquad \times w^{\mathrm{sat}}_{l^*_{2}}(M,z)  \left[2 \frac{w_{l^*_{3}}^{\mathrm{cen}}(M,z)}{u(k_3,M,z)} + w^{\mathrm{sat}}_{l^*_{3}}(M,z) \right] \nonumber \\
    & + 2 P_{\mathrm{lin}}\left(k_3,z\right)\int \mathrm{d}M'\, n(M',z)b(M',z) u(k_3,M',z) \nonumber \\
    & \hphantom{+ 2 P_{\mathrm{lin}}\left(k_3,z\right)\int \mathrm{d}M'}\times \left[\frac{w_{l^*_{3}}^{\mathrm{cen}}(M',z)}{u(k_3,M',z)} + w^{\mathrm{sat}}_{l^*_{3}}(M',z) \right] \nonumber \\
    & \quad \times \int \mathrm{d}M\, n(M,z) b(M,z) \frac{M}{\bar{\rho}}u(k_1,M,z) u(k_2,M,z)\nonumber \\
    &\hphantom{+ 2 P_{\mathrm{lin}}\left(k_3,z\right)\int \mathrm{d}M'}\times  \left[\frac{w_{l^*_{2}}^{\mathrm{cen}}(M,z)}{u(k_2,M,z)} + w^{\mathrm{sat}}_{l^*_{2}}(M,z) \right]\,.
\end{align}
In these expressions, we have assumed that the spatial distribution of galaxies in a host halo follows an NFW profile, as is common in the literature.

For the mixed biases involving both tSZ and CIB legs, we implement
\begin{align}
    B_{\kappa I y}^{\mathrm{1h}}(\bm{k}_1,\bm{k}_2,\bm{k}_3;z) &= \int \mathrm{d}M\, n(M,z) \frac{M}{\bar{\rho}} u(k_1,M,z)u(k_2,M,z)y_{\mathrm{3D}}(k_3,M,z) \nonumber \\
    & \hphantom{= \int \mathrm{d}M\, n(M} \times \left[\frac{w_{l^*_{2}}^{\mathrm{cen}}(M,z)}{u(k_2,M,z)} + w^{\mathrm{sat}}_{l^*_{2}}(M,z) \right] \,, 
\end{align}
and
\begin{align}
    B_{\kappa I y}^{\mathrm{2h}}(\bm{k}_1,\bm{k}_2,\bm{k}_3;z) &= P_{\mathrm{lin}}\left(k_1,z\right)\int \mathrm{d}M'\, n(M',z)b(M',z)\frac{M'}{\bar{\rho}}u(k_1,M',z) \nonumber \\
    & \quad \times \int \mathrm{d}M\, n(M,z) b(M,z)u(k_2,M,z)y_{\mathrm{3D}}(k_3,M,z)\nonumber\\
    & \qquad \times \left[\frac{w_{l^*_{2}}^{\mathrm{cen}}(M,z)}{u(k_2,M,z)} + w^{\mathrm{sat}}_{l^*_{2}}(M,z) \right]\nonumber \\
    &+ P_{\mathrm{lin}}\left(k_2,z\right)\int \mathrm{d}M'\, n(M',z)b(M',z) u(k_2,M',z) \nonumber \\
    & \hphantom{P_{\mathrm{lin}}\left(k_2,z\right)\int \mathrm{d}M'\,} \times\left[\frac{w_{l^*_{2}}^{\mathrm{cen}}(M',z)}{u(k_2,M',z)} + w^{\mathrm{sat}}_{l^*_{2}}(M',z) \right] \nonumber \\
    & \quad \times \int \mathrm{d}M\, n(M,z) b(M,z)\frac{M}{\bar{\rho}}u(k_1,M,z)y_{\mathrm{3D}}(k_3,M,z)\nonumber\\
    &+
    P_{\mathrm{lin}}\left(k_3,z\right)\int \mathrm{d}M'\, n(M',z)b(M',z) y_{\mathrm{3D}}(k_3,M',z) \nonumber \\
    & \quad \times \int \mathrm{d}M\, n(M,z) b(M,z)\frac{M}{\bar{\rho}}u(k_1,M,z)u(k_2,M,z) \nonumber \\
    & \hphantom{P_{\mathrm{lin}}\left(k_2,z\right)\int \mathrm{d}M'\,} \times\left[\frac{w_{l^*_{2}}^{\mathrm{cen}}(M,z)}{u(k_2,M,z)} + w^{\mathrm{sat}}_{l^*_{2}}(M,z) \right] 
    \,.
\end{align}
When evaluating equation~\eqref{eqn:prim_bispec_bias}, a permutation factor of 2 needs to be tacked onto these terms to account for the symmetry under exchange of the two QE legs. This is in addition to the factor of 2 coming from the exchange of the two QEs, which is already included in equation~\eqref{eqn:prim_bispec_bias}.

Some very simple replacements in the expressions above will allow us to calculate biases to CMB lensing cross-correlations with a matter tracer, which we will call $\mathcal{G}$. The starting point will be the discussion of the halo model for galaxies introduced in section~\ref{sec:halo_model_for_galaxies}. For now, we consider the case where this matter tracer is a galaxy survey which we parametrize via a halo occupation distribution (HOD); this could be easily generalized to other tracers, such as galaxy weak lensing maps, and we intend to do so in future. To obtain $B_{\mathcal{G} y y}$, $B_{\mathcal{G} I I}$ and $B_{\mathcal{G} I y}$, we simply replace $\kappa \rightarrow \mathcal{G}$ and $ M/\bar{\rho} \rightarrow \langle \Ngal\rangle/\bar{n}_{\mathrm{gal}}$ in the expressions above. We calculate $\langle \Ngal\rangle$ and $\bar{n}_{\mathrm{gal}}$ using the HOD infrastructure in \texttt{hmvec}, implemented following~\cite{ref:smith_et_al_18}. This is built around the galaxy HOD model of~\cite{ref:leauthaud_et_al_11a, ref:leauthaud_11b}, extended to $z\sim4$ using the stellar mass to halo mass relation from~\cite{ref:behroozi_et_al_10} -- altogether, the baseline model of~\cite{ref:hearin_et_al_16}.

In this framework, galaxy samples can be specified from a user-provided redshift distribution and either a  minimum-stellar-mass threshold, or a comoving number density of galaxies, at every redshift.  In the latter case, the underlying \texttt{hmvec} code converts comoving number density to minimum stellar mass thresholds for centrals and satellites at every redshift\footnote{ The latter step is performed as a bisection search for the input of the inverse function -- that is, the function that maps minimum mass thresholds to galaxy number density, a straight-forward computation which is one-to-one and monotonic -- that returns the correct galaxy number density.}. These thresholds then set the mean number of galaxies of each type in halos of a given mass (see appendix B.3 of~\cite{ref:munchmeyer_et_al_19} for details). This simple prescription works well for highly complete galaxy samples such as the DESI LRGs of Ref.~\cite{ref:zhou_et_al_22}. Extending it to accommodate more complex selections is left to future work.

\section[Halo model for trispectrum biases]{Halo-model for trispectrum biases}\label{appendix:hm_trispectra}
Let us start with the pure-tSZ trispectrum. We implement a one-halo contribution
\begin{align}
    T_{y y y y}^{\mathrm{1h}}&(\bm{k}_1,\bm{k}_2,\bm{k}_3, \bm{k}_4;z) \nonumber \\
    &=\int \mathrm{d}M\, n(M,z)y_{\mathrm{3D}}(k_1,M,z)y_{\mathrm{3D}}(k_2,M,z)y_{\mathrm{3D}}(k_3,M,z)y_{\mathrm{3D}}(k_4,M,z)\,. 
\end{align}
As per the two-halo terms, we calculate a 1-3 term
\begin{align}
    T_{y y y y}^{\mathrm{2h},\, 1\text{-}3}&(\bm{k}_1,\bm{k}_2,\bm{k}_3, \bm{k}_4;z) \nonumber \\
    &=4 P_{\mathrm{lin}}(k_1, z) \int dM \,n(M, z) b(M, z) y_{\mathrm{3D}} (k_1, M, z) \nonumber \\
    & \qquad \times \int dM'\,n(M', z)  b(M', z) \,y_{\mathrm{3D}} (k_2, M', z) y_{\mathrm{3D}} (k_3, M', z) y_{\mathrm{3D}} (k_4, M', z)  \,,
\end{align}
(note the permutation factor, which comes from the four choices of $k_i$ that can be associated with the pressure profile that goes by itself in one of the haloes) and also a 2-2 term
\begin{align}
    T_{y y y y}^{\mathrm{2h},\, 2\text{-}2}&(\bm{k}_1,\bm{k}_2,\bm{k}_3, \bm{k}_4;z) \nonumber \\
    &=2 P_{\mathrm{lin}}\left(|\bm{k}_1+\bm{k}_2|,z\right)\int \mathrm{d} M' \, n(M',z)b(M',z)y_{\mathrm{3D}}(k_1,M',z)y_{\mathrm{3D}}(k_2,M',z) \nonumber \\
    & \hphantom{=P_{\mathrm{lin}}(|\bm{k}_1 }\times \int \mathrm{d}M \,n(M,z)b(M,z)y_{\mathrm{3D}}(k_3,M,z)y_{\mathrm{3D}}(k_4,M,z)\,,
\end{align}
where now the permutation factor comes from swapping $\bm{k}_1+\bm{k}_2\rightarrow \bm{k}_3+\bm{k}_4$.

For the CIB-only case, we calculate a one-halo term
\begin{align}
    T_{I I I I}^{\mathrm{1h}}&(\bm{k}_1,\bm{k}_2,\bm{k}_3, \bm{k}_4;z) \nonumber \\
    &=\int \mathrm{d}M\, n(M,z) u(k_1,M,z)u(k_2,M,z)u(k_3,M,z)u(k_4,M,z) \nonumber \\
    & \qquad \times w^{\mathrm{sat}}_{l^*_{1}}(M,z) w^{\mathrm{sat}}_{l^*_{2}}(M,z) w^{\mathrm{sat}}_{l^*_{3}}(M,z) \left[4 \frac{w_{l^*_{4}}^{\mathrm{cen}}(M,z)}{u(k_4,M,z)} + w^{\mathrm{sat}}_{l^*_{4}}(M,z) \right]\,,
\end{align}
where $l^*_i = k_i  \chi(z)$, and a 1-3 two-halo term
\begin{align}
    T_{I I I I}^{\mathrm{2h},\, 1\text{-}3}&(\bm{k}_1,\bm{k}_2,\bm{k}_3, \bm{k}_4;z) \nonumber \\
    &=4 P_{\mathrm{lin}}(k_1, z) \int dM \,n(M, z) b(M, z) u(k_1,M,z) \left[\frac{w_{l^*_{1}}^{\mathrm{cen}}(M,z)}{u(k_1,M,z)} + w^{\mathrm{sat}}_{l^*_{1}}(M,z) \right] \nonumber \\
    & \qquad \times \int dM'\,n(M', z)  b(M', z) \,u(k_2,M',z)u(k_3,M',z)u(k_4,M',z) \nonumber \\
    & \qquad \qquad \times \bigg\{ w^{\mathrm{sat}}_{l^*_{3}}(M',z) w^{\mathrm{sat}}_{l^*_{4}}(M',z) \left[\frac{w_{l^*_{2}}^{\mathrm{cen}}(M',z)}{u(k_2,M',z)} \right] \nonumber \\
    & \qquad \qquad \qquad +  w^{\mathrm{sat}}_{l^*_{2}}(M',z) w^{\mathrm{sat}}_{l^*_{3}}(M',z) \left[2 \frac{w_{l^*_{4}}^{\mathrm{cen}}(M',z)}{u(k_4,M',z)} + w^{\mathrm{sat}}_{l^*_{4}}(M',z) \right] \bigg\}  \,.
\end{align}
As before, the permutation factor of 4 comes from choosing which $l^*$ goes alone in the halo with mass $M$. Once this is fixed, there remains a choice of whether the contribution from this halo enters a QE accompanied by a central or a satellite. When it's a satellite, there are two remaining ways of putting the central in the other QE (last line). We also impletement a 2-2 term
\begin{align}
    T_{I I I I}^{\mathrm{2h},\, 2\text{-}2}&(\bm{k}_1,\bm{k}_2,\bm{k}_3, \bm{k}_4;z) \nonumber \\
    &=2 P_{\mathrm{lin}}\left(|\bm{k}_1+\bm{k}_2|,z\right)\int \mathrm{d} M' \, n(M',z)b(M',z)u(k_1,M',z)u(k_2,M',z) \nonumber \\
    & \hphantom{P_{\mathrm{lin}}\left(|\bm{k}_1+\bm{k}_2|,z\right)\int \mathrm{d} M'} \times w^{\mathrm{sat}}_{l^*_{1}}(M',z)  \left[2 \frac{w_{l^*_{2}}^{\mathrm{cen}}(M',z)}{u(k_2,M',z)} + w^{\mathrm{sat}}_{l^*_{2}}(M',z) \right] \nonumber \\
    & \hphantom{=P_{\mathrm{lin}}(|\bm{k}_1-\bm{k}_2|z }\times \int \mathrm{d}M \,n(M,z)b(M,z)u(k_3,M,z)u(k_4,M,z)\nonumber \\
    & \hphantom{P_{\mathrm{lin}}\left(|\bm{k}_1+\bm{k}_2|,z\right)\int \mathrm{d} M'} \times w^{\mathrm{sat}}_{l^*_{3}}(M,z)  \left[2 \frac{w_{l^*_{4}}^{\mathrm{cen}}(M,z)}{u(k_4,M,z)} + w^{\mathrm{sat}}_{l^*_{4}}(M,z) \right]\,.
\end{align}

We can also calculate mixed biases involving both tSZ and CIB emission. The structure of the calculation, given by~\eqref{eqn:trispec_bias}, is such that we have two quadratic estimators acting on trispectra of the form $T_{abcd}$: the first estimator taking in legs $a$ and $b$, the second $c$ and $d$. We therefore have a symmetry under the exchange of pair $(a,b)$ with $(c,d)$, as well as to exchanges of legs within each pair. The total contribution from tSZ and CIB to the CMB trispectrum can therefore be written schematically as
\begin{align}\label{eqn:trispec_perms}
    Q^2[T^{\mathrm{tot}}] = Q^2[T_{IIII} + T_{yyyy} + 4T_{Iyyy} + 2T_{IIyy} + 4T_{IyIy} + 4T_{yIII}]\,,
\end{align}
where $Q^{2}$ denotes the action of the two quadratic estimators. Note that trispectra featuring two foregrounds of each type have been separated into two terms, depending on whether a given QE takes in foregrounds of a single or both kinds.

%The way we evalute these biases -- recall, by `pushing' the lensing reconstructions inside the redshift and mass integrals -- lets us simplify the expressions significantly.

We implement a subset of the terms above. These include a one-halo term of the form
\begin{align}
    T_{I y y y}^{\mathrm{1h}}&(\bm{k}_1,\bm{k}_2,\bm{k}_3, \bm{k}_4;z) \nonumber \\
    &=\int \mathrm{d}M\, n(M,z) u(k_1,M,z)y_{\mathrm{3D}}(k_2,M,z)y_{\mathrm{3D}}(k_3,M,z)y_{\mathrm{3D}}(k_4,M,z) \nonumber \\
    & \qquad \times \left[\frac{w_{l^*_{1}}^{\mathrm{cen}}(M,z)}{u(k_1,M,z)} + w^{\mathrm{sat}}_{l^*_{1}}(M,z) \right]\,,
\end{align}
a 1-3 two-halo term
\begin{align}
    T_{I y y y}^{\mathrm{2h},\, 1\text{-}3}&(\bm{k}_1,\bm{k}_2,\bm{k}_3, \bm{k}_4;z) \nonumber \\
    &=P_{\mathrm{lin}}(k_1, z) \int dM \,n(M, z) b(M, z) u(k_1,M,z) \left[\frac{w_{l^*_{1}}^{\mathrm{cen}}(M,z)}{u(k_1,M,z)} + w^{\mathrm{sat}}_{l^*_{1}}(M,z) \right]  \nonumber \\
    & \qquad \times \int dM'\,n(M', z)  b(M', z) \,y_{\mathrm{3D}} (k_2, M', z) y_{\mathrm{3D}} (k_3, M', z) y_{\mathrm{3D}} (k_4, M', z) \nonumber \\
    & \quad + 3 P_{\mathrm{lin}}(k_4, z) \int dM \,n(M, z) b(M, z) y_{\mathrm{3D}} (k_4, M, z) \nonumber \\
    & \qquad \times \int dM'\,n(M', z)  b(M', z) \,u(k_1,M',z)  y_{\mathrm{3D}}(k_2,M',z)y_{\mathrm{3D}}(k_3,M',z)\nonumber \\
    & \qquad \qquad \times  \left[\frac{w_{l^*_{1}}^{\mathrm{cen}}(M',z)}{u(k_1,M',z)} + w^{\mathrm{sat}}_{l^*_{1}}(M',z) \right]\,.
\end{align}
and a 2-2 contribution
\begin{align}
    T_{I y y y}^{\mathrm{2h},\, 2\text{-}2}&(\bm{k}_1,\bm{k}_2,\bm{k}_3, \bm{k}_4;z) \nonumber \\
    &=2 P_{\mathrm{lin}}\left(|\bm{k}_1+\bm{k}_2|,z\right)\int \mathrm{d} M' \, n(M',z)b(M',z)u(k_1,M',z)y_{\mathrm{3D}}(k_2,M',z) \nonumber \\
    & \hphantom{P_{\mathrm{lin}}\left(|\bm{k}_1+\bm{k}_2|,z\right)\int \mathrm{d} M'} \times \left[\frac{w_{l^*_{1}}^{\mathrm{cen}}(M',z)}{u(k_1,M',z)} + w^{\mathrm{sat}}_{l^*_{1}}(M',z) \right] \nonumber \\
    & \hphantom{=P_{\mathrm{lin}}(|\bm{k}_1-\bm{k}_2|z }\times \int \mathrm{d}M \,n(M,z)b(M,z)y_{\mathrm{3D}}(k_3,M,z)y_{\mathrm{3D}}(k_4,M,z)\,.
\end{align}
As seen from equation~\eqref{eqn:trispec_perms}, these $T_{I y y y}$ terms feature with an additional permutation factor of 4.

Another coupling is
\begin{align}
    T_{I I y y}^{\mathrm{1h}}&(\bm{k}_1,\bm{k}_2,\bm{k}_3, \bm{k}_4;z) \nonumber \\
    &=\int \mathrm{d}M\, n(M,z) u(k_1,M,z)u(k_2,M,z)y_{\mathrm{3D}}(k_3,M,z)y_{\mathrm{3D}}(k_4,M,z) \nonumber \\
    & \qquad \times w^{\mathrm{sat}}_{l^*_{1}}(M,z)  \left[2 \frac{w_{l^*_{2}}^{\mathrm{cen}}(M,z)}{u(k_2,M,z)} + w^{\mathrm{sat}}_{l^*_{2}}(M,z) \right]\,,
\end{align}
with a 1-3 two-halo contribution
\begin{align}
    T_{I I y y}^{\mathrm{2h},\, 1\text{-}3}&(\bm{k}_1,\bm{k}_2,\bm{k}_3, \bm{k}_4;z) \nonumber \\
    &= 2 P_{\mathrm{lin}}(k_1, z) \int dM \,n(M, z) b(M, z) u(k_1,M,z) \left[\frac{w_{l^*_{1}}^{\mathrm{cen}}(M,z)}{u(k_1,M,z)} + w^{\mathrm{sat}}_{l^*_{1}}(M,z) \right]  \nonumber \\
    & \qquad \times \int dM'\,n(M', z)  b(M', z) \,u(k_2,M,z)  y_{\mathrm{3D}} (k_3, M', z) y_{\mathrm{3D}} (k_4, M', z) \nonumber \\
    & \qquad \qquad \times \left[\frac{w_{l^*_{2}}^{\mathrm{cen}}(M',z)}{u(k_2,M',z)} + w^{\mathrm{sat}}_{l^*_{2}}(M',z) \right] \\
    & \quad + 2 P_{\mathrm{lin}}(k_4, z) \int dM \,n(M, z) b(M, z) y_{\mathrm{3D}} (k_4, M, z) \nonumber \\
    & \qquad \times \int dM'\,n(M', z)  b(M', z) \,u(k_1,M',z)  u(k_2,M',z)y_{\mathrm{3D}}(k_3,M',z)\nonumber \\
    & \qquad \qquad \times  w^{\mathrm{sat}}_{l^*_{1}}(M',z)  \left[2 \frac{w_{l^*_{2}}^{\mathrm{cen}}(M',z)}{u(k_2,M',z)} + w^{\mathrm{sat}}_{l^*_{2}}(M',z) \right]\,.
\end{align}
and a 2-2 term
\begin{align}
    T_{I I y y}^{\mathrm{2h},\, 2\text{-}2}&(\bm{k}_1,\bm{k}_2,\bm{k}_3, \bm{k}_4;z) \nonumber \\
    &=2 P_{\mathrm{lin}}\left(|\bm{k}_1+\bm{k}_2|,z\right)\int \mathrm{d} M' \, n(M',z)b(M',z)u(k_1,M',z)u(k_2,M',z) \nonumber \\
    & \hphantom{P_{\mathrm{lin}}\left(|\bm{k}_1+\bm{k}_2|,z\right)\int \mathrm{d} M'} \times w^{\mathrm{sat}}_{l^*_{1}}(M',z)  \left[2 \frac{w_{l^*_{1}}^{\mathrm{cen}}(M',z)}{u(k_1,M,z)} + w^{\mathrm{sat}}_{l^*_{1}}(M',z) \right] \nonumber \\
    & \hphantom{=P_{\mathrm{lin}}(|\bm{k}_1-\bm{k}_2|z }\times \int \mathrm{d}M \,n(M,z)b(M,z)y_{\mathrm{3D}}(k_3,M,z)y_{\mathrm{3D}}(k_4,M,z)\,.
\end{align}
These $T_{I I y y}$ terms contribute to the trispectrum bias with a permutation factor of 2.

Also,
\begin{align}
    T_{I y I y}^{\mathrm{1h}}&(\bm{k}_1,\bm{k}_2,\bm{k}_3, \bm{k}_4;z) \nonumber \\
    &=\int \mathrm{d}M\, n(M,z) u(k_1,M,z)y_{\mathrm{3D}}(k_2,M,z)u(k_3,M,z)y_{\mathrm{3D}}(k_4,M,z) \nonumber \\
    & \qquad \times w^{\mathrm{sat}}_{l^*_{1}}(M,z)  \left[2 \frac{w_{l^*_{3}}^{\mathrm{cen}}(M,z)}{u(k_3,M,z)} + w^{\mathrm{sat}}_{l^*_{3}}(M,z) \right]\,,
\end{align}
and
\begin{align}
    T_{I y I y}^{\mathrm{2h},\, 1\text{-}3}&(\bm{k}_1,\bm{k}_2,\bm{k}_3, \bm{k}_4;z) \nonumber \\
    &= 2 P_{\mathrm{lin}}(k_1, z) \int dM \,n(M, z) b(M, z) u(k_1,M,z) \left[\frac{w_{l^*_{1}}^{\mathrm{cen}}(M,z)}{u(k_1,M,z)} + w^{\mathrm{sat}}_{l^*_{1}}(M,z) \right]  \nonumber \\
    & \qquad \times \int dM'\,n(M', z)  b(M', z) y_{\mathrm{3D}} (k_2, M', z) u(k_3,M',z) y_{\mathrm{3D}} (k_4, M', z) \nonumber \\
    & \qquad \qquad \times \left[\frac{w_{l^*_{3}}^{\mathrm{cen}}(M',z)}{u(k_3,M',z)} + w^{\mathrm{sat}}_{l^*_{3}}(M',z) \right] \\
    & \quad + 2 P_{\mathrm{lin}}(k_4, z) \int dM \,n(M, z) b(M, z) y_{\mathrm{3D}} (k_4, M, z) \nonumber \\
    & \qquad \times \int dM'\,n(M', z)  b(M', z) \,u(k_1,M',z)  y_{\mathrm{3D}}(k_2,M',z) u(k_3,M',z) \nonumber \\
    & \qquad \qquad \times  w^{\mathrm{sat}}_{l^*_{1}}(M',z)  \left[2 \frac{w_{l^*_{3}}^{\mathrm{cen}}(M',z)}{u(k_3,M',z)} + w^{\mathrm{sat}}_{l^*_{3}}(M',z) \right]\,.
\end{align}
and
\begin{align}
    T_{I y I y}^{\mathrm{2h},\, 2\text{-}2}&(\bm{k}_1,\bm{k}_2,\bm{k}_3, \bm{k}_4;z) \nonumber \\
    &=2 P_{\mathrm{lin}}\left(|\bm{k}_1+\bm{k}_2|,z\right)\int \mathrm{d} M' \, n(M',z)b(M',z)u(k_1,M',z)y_{\mathrm{3D}}(k_2,M',z) \nonumber \\
    & \hphantom{P_{\mathrm{lin}}\left(|\bm{k}_1+\bm{k}_2|,z\right)\int \mathrm{d} M'} \times \left[\frac{w_{l^*_{1}}^{\mathrm{cen}}(M',z)}{u(k_1,M',z)} + w^{\mathrm{sat}}_{l^*_{1}}(M',z) \right] \nonumber \\
    & \hphantom{=P_{\mathrm{lin}}(|\bm{k}_1-\bm{k}_2|z }\times \int \mathrm{d}M \,n(M,z)b(M,z)u(k_3,M,z)y_{\mathrm{3D}}(k_4,M,z) \nonumber\\
    & \hphantom{P_{\mathrm{lin}}\left(|\bm{k}_1+\bm{k}_2|,z\right)\int \mathrm{d} M} \times \left[\frac{w_{l^*_{3}}^{\mathrm{cen}}(M,z)}{u(k_3,M,z)} + w^{\mathrm{sat}}_{l^*_{3}}(M,z) \right] \,.
\end{align}
These $T_{I y I y}$ terms ultimately appear with a permutation factor of 4.

Finally, we have
\begin{align}
    T_{y I I I}^{\mathrm{1h}}&(\bm{k}_1,\bm{k}_2,\bm{k}_3, \bm{k}_4;z) \nonumber \\
    &=\int \mathrm{d}M\, n(M,z) y_{\mathrm{3D}}(k_1,M,z)u(k_2,M,z)u(k_3,M,z)u(k_4,M,z) \nonumber \\
    & \qquad \times  \bigg\{w^{\mathrm{sat}}_{l^*_{3}}(M,z) w^{\mathrm{sat}}_{l^*_{4}}(M,z) \left[\frac{w_{l^*_{2}}^{\mathrm{cen}}(M,z)}{u(k_2,M,z)}\right] \nonumber \\
    & \qquad \qquad +  w^{\mathrm{sat}}_{l^*_{2}}(M,z) w^{\mathrm{sat}}_{l^*_{3}}(M,z) \left[2 \frac{w_{l^*_{4}}^{\mathrm{cen}}(M,z)}{u(k_4,M,z)} + w^{\mathrm{sat}}_{l^*_{4}}(M,z) \right] \bigg\}\,.
\end{align}
There is also a 1-3 term
\begin{align}
    T_{y I I I}^{\mathrm{2h},\, 1\text{-}3}&(\bm{k}_1,\bm{k}_2,\bm{k}_3, \bm{k}_4;z) \nonumber \\
    &= P_{\mathrm{lin}}(k_1, z) \int dM \,n(M, z) b(M, z) y_{\mathrm{3D}}(k_1,M,z)  \nonumber \\
    & \qquad \times \int dM'\,n(M', z)  b(M', z) \,u(k_2,M',z)  u(k_3,M',z) u(k_4,M',z) \nonumber \\
    & \qquad \qquad \times  \bigg\{w^{\mathrm{sat}}_{l^*_{3}}(M,z) w^{\mathrm{sat}}_{l^*_{4}}(M,z) \left[\frac{w_{l^*_{2}}^{\mathrm{cen}}(M,z)}{u(k_2,M,z)}\right] \nonumber \\
    & \qquad \qquad \qquad +  w^{\mathrm{sat}}_{l^*_{2}}(M,z) w^{\mathrm{sat}}_{l^*_{3}}(M,z) \left[2 \frac{w_{l^*_{4}}^{\mathrm{cen}}(M,z)}{u(k_4,M,z)} + w^{\mathrm{sat}}_{l^*_{4}}(M,z) \right] \bigg\} \\
    & \quad + P_{\mathrm{lin}}(k_2, z) \int dM \,n(M, z) b(M, z) u(k_2,M,z) \left[\frac{w_{l^*_{2}}^{\mathrm{cen}}(M,z)}{u(k_2,M,z)} + w^{\mathrm{sat}}_{l^*_{2}}(M,z) \right]  \nonumber \\
    & \qquad \times \int dM'\,n(M', z)  b(M', z) \,y_{\mathrm{3D}}(k_1,M',z)  u(k_3,M',z) u(k_4,M',z) \nonumber \\
    & \qquad \qquad \times  w^{\mathrm{sat}}_{l^*_{3}}(M',z) \left[2 \frac{w_{l^*_{4}}^{\mathrm{cen}}(M',z)}{u(k_4,M',z)} + w^{\mathrm{sat}}_{l^*_{4}}(M',z) \right]\\
    & \quad + 2 P_{\mathrm{lin}}(k_4, z) \int dM \,n(M, z) b(M, z) u(k_4,M,z) \left[\frac{w_{l^*_{4}}^{\mathrm{cen}}(M,z)}{u(k_4,M,z)} + w^{\mathrm{sat}}_{l^*_{4}}(M,z) \right]  \nonumber \\
    & \qquad \times \int dM'\,n(M', z)  b(M', z) \,y_{\mathrm{3D}}(k_1,M',z)  u(k_2,M',z) u(k_3,M',z) \nonumber \\
    & \qquad \qquad \times   \left[ w^{\mathrm{sat}}_{l^*_{2}}(M',z)\frac{w_{l^*_{3}}^{\mathrm{cen}}(M',z)}{u(k_3,M',z)} + w^{\mathrm{sat}}_{l^*_{3}}(M',z)\frac{w_{l^*_{2}}^{\mathrm{cen}}(M',z)}{u(k_2,M',z)} + w^{\mathrm{sat}}_{l^*_{2}}(M',z) w^{\mathrm{sat}}_{l^*_{3}}(M',z) \right] \,.
\end{align}
In the first of the three terms, halo $M$ contains the tSZ profile. We further split this into contributions where the integral over $M$ enters the QE along with a central or satellites living in halo $M'$. On the other hand, there are two other terms where halo $M$ contains a CIB profile: one where this enters the same QE as the tSZ profile, and one where in goes in the other QE along with another CIB profile from $M'$. Finally, there is a 2-2 term
\begin{align}
    T_{y I I I}^{\mathrm{2h},\, 2\text{-}2}&(\bm{k}_1,\bm{k}_2,\bm{k}_3, \bm{k}_4;z) \nonumber \\
    &=2 P_{\mathrm{lin}}\left(|\bm{k}_1+\bm{k}_2|,z\right)\int \mathrm{d} M' \, n(M',z)b(M',z)y_{\mathrm{3D}}(k_1,M',z)u(k_2,M',z) \nonumber \\
    & \hphantom{P_{\mathrm{lin}}\left(|\bm{k}_1+\bm{k}_2|,z\right)\int \mathrm{d} M'} \times \left[\frac{w_{l^*_{2}}^{\mathrm{cen}}(M',z)}{u(k_2,M',z)} + w^{\mathrm{sat}}_{l^*_{2}}(M',z) \right] \nonumber \\
    & \hphantom{=P_{\mathrm{lin}}(|\bm{k}_1-\bm{k}_2|z }\times \int \mathrm{d}M \,n(M,z)b(M,z)u(k_3,M,z)u(k_4,M,z) \nonumber\\
    & \hphantom{P_{\mathrm{lin}}\left(|\bm{k}_1+\bm{k}_2|,z\right)\int \mathrm{d} M} \times  w^{\mathrm{sat}}_{l^*_{3}}(M,z)  \left[2 \frac{w_{l^*_{4}}^{\mathrm{cen}}(M,z)}{u(k_4,M,z)} + w^{\mathrm{sat}}_{l^*_{4}}(M,z) \right]\,.
\end{align}
These $T_{y I I I}$ terms enter with a permutation factor of 4.

\section[Constructing a curved-sky, first-order, lensed $T$ template]{Constructing a curved-sky, first-order, lensed temperature template}\label{appendix:curved_sky_T_template}

We describe here the construction of a map-level template for the first-order lensing correction to the temperature anisotropies, working in the curved-sky formalism (see, e.g.,~\cite{ref:hu_2000}). Ultimately, we arrive at a fast position-space implementation which is made publicly-available on \texttt{GitHub}\footnote{\texttt{https://github.com/abaleato/curved\_sky\_B\_template}}. 

Our goal is to compute\footnote{In this notation summation is implicit over matching pairs of indices.}
\begin{align}
    \tilde{T}_{lm} &= \sum_{(lm)_1} \sum_{(lm)_2} \phi_{(lm)_1} T_{(lm)_2} I_{l l_1 l_2}^{m m_1 m_2}\nonumber\\
    &= \frac{(-1)^m}{2} \sum_{(lm)_1} \sum_{(lm)_2} [l_1(l_1+1)+l_2(l_2+1)-l(l+1)]\sqrt{\frac{(2l+1)(2l_1+1)(2l_2+1)}{4\pi}} \nonumber\\
    & \quad \times \phi_{(lm)_1} T_{(lm)_2} \begin{pmatrix}l_1&l_2&l \\ m_1&m_2&-m \end{pmatrix} \begin{pmatrix}l_1&l_2&l \\ 0&0&0 \end{pmatrix}\,.
    \label{eq:harmonic_firstorder}
\end{align}
Following the implementation of the quadratic estimators of lensing in the \texttt{QuickLens} code, we write
\begin{align}
    \hat{\tilde{T}}_{lm} = \frac{(-1)^{m}}{2} \sum_{(lm)_1}\sum_{(lm)_2} \begin{pmatrix}l_1&l_2&l \\ m_1&m_2&-m \end{pmatrix} Z_{l_1 l_2 l} \hat{\phi}_{(lm)_1} \hat{T}_{(lm)_2}\,,
\end{align}
where the weights
\begin{align}
    Z_{l_1 l_2 l}  = & [l_1(l_1+1)+l_2(l_2+1)-l(l+1)]
    \sqrt{\frac{(2l+1)(2l_1+1)(2l_2+1)}{4\pi}}   \begin{pmatrix}l_1&l_2&l \\ 0&0&0 \end{pmatrix}
\end{align}
can be separated into three terms
\begin{align}
    Z_{l_1 l_2 l} = \sum_i Z^{i}_{l_1 l_2 l}\,,
\end{align}
with 
\begin{align}
    Z^{i}_{l_1 l_2 l} = \sqrt{\frac{(2l+1)(2l_1+1)(2l_2+1)}{4\pi}} \begin{pmatrix}l_1&l_2&l \\ 0&0&0 \end{pmatrix} z^i_{l_1} z^i_{l_2} z^i_{l}\,.
\end{align}
The value for the separable weights $z^i_{l_j}$ can be found in Table~\ref{tab:weights_for_temp}. 

We note that this template may also be evaluated (perhaps more efficiently) in real space as $\boldsymbol{\nabla}\phi \cdot \boldsymbol{\nabla}T$, which can be written as products of spin $\pm 1$ fields with multipoles proportional to $\phi_{lm}$ and $T_{lm}$.

\begin{table}
    \centering
    \begin{tabular}{c c  c  c }
        \toprule
        $i$& $z_{l_1}^i$ &$z_{l_2}^i$ & $z_{l}^i$\\ \midrule
        $1$ & $l_1(l_1+1)$ & $1$ & $1$ \\ 
        $2$ & $1$ & $l_2(l_2+1)$ & $1$ \\ 
        $3$ &  $-1$ & $1$ & $l(l+1)$ \\ 
        \bottomrule
    \end{tabular}
    \caption{Weights for a fast separable calculation of the first-order lensing correction to the CMB temperature anisotropies}\label{tab:weights_for_temp}
\end{table}

\bibliographystyle{JHEP}
\bibliography{main}
\end{document}